\documentclass[fleqn,usenatbib]{mnras}

\usepackage{amssymb}
\usepackage{newtxtext,newtxmath}
\usepackage{ upgreek }
\usepackage{ amssymb }
\usepackage{dsfont}
\usepackage{mathrsfs}
\usepackage{comment}
\usepackage{float}
\usepackage{subfigure}
\usepackage[T1]{fontenc}

\DeclareRobustCommand{\VAN}[3]{#2}
\let\VANthebibliography\thebibliography
\def\thebibliography{\DeclareRobustCommand{\VAN}[3]{##3}\VANthebibliography}

\usepackage{graphicx}	% Including figure files
\usepackage{amsmath}	% Advanced maths commands

\DeclareRobustCommand{\loongrightarrow}{%
  \DOTSB\relbar\joinrel\relbar\joinrel\relbar\joinrel\rightarrow
}

\DeclareRobustCommand{\looongrightarrow}{%
  \DOTSB\relbar\joinrel\relbar\joinrel\relbar\joinrel\relbar\joinrel\relbar\joinrel\relbar\joinrel\relbar\joinrel\relbar\joinrel\relbar\joinrel\rightarrow
}

\renewcommand\vec{\mathbf}
\title[Jets with finite Poynting flux]{On the collimation properties of jets with finite Poynting flux launched from Keplerian accretion discs}

\author[T. Jannaud et al.]{
Thomas Jannaud,$^{1,2}$\thanks{E-mail: taj30@cam.ac.uk}
Jonathan Ferreira,$^{2}$
and Claudio Zanni$^{3}$
\\
$^{1}$DAMTP, University of Cambridge, CMS, Wilberforce Road, Cambridge CB3 0WA, United Kingdom\\
$^{2}$Univ. Grenoble Alpes, CNRS, IPAG, 38000 Grenoble, France\\
$^{3}$INAF - Osservatorio Astrofisico di Torino, Strada Osservatorio 20, Pino Torinese 10025, Italy
}

\date{Accepted XXX. Received YYY; in original form ZZZ}

\pubyear{\the\year{}}

\begin{document}
\label{firstpage}
\pagerange{\pageref{firstpage}--\pageref{lastpage}}
\maketitle

\defcitealias{jannaud2023}{JZF23}

\begin{abstract}
It is generally accepted that the launching of astrophysical jets requires a large-scale magnetic field threading a central object (black hole or star) and/or its surrounding accretion disc. However, the collimation mechanism far away from the central object has not yet been fully understood. In a previous work we investigated a mechanism in which the jet is self-collimated due to a dominant hoop stress. We ran numerical simulations in which a Jet-Emitting disc (JED) spans the entire lower computational boundary. Those were the first of their kind to showcase the steady recollimation shocks predicted by steady-state analytical studies of jets. However, the huge size of the JED prevented a complete study of the connection between the accelerating and asymptotic electric circuits, as well as the influence of the outer medium. We performed a set of axisymmetric ideal MagnetoHydroDynamics (MHD) non-relativistic jet simulations. In those, only the innermost region of the accretion disc is a jet-launching zone. The jets of finite radial extent in those simulations also produce steady recollimation shocks at large distances from the central object. Standing recollimation shocks are not a bias of self-similarity, but a generic feature of jets emitted from magnetized Keplerian accretion discs. They may produce observable features, such as a standing emission knots, a decrease of the rotation rate or a change in polarisation. We also recover previous results on the influence of external pressure on jet confinement, such as the relation between pressure profile and jet shape, and jet acceleration efficiency.

\end{abstract}


%%%%%%%%%%%%%%%%%%%%%%%%%%%%%%%%%%%%%%%%%%%%%%%%%%

%%%%%%%%%%%%%%%%% BODY OF PAPER %%%%%%%%%%%%%%%%%%

%%%%%%%%%%%%%%%%%%%%%%%%%%%%%%%%%%%%%%%%%%%%%%%%%%%%%%%%%%%%%%%%%%%
\section{Introduction}
%%%%%%%%%%%%%%%%%%%%%%%%%%%%%%%%%%%%%%%%%%%%%%%%%%%%%%%%%%%%%%%%%%%

Astrophysical jets are ubiquitous in the universe. They are routinely observed around most if not all types of accreting objects: AGNs \citep{boccardi2016,Blandford2019}, YSOs \citep{bally2007,ray2007,frank2014,ray2021}, X-ray binaries \citep{fender2014,tudor2017}, and even post-AGB stars \citep[Asymptotic Giant Branch;][]{bollen2017}. Even though the central objects are very different (black hole, protostar, neutron star or white dwarf), all those jets share some striking properties: (a) they are supersonic and collimated (small opening angle), (b) their asymptotic speed far away from the accreting object scales with its escape velocity, (c) they transport a sizeable fraction of the accreting power and mass. For all these objects this fraction is also consistent, as shown by the accretion-ejection correlations \citep[see][and references therein]{Ghisellini2014,merloni2003,corbel2003,gallo2004,coriat2011,Ferreira2006b,cabrit2007,ellerbroeck2013}. The only characteristics shared by all these objects being the presence of an accretion disc and a jet, it is reasonable to seek a jet model where the ejection is powered by the disc and not the central object.

Around all object types, jets show a variety of different collimation properties. AGN jets are generally classified following \cite{fanaroff1974}, hereafter FR. Edge-darkened FRI jets appear conical and show large-scale bending \citep{laing2011,laing2014} while edge-brightened FRII jets appear nearly cylindrical \citep{laing1994,boccardi2017}. These differences were historically only linked to jet power: above a certain critical power all jets would be FRII while below all would be FRI. However, observations of low-luminosity FRIIs \citep{mingo2019} recently showed that the jet interaction with its environment should also play a major role. This had been hinted by the General Relativity MagnetoHydroDynamics (GRMHD) simulations of \cite{tchekhovskoy2016}: the critical jet power should strongly depend on the density profile of the material in which the jet propagates. X-ray binaries consist in a compact object (black hole or neutron star) accreting plasma from a stellar companion. As their angular size is most often sub-nano-arcsecond, they cannot be directly imaged. On the other hand, there has been plenty of images of protostellar jets, around stars of Class 0 such as HH212 \citep{lee2020}, Class I such as DG Tau B \citep{Zapata2015} and Class II such as HH30 \citep{burrows1996,louvet2018}. Although there is no equivalent of the FR dichotomy in protostellar jets, most appear to be very collimated, with opening angles of only a few degrees \citep{ray2007,Davis2011}.

Among both protostellar and AGN jets, peculiar features appear along the flow. For AGN jets some of those knots are stationary \citep{lister2009,lister2013,walker2018,doi2018,park2019}. Around M87* a standing shock appears when the jet is recollimating towards its axis, and has thus been coined `recollimation shock' \citep{cheung2007,asada2012}. Even though the expression came from AGN jets, there is growing evidence that such recollimation shocks are also present in protostellar jets \citep{white2014,RodriguezKamenetzky2017,Moscadelli2021}. Note however that in protostellar jets knots are mostly moving with the jet \citep[although see][]{bonito2011} and can take the form of bow shocks (see \citealp{lee2017} for HH212 or \citealp{louvet2018} for HH30) whose origin is still debated.

In their seminal work, \cite{blandford1982} showed that if a large-scale vertical magnetic field is threading an accretion disc, its material can be flung at super-fast magnetosonic (super-FM) speeds. The rotation of the disc creates a strong toroidal magnetic field that collimates the ejected matter close to the jet axis. This collimating (hoop-stress) mechanism is called Z-pinch, and is also used to confine plasma in tokamak devices. The work of \cite{blandford1982} was later extended by \cite{Ferreira1993a,Ferreira1993b,ferreira1995} to study the interdependence between accretion and ejection. This lead to the Jet-Emitting disc (JED) model of \cite{ferreira1997}, where the jet is connected to a turbulent quasi-Keplerian accretion disc. In this model, super-FM jets systematically undergo a recollimation towards its axis, potentially leading to a recollimation shock \citep[see also][]{polko2010}. This property was later verified for warm jets \citep{casse2000a} and weak magnetic fields \citep{jacquemin-ide2019}.

The works mentioned in the above paragraph made the assumption of radial self-similarity. Other self-similar works extended that of \cite{blandford1982} by exploring different magnetic field distributions \citep{contopoulos1994,ostriker1997} or thermal effects \citep{vlahakis2000,ceccobello2018}. However, the self-similar approach induces some biases: the conditions on the axis are unphysical and the interaction between the jet and the ambient medium cannot be taken into account. The jet collimation properties and the presence of shocks might then be a result of the self-similar geometry.

Partly to probe whether the self-similar results hold without some of the biases, in \cite{jannaud2023} (hereafter JZF23) we performed 2.5D ideal MHD simulations of jets using the PLUTO code \citep{mignone2007}. In those, on most of the lower boundary ($r_0 \in [1;5650]$) we impose JED ejection conditions. On the rest of the boundary ($r_0 \in [0;1]$) we smoothly transition from the conditions wanted at the inner edge of the disc ($r_0 = 1$) to those required on the axis ($r_0 = 0$). This could model both a protostellar jet where the inner edge of the disc would be around 0.1 astronomical units (au), or a jet emitted from the accretion disc of a Schwarzchild black hole where the inner edge of the disc would be the Innermost Stable Circular Orbit (ISCO)\footnote{For a Schwarzchild black hole of spin parameter $a=0$ the ISCO radius is three times the Schwarzchild radius.}. In these simulations, a super-FM collimated jet is launched from a sub-Alfvénic Keplerian disc. They also contain standing recollimation shocks, beyond a thousand times the innermost disc radius. To our knowledge, these simulations were the first of their kind to show such shocks, surely because of the large scales required in space and time. They necessitated the development of a specific algorithm to accelerate their convergence. These large scales enabled us to narrow the gap between analytical and numerical approaches, showing that recollimation shocks are not a bias of self-similarity, but are intrinsic to the process of MHD self-collimation. In addition, a parameter space study enabled us to probe the influence of launch conditions on the shocks. We showed that they follow qualitatively the behavior demonstrated by self-similar studies, i.e. that they get closer to the disc as the mass load $\kappa$ increases. We also confirmed that the magnetic field distribution in the disc ($\alpha$ in $B_z \propto r^{\alpha - 2}$) is the key quantity shaping the collimation of the jet. It follows the trend of potential fields: the greater $\alpha$, the stronger the collimation.

Still, a few important questions were left unanswered by these extended jet simulations. Even though the mass and power ejected by the central object ($r_0 \in [0;1]$) on the axial spine was minimized, it still was shown to influence shock position and jet collimation. The full extent of that impact should be explored. Moreover, the JED self-similar boundary condition of the disc extending until the end of the simulation box ($r_0 \in [0;5650]$), the impact of an outer medium could not be probed.

More importantly, the main objective of \citetalias{jannaud2023} was to showcase the asymptotic jet regions (beyond all recollimation shocks) in a simulation where the jet is launched from a consistent sub-Alfvénic disc. We wanted to probe if the ejection conditions could have an impact on the asymptotic jet collimation properties, and thus its observable shape. The pioneering work of \cite{heyvaerts1989} predicted the asymptotic morphology of stationary axisymmetric magnetized jets using their poloidal electric current remaining at infinity. They showed that depending on whether this current vanishes or not, the jet would eventually collimate to a parabola, cone or cylinder. This generic result has later been extended, taking into account current closure \citep{heyvaerts2003a,heyvaerts2003b,heyvaerts2003c}, as well as the geometry of the solution \citep[collimating or decollimating, see][]{okamoto2001,okamoto2003}. In particular, \cite{heyvaerts2003a} showed that kinetic energy-dominated jets should have a vanishing asymptotic current and collimate into paraboloids. However those results only focused on the asymptotic current, and the question of its link to the jet source (central object and/or disc) was left open. Even though they were performed at unprecedented scales, the simulations of \citetalias{jannaud2023} did not show true asymptotic circuits beyond the recollimation shocks. Because of the extended JED at their lower boundary, some electric circuits emerging from the disc at large radii would circumvent the shocks before reaching the asymptotic zone. Thus, the `asymptotic' current beyond the shocks would be directly determined by the current at the disc surface.

The logical step to tackle the issues addressed in the two paragraphs above is to limit the jet ejection to innermost disc. This is also in line with direct imaging of both protostellar and AGN jets. In protostellar discs the outermost launch radius varies a lot from object to object, but seems to be no greater than a few tens of au \citep[see][and references therein]{Ferreira2006b,lee2020,tabone2020}. In AGNs, radio-emitting (i.e. jet-emitting) regions seem limited to a few thousand Schwarzchild radii from the black hole. For X-ray binaries, as the jets cannot be directly imaged, the outermost launch radius has to be inferred by fitting X-ray and radio spectra to ejection models. One such model is the JED-SAD \citep{Ferreira2006a,marcel2018,Marcel2018a}. It consists in a hybrid disc, with an inner highly magnetized Jet-Emitting disc (JED, \citealp{ferreira1997}) surrounded by an outer Standard Accretion disc (SAD, \citealp{Shakura1973}). The transition from the jet-launching JED to the non-jet launching SAD happens at a finite radius $r_{\text{j}}$. Using this model, \cite{barnier2022,Marcel2019,Marcel2022} managed to fit X-ray and radio spectra during the jetted hard states of GX 339-4. The transition radius $r_{\text{j}}$ is then expected to vary during the outburst, but should not exceed a few tens of ISCO radii.

In summary, performing simulations of jets emitted from a smaller JED would enable us to probe the influence of the outer medium on the collimation, and to better understand the connection between the launching conditions and the observable jet shape. It also better suits observational evidence of jets among all objects. The paper is organized as follows. Section \ref{sec:Framework} describes the numerical setup and boundary conditions. The latter mimic the ejection of: (a) an axial spine emitted by the central object and its interaction with the disc ($r_0 \in [0;1]$), (b) a cold jet emitted from a self-similar Keplerian disc ($r_0 \in [1;r_{\text{j}}]$), (c) a sparse outflowing atmosphere lying above an outer disc where the ejection is vanishing ($r_0 \in [r_{\text{j}};5650]$). Our fiducial simulation with a non-rotating spine is described in section \ref{sec:ReferenceStationaryCase}. Recollimation shocks are obtained, with a structure similar to those of \citetalias{jannaud2023}. This further proves that recollimation shocks are not a bias of the self-similar ansatz, but an intrinsic property of self-collimated jets. A parametric study is presented in section \ref{sec:ParametricStudies}, showing the impact of small magnetic pressures in the external medium on jet collimation. We also show that the rotation of the axial spine, even when it is much sparser than the jet, has a considerable impact on the altitude of the recollimation shocks. We discuss our results in section \ref{sec:Discussions}, highlighting the limitations of the setup and comparing or results with previous numerical works. We then consider possible observational consequences of self-collimation and standing recollimation shocks, before concluding in section \ref{sec:Conclusion}.

\begin{comment}
    Note however that detailed observations with radio interferometry have recently revisited the FR classification, showing the presence of another type, called FR0 (\cite{Garofalo2019}) and that the FRI/FRII dichotomy is not as rigid as previously thought
\end{comment}

%%%%%%%%%%%%%%%%%%%%%%%%%%%%%%%%%%%%%%%%%%%%%%%%%%%%%%%%%%%%%%%%%
\section{Physical and numerical framework}\label{sec:Framework}
%%%%%%%%%%%%%%%%%%%%%%%%%%%%%%%%%%%%%%%%%%%%%%%%%%%%%%%%%%%%%%%%%

\subsection{Governing equations and numerical setup}%%%%%%%%%%%%%%%%%%%%%%%%%%%

\begin{figure}
\centering
\includegraphics[trim=10 10 440 10, clip, width = .8\linewidth]{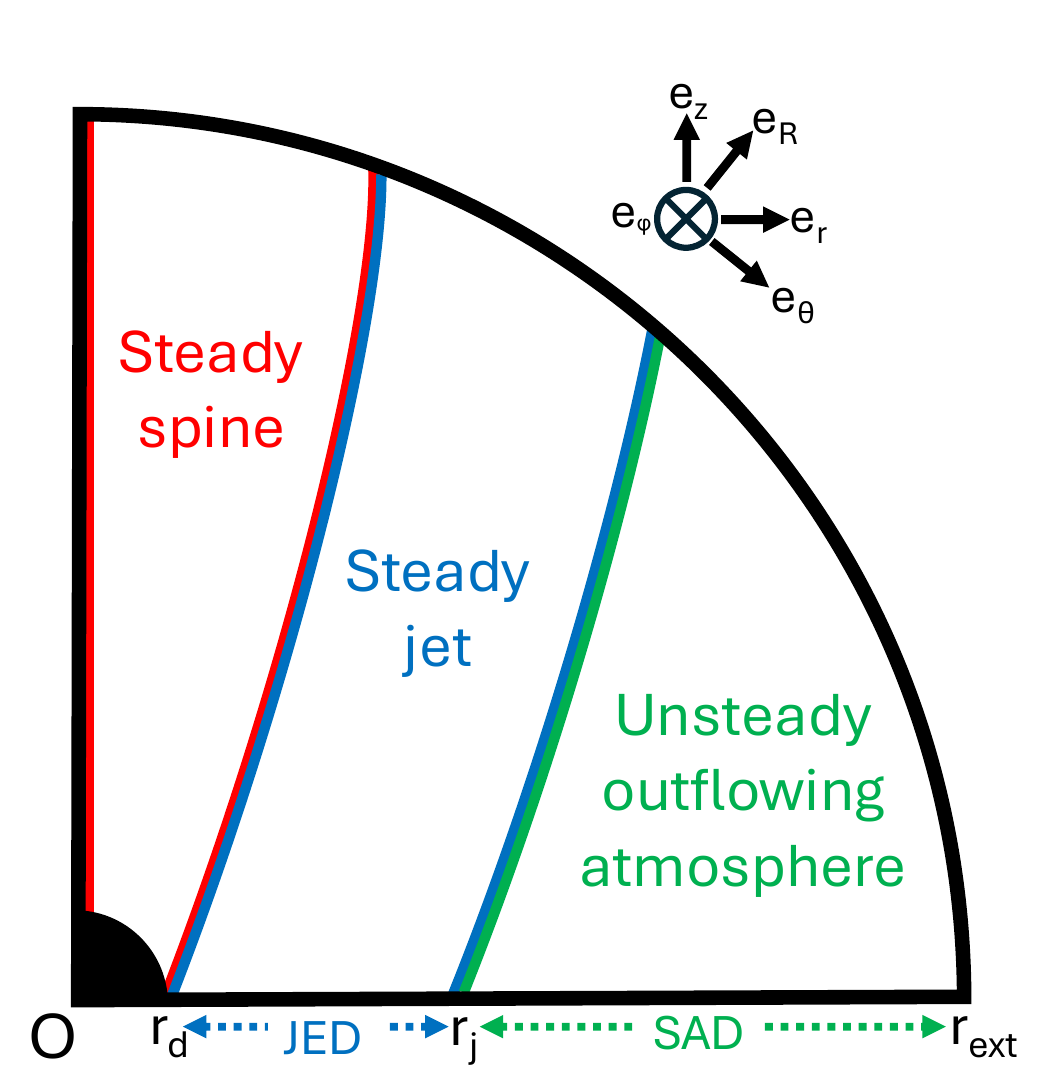}
\caption{Sketch of the computational domain.}
\label{fig:SchemaGeometrie}
\end{figure}

The intent of this work is to study jets ejected from Keplerian accretion discs of finite size. A scheme of the geometry is shown on Fig. \ref{fig:SchemaGeometrie}. The central object (black hole or star) of mass $M$ is the origin of the two systems of coordinates we use, spherical ($R$, $\theta$, $\upvarphi$) and cylindrical ($r$, $\upvarphi$, $z$). The central object is surrounded by an accretion disc, settled from an inner radius $r_{\text{d}}$ to an outer radius $r_{\text{ext}} = 5650 r_{\text{d}}$. As in \citetalias{jannaud2023}, the disc itself is not computed. However, here only the inner disc ($r \in [r_{\text{d}};r_{\text{j}}]$) behaves like a JED. The outer disc ($r \in [r_{\text{j}};r_{\text{ext}}]$) emits very little mass, behaving like a SAD. We further assume the presence of a large-scale magnetic field threading both the central object and the disc. This allows the presence of three outflows :

\begin{itemize}
    \item The spine, emitted from the inner region ($r \in [0; r_{\text{d}}]$).
    \item The jet, emitted from the JED ($r \in [r_{\text{d}}; r_{\text{j}}]$).
    \item The atmosphere, emitted from the SAD ($r \in [r_{\text{j}};r_{\text{ext}}]$).
\end{itemize}

We focus on the dynamics of the jet. The influence of the spine is limited as much as possible, and mass ejected from the SAD into the outflowing atmosphere is minimized. All three outflows are assumed to be in ideal MHD, and we numerically solve the usual set of MHD equations :

\begin{align}
&\frac{\partial \rho}{\partial t} + \nabla \cdot \left( \rho \bf{u} \right) = 0 ,\\
&\frac{\partial \rho {\bf u}}{\partial t} + \nabla \cdot \left[ \rho {\bf u} \otimes {\bf u} + \left( P + \frac{ \bold{B \cdot B}  }{2 \mu_0}\right) \bar{\bar{\bf I}} - \frac{\bold{B \otimes B}}{\mathnormal{\mu_0}} \right] =- \rho \nabla \Phi_G,\\
&\frac{\partial \bf B}{\partial t} + \nabla  \times \left(\bf B \times \bf u \right) = \vec{0},\\
&\frac{\partial S}{\partial t} + \nabla \cdot \left( S \bf u \right) = 0,
\end{align}

where $\rho$ is the density, $P$ the thermal pressure, $\vec{u}$ the flow velocity and $\vec{B}$ the magnetic field. The gravitational potential due to the central mass is $\Phi_G = - GM/R $. The entropy is $S = P / \rho^\Gamma$ where $\Gamma = 1.25$ is the polytropic index.

The setup is 2.5D. This means that axisymmetry is assumed along the polar axis ($\theta = 0$), but velocity and magnetic field in the toroidal direction (along $\vec{e}_\upvarphi$) are taken into account, as they are critical in the jet acceleration and collimation processes. Thanks to axisymmetry, the poloidal magnetic field can be computed using the magnetic flux function $\Psi$:

\begin{equation}
B_R = \frac{1}{R^2 \sin\theta} \frac{\partial \Psi}{\partial \theta} \qquad
B_\theta = -\frac{1}{R\sin\theta}\frac{\partial \Psi}{\partial R}
,\end{equation}

The ideal MHD equations equations are solved using the PLUTO code \citep{mignone2007}. We use a spherical grid ($R, \theta$), encompassing a spherical domain going from $R = r_{d}$ (inner boundary) to $R = r_{\text{ext}}$ (outer boundary), and from $\theta = 0$ (jet axis) to $\theta = \pi / 2 - \epsilon$ (disc surface). Here $\epsilon \equiv c_{S_{\text{d}}} / V_{K_{\text{d}}}$ is the disc scale height. We assume a very thin disc with $\epsilon = 10^{-2} \ll 1$ and the disc surface is very close to its midplane $\theta  = \pi /2$.

In the radial direction, the domain is discretized on $N_R = 704$ cells with logarithmic spacing ($\Delta R \propto R$). In the orthoradial direction, it is discretized on $N_\theta = 266$ cells. The cell size along $\vec{e}_\theta$ is mostly uniform but decreases close to the jet axis to allow sufficient resolution in this zone where the recollimation shocks form. The resolution is smaller than that of the setup from \citetalias{jannaud2023}. It was mandated by a loss in efficency on our specific acceleration scheme on the Courant–Friedrichs–Lewy condition\footnote{See section 3.1.5 of \cite{janna2023} for an explanation of the acceleration scheme or Appendix A of \citetalias{jannaud2023} for a quick overview.}. This setup sits further away from self-similar steady-state jet studies than that of \citetalias{jannaud2023}, and the acceleration scheme relies on a large part of the computational domain reaching a steady state. Still, this acceleration scheme enabled us to reach physical times $10^3$ greater than what would have been possible using usual time-stepping methods.

The same set of Riemann solvers, slope limiters and reconstruction schemes as in \citetalias{jannaud2023} is used. In most of the computational domain, we use the HLLD solver of \cite{miyoshi2005} and the default reconstruction scheme of PLUTO: Monotonized Center \citep{VanLeer1977} for $\rho$, Van Leer \citep{VanLeer1974} for $\vec{u}$ and $\vec{B}$ and Minmod \citep{Roe1986} for $P$. In the regions closest to the axis, as well as in those of very low density and high Alfv\'en speed, we use a HLL solver with a flat reconstruction scheme and a flat slope limiter. This is done to avoid prohibitively high computational costs.

\subsection{Initial conditions}

In this section and in all the following, quantities with a subscript $a$ refer the values on the axis directly over the central sphere ($R = r_{\text{d}}$ , $\theta = 0$), while quantities with a subscript $d$ refer to the values at the inner edge of the disc ($R = r_{\text{d}}$ , $\theta = \pi/2 - \epsilon$).

The initial magnetic field is assumed to be potential, leading to second order partial differential equations on $\Psi (R, \theta)$. We further assume radial self-similarity, and

\begin{equation}
    \Psi = \underbrace{ \Psi_{\text{d}} (R/r_{\text{d}})^\alpha \Phi(\theta) }_{\text{Blandford \& Payne}} + \underbrace{ B_{\text{ext}} \frac{R^2 \sin \theta}{2} }_{\text{external field}} = \Psi_{\text{BP}} + \Psi_{\text{ext}},
\end{equation}

The first term $\Psi_{\text{BP}}$ is the same as in \citetalias{jannaud2023}, where radial self-similarity à la \cite{blandford1982} was assumed. $\Phi(\theta)$ has been determined assuming the initial magnetic field is potential (current-free and force-free). The exponent $\alpha$ is a free parameter of the model, setting the magnetic field profile: $B_R \propto (B_\theta + B_{\text{ext}}) \propto R^{\alpha -2}$. All simulations in this work are performed with the same magnetic field profile as in the seminal work of \cite{blandford1982}, that is with $\alpha = 3/4$.

The second term $\Psi_{\text{ext}}$ adds an extra constant magnetic field $-B_{\text{ext}} \vec{e}_\theta$, of small amplitude $B_{\text{ext}}$ compared to that of the self-similar solution. It makes the initial magnetic field consistent with its boundary condition on the disc (equations \ref{eq:InjectionCondition} and \ref{eq:BthetaTronque}).

The initial magnetic field is thus purely poloidal, and

\begin{equation} \label{eq:initialBp}
    \vec{B}_p = \frac{1}{2 \pi r} \nabla \Psi \times \vec{e}_\upvarphi
\end{equation}

This magnetic field being potential, there is no magnetic force imposed on the plasma. Therefore, a spherically symmetric hydrostatic equilibrium is assumed: $\vec{u} = 0$ and $d P / d R = - \rho GM / R^2$. We choose the same solution as in \citetalias{jannaud2023}:

\begin{equation}
\begin{aligned}
\rho & = \rho_{a} \left(\frac{R}{r_{\text{d}}}\right)^{2\alpha-3} \\
P    & = \frac{1}{4-2\alpha}\frac{\rho_{\text{a}} GM}{r_{\text{d}}}\left(\frac{R}{r_{\text{d}}}\right)^{2\alpha-4} \, ,
\end{aligned}
\end{equation}

where the sound speed $c_s$ is defined as $c_s^2 = \partial P / \partial \rho = \Gamma P / \rho$. The value of $\rho_{\text{a}}$ is fixed in accordance with the boundary conditions.

We chose this simple potential solution as our initial conditions to ensure that the eventual appearance of a collimated super-FM jet would be the result of the ejection conditions (boundary conditions at $R=r_{\text{d}}$ and $\theta = \pi /2 - \epsilon$) and not a slight variation from the initial conditions.

\subsection{Boundary conditions} \label{sec:BoundaryConditions}

There are four boundaries to the simulation domain:

\begin{itemize}
    \item The polar axis ($\theta=0$, $R \in [r_{\text{d}}; r_{\text{ext}}]$)
    \item The outer boundary ($R=r_{\text{ext}}$, $\theta \in [0; \pi / 2 - \epsilon]$)
    \item The accretion disc ($\theta = \pi /2 - \epsilon$, $R \in [r_{\text{d}}; r_{\text{ext}}]$)
    \item The central object ($R=r_{\text{d}}$, $\theta \in [0; \pi /2 - \epsilon]$)
\end{itemize}

At these four boundaries, conditions have to be set on the eight evolving quantities $\rho$, $P$, $\vec{u}$ and $\vec{B}$. On the polar axis, usual reflecting conditions are applied on all quantities. At the outer boundary, `outflow' conditions are imposed: the gradient along $\vec{e}_R$ of $\rho$, $P$, $B_R$, $B_\theta$, $R B_\upvarphi$, $u_R$, $u_\theta$ and $u_\upvarphi$ are conserved. The Van Leer slope limiter is used to avoid spurious oscillations. Additionally, a positive Lorentz force is enforced on the subalfv\'{e}nic parts of the outer boundary. This last condition is not strictly needed for the convergence of the simulation, but slightly decreases the computational cost.

\subsubsection{Spine generation: $R = r_{\text{d}}$}

Along the $R=r_{\text{d}}$ inner boundary, the conditions are set as in \citetalias{jannaud2023}. As the spine is launched super-slow-magnetosonic (super-SM), two of the eight quantities have to be let free, $B_\upvarphi$ and $B_\theta$. For them, we use an outflow condition, and their gradient along $\vec{e}_R$ is conserved through the boundary. For the other six ($B_R$, $u_R$, $u_\theta$, $u_\upvarphi$, $\rho$ and $P$), conditions should be specified on the whole boundary.

To keep the magnetic field flowing in the spine constant with time, $B_R$ is fixed to its initial value (equation \ref{eq:initialBp}). The conditions are set consistently with steady-state ideal MHD, which requires $\vec{u}_p \parallel \vec{B}_p$. Thus, the orthoradial velocity is set by $u_\theta = u_R (B_\theta / B_R)$.

We define the smooth spline function

\begin{equation}
    \begin{aligned}\label{eq:SplineAutosimilaire}
    f \colon  \theta \looongrightarrow & \left( 3 \sin^2 \theta - 2 \sin^3 \theta \right)^{3/2}\\
      [0;\pi / 2 - \epsilon] \mapsto &  [0;1] 
\end{aligned}
\end{equation}

chosen such that $f(\theta=0)=0$ on the axis and $f(\theta = \pi /2 - \epsilon)=1$ on the disc. This function helps us set the distributions of $u_\upvarphi$, $u_R$, $\rho$ and $P$, by passing from their desired values on the axis (subscript $a$) to their desired values on the disc (subscript $d$).

The plasma rotation $u_\upvarphi = \Omega_* r + u_R (B_\upvarphi / B_R)$ is fixed via the rotation rate of the magnetic surfaces $\Omega_*$:

\begin{equation} \label{eq:OmegaEtoileConditionsAutosimilaires}
    \Omega_* = \Omega_{*_{\text{a}}} \left(1- f(\theta) \right) +\Omega_{K_{\text{d}}} f(\theta)
\end{equation}

where $\Omega_{*_{\text{a}}}$ is the central object rotation on the axis. It will be used as a free parameter. In the reference simulation $\Omega_{*_{\text{a}}} = 0$ in order to limit the influence of the spine on jet dynamics.

The radial speed $u_R$ is fixed through the sonic Mach number $M_S$ via $u_R = M_{S_R} c_S$. The sonic Mach number is constant with $\theta$. Its orthoradial component is set as $M_{S_\theta} = 10$ because of the disc conditions. Its radial component is set as $M_{S_R} = M_{S_\theta} \lvert B_R / B_\theta \rvert_{\text{d}}$ where $\lvert B_R / B_\theta \rvert_{\text{d}}$ is the inclination of the magnetic field lines at ($R=r_{\text{d}}$, $\theta = \pi / 2 - \epsilon$), initially $1/ \alpha$. As $B_R$ is fixed at the $R = r_{\text{d}}$ boundary and $B_\theta$ is fixed at the $\theta = \pi / 2 - \epsilon$ boundary, at this point the inclination of the field lines is fixed and for the whole duration of the simulation $\lvert B_R / B_\theta \rvert_{\text{d}} = 1 / \alpha$. The radial component of the sonic Mach number then becomes $ M_{S_R} = 10 / \alpha  = 40/3 \gg 1$

\begin{comment}

Its radial component is  $M_{S_R}=M_{S_\theta} \lvert B_R / B_\theta \rvert_{\text{d}}$ where $\lvert B_R / B_\theta \rvert_{\text{d}} = 1/ \alpha$ as in the initial conditions, and we set $M_{S_\theta} = 10$. Thus $ M_{S_R} = 10 \lvert B_R / B_\theta \rvert_{\text{d}} > 1$. As the magnetic field inclination $\lvert B_R / B_\theta \rvert $ is constant at the inner disc boundary, $M_{S_R}$ does not change with time. However, it may change from simulation to simulation with the inclination of the field lines (in our parameter space with the value of $\alpha$), while always staying higher than unity. The flow is thus always injected at supersonic speeds.

\end{comment}

The sound speed on the central object is computed using

\begin{equation}
    c_S = c_{S_{\text{a}}} (1-f(\theta)) + c_{S_{\text{d}}} f(\theta)
\end{equation}

The sound speed on the axis $c_{S_{\text{a}}}$ is fixed thanks to the Bernoulli invariant on the axis $E_{\text{a}}$. The MHD term $\Omega_* r B_\upvarphi / \eta$ vanishing on the axis, one directly obtains

\begin{equation}
c_{S_{\text{a}}}^2 = \frac{GM}{r_{\text{d}}} \frac{1 + e_{\text{a}}}{\frac{1}{2}M_{S_R}^2+\frac{1}{\Gamma-1}} = V_{K_{\text{d}}}^2 \frac{1 + e_{\text{a}}}{\frac{1}{2}M_{S_R}^2+\frac{1}{\Gamma-1}}
\end{equation}

where $e_{\text{a}} = E_{\text{a}} r_{\text{d}} / (GM)$ is the Bernoulli invariant normalized to the gravitational energy at $R=r_{\text{d}}$. We choose $e_{\text{a}} = 2$ to produce a spine of lower energy than the jet. This also gives the spine and the jet similar asymptotic speeds and avoids introducing discontinuities.

The sound speed on the disc $c_{S_{\text{d}}}$ is directly fixed by the disc scale height $\epsilon \equiv c_{S_{\text{d}}} / V_{K_{\text{d}}} = 10^{-2}$.

The plasma density $\rho$ is connected from its value on the axis $\rho_{\text{a}}$ to its value on the disc $\rho_{\text{d}} = \delta \rho_{\text{a}}$ via the MHD invariant $\eta$, as $\rho (\theta) = \eta B_R / (\mu_0 u_R)$, and $\eta$ follows:

\begin{equation}
    \eta = \eta_{\text{a}} (1 - f(\theta)) + \eta_{\text{d}} f(\theta)
\end{equation}

The density on the axis is fixed at $\rho_{\text{a}} = 1$, normalizing the density in the simulation. It also ensures a super-SM but sub-Alfvénic ejection.

As both the sound speed and the density are determined, the pressure $P = c_S^2 \rho / \Gamma$ is also determined.

The lower boundary at $\theta = \pi / 2 - \epsilon$ is the disc injection condition. As shown on Fig. \ref{fig:SchemaGeometrie}, this boundary is divided in two parts: the JED that launches the jet for $r_{\text{d}} < r \lesssim r_{\text{j}}$ and the SAD with minimized ejection for $r_{\text{j}} \lesssim r < r_{\text{ext}}$. We will see in this section how the boundary conditions are set in these two zones, then how the transition between the two is done.

\subsubsection{Jet launching: JED}

\hfill\break

In the innermost disc, the boundary conditions are set as in \citetalias{jannaud2023}, consistently with a Jet-Emitting disc \citep{ferreira1997}. As on the central object, the flow is launched super-SM but sub-Alfvénic. Thus two quantities have to be let free: $B_R$ and $B_\upvarphi$. For them, the gradient along $\vec{e}_\theta$ is conserved through the boundary. For the remaining six quantities, we choose the following conditions:

\begin{equation}
\begin{aligned}
\rho_{\text{JED}} &= \rho_{\text{d}} \left(\frac{R}{r_{\text{d}}} \right)^{2\alpha-3} \\
P_{\text{JED}} &= \rho_{\text{d}} \frac{c_{s_{\text{d}}}^2}{\Gamma} \left(\frac{R}{r_{\text{d}}} \right)^{2\alpha-4} \\
B_{\theta_{\text{JED}}} & = -B_{\text{d}} \left(\frac{R}{r_{\text{d}}} \right)^{\alpha-2} - B_{\text{ext}} \\
u_{\theta_{\text{JED}}} & = -u_{\text{d}} \left(\frac{R}{r_{\text{d}}} \right)^{-1/2} \\
u_{R_{\text{JED}}} & = u_\theta \frac{B_R}{B_\theta} \\
u_{\upvarphi_{\text{JED}}} & = \Omega_* r\, +\,  u_\theta\frac{B_\upvarphi}{B_\theta},
\end{aligned}
\label{eq:InjectionCondition}
\end{equation}

The exponent remains $\alpha = 3/4$. The disc is Keplerian, and the rotation of the magnetic surfaces is $\Omega_* = \Omega_K = \sqrt{GM/r^3}$. The radial speed is set so that $\vec{u}_p \parallel \vec{B}_p$, consistently with a steady-state ideal MHD jet. Like on the central object, this sets the toroidal component of the injected electric field to zero thus enforcing that the injected magnetic flux stays constant with time. These conditions let four dimensionless parameters to be specified at $r=r_{\text{d}}$, acting as normalizing quantities: $\rho_{\text{d}}$, $c_{S_{\text{d}}}$, $B_{\text{d}}$ and $u_{\text{d}}$. They are set consistently with the JED self-similar solutions.

The jet density $\rho_{\text{d}}$ is fixed relative to the density at the polar axis with the parameter $\delta \equiv \rho_{\text{d}} / \rho_{\text{a}}$. This $\delta$ parameter defines the density contrast between the spine and the jet. It is chosen as $\delta = 10^2$ to lower the mass flux of the spine compared to that of the jet.

The disc sound speed $c_{S_{\text{d}}}$ is fixed relative to the Keplerian speed with the disc scale height $\epsilon = c_{S_{\text{d}}}/V_{K_{\text{d}}}$. The disc sound speed sets the jet temperature, and since $\epsilon  = 10^{-2} \ll 1$ the jet is cold.

The disc magnetic field strength $B_{\text{d}}$ is fixed with the parameter $\mu \equiv V_{A_{\text{d}}}/V_{K_{\text{d}}}=B_{\text{d}}/\sqrt{\mu_0 \rho_{\text{d}} G M}$, vertical component of the Alfv\'{e}n speed over the Keplerian speed. The vertical injection speed $u_{\text{d}}$ is related to the mass-loading parameter $\kappa$ introduced by \cite{blandford1982}
%%%%%%%
\begin{equation} \label{eq:DefintionKappa}
    \kappa= \frac{\mu_0 \rho_{\text{d}} u_{\text{d}} V_{K_{\text{d}}}}{B_{\text{d}}^2} = \frac{u_{\text{d}} V_{K_{\text{d}}}}{V_{\text{Ad}}^2} = \frac{u_{\text{d}}}{V_{K_{\text{d}}}} \frac{1}{\mu^2} \ .
\end{equation}

We choose $\mu = 1$ and $\kappa = 0.1$. Those values, along with $\alpha = 3/4$, are the same for all simulations and define our Blandford \& Payne jet. As $\kappa \mu = u_{\text{d}} / V_{A_{\text{d}}}$ this enforces a sub-Alfvénic ejection.

The constant magnetic field $B_{\text{ext}}$ is also applied on the SAD (equation \ref{eq:BthetaTronque}). For all simulations, its amplitude is chosen such that $\rvert B_{\text{ext}}/B_{\text{d}} \rvert < 10^{-3}$ (see Table \ref{tab:ParametresSimusTronquees}). As a consequence, its influence is imperceptible on the JED and only noticeable in the SAD.

The sonic Mach number $M_S$ is constant along the whole boundary. Its vertical component is $M_{S_\theta}=u_{\text{d}}/c_{S_{\text{d}}}=u_{\text{d}}/(V_{K_{\text{d}}} \epsilon)=10$. The poloidal speed on the inner edge $u_{p_{\text{d}}}$ being greater than its vertical component $u_{\text{d}}$, the sonic mach number gives $M_{S} = u_{p_{\text{d}}} / c_{S_{\text{d}}}>u_{\text{d}}/c_{S_{\text{d}}}=M_{S_\theta}=10$. As the Alfv\'{e}n speed is much greater than the sound speed ($\mu \gg \epsilon$), the sound speed is greater than the SM speed, and thus the injection is super-SM.

Note that the values of these five parameters ($\alpha = 3/4$, $\epsilon = 10^{-2}$, $\delta = 10^2$, $\mu = 1$ and $\kappa = 0.1$) are the same as in the reference simulation of \citetalias{jannaud2023}.

\subsubsection{Outflowing atmosphere ejection: SAD-like}

\hfill\break

In the SAD, quantities are set to limit the ejection as much as possible. The density and pressure are fixed according to:

\begin{equation}
    \begin{aligned}
        \rho_{\text{SAD}} &= \rho_{\text{a}} \left( \frac{R}{r_{\text{d}}} \right)^{2 \alpha -3}\\
        P_{\text{SAD}} &= \frac{1}{4 - 2 \alpha}\frac{\rho_{\text{a}} G M}{r_{\text{d}}} \left( \frac{R}{r_{\text{d}}} \right)^{2 \alpha -4}
    \end{aligned}
\end{equation}

The pressure is consistent with the initial condition. The density distribution follows the same power law as in the inner JED, normalized at the axis density $\rho_{\text{a}}$ instead of $\rho_{\text{d}}$. As $\delta  \equiv \rho_{\text{d}} / \rho_{\text{a}} = 10^2$, it drops by two orders of magnitude between the JED and the SAD.

\begin{figure}
    \centering
\includegraphics[width=\linewidth]{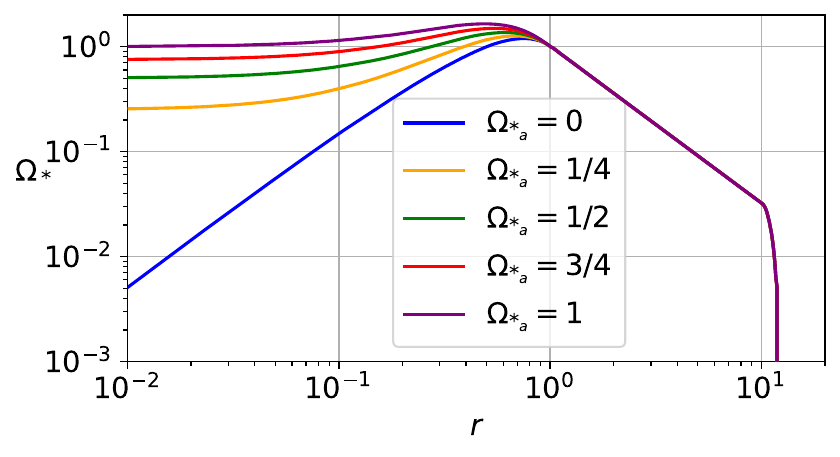}
    \caption{Radial distributions of the rotation rate $\Omega_*$ of the magnetic surfaces at the ejection boundary for different values of the rotation at the axis  $\Omega_{*_{\text{a}}}$. The JED follows a keplerian distribution $r^{-3/2}$ and is established from $r =1$ to $r=10$, with a SAD beyond $r =12$ assumed to launch no jet ($\Omega_*=0$) and a short continuous transition region between the two. The reference simulation O1 is the blue one, with $\Omega_{*_{\text{a}}}=0$ (see Table~\ref{tab:ParametresSimusTronquees}).}
        \label{fig:FrontiereOmegaTronque}
\end{figure}

We set $\Omega_{*_{\text{SAD}}} = 0$ to suppress the MHD Poynting flux $- ( 1 / \mu_0 ) \Omega_* r B_\upvarphi \vec{B}_p$.  The vanishing of magnetic surfaces rotation $\Omega_*$ greatly reduces the plasma rotation $u_{\upvarphi_{\text{SAD}}} = \Omega_* r + u_\theta B_\upvarphi / B_\theta$, and consequently any magnetocentrifugal ejection. On Fig. \ref{fig:FrontiereOmegaTronque} are represented rotation profiles at the ejection boundary (JED+SAD). We do so for the five simulations exploring the influence of the rotation of the axis $\Omega_{*_{\text{a}}}$: from $\Omega_{*_{\text{a}}} = 0$ (reference simulation, later called O1) to $\Omega_{*_{\text{a}}} = 1$ (solid-body rotation, later called O5). Four regions are clearly seen. On the disc ($r \geq 1$), the conditions are the same for all simulations. In the ejecting region (JED, $1 \leq r \leq 10$) the rotation is Keplerian: $\Omega_{*_{\text{JED}}} = \sqrt{GM/r^3}$. In the transition region ($10 \leq r \leq 12$) it quickly drops to zero and in the non-ejecting region (SAD, $12 \leq r \leq 5650$) there is no rotation: $\Omega_{*_{\text{SAD}}} = 0$. It is only at the central object boundary ($r \leq 1$) that the rotation varies from simulation to simulation, in order to reach its value on the axis as set by equations \ref{eq:SplineAutosimilaire} and \ref{eq:OmegaEtoileConditionsAutosimilaires}.

%%%%%%%%
%%%%%%%

The radial velocity is simply set to zero: $u_{R_{\text{SAD}}} = 0$. Setting the orthoradial velocity $u_{\theta_{\text{SAD}}}$ to zero would naturally limit the ejected mass flux, but also overconstrain the problem, as the ejected flow would then be sub-SM. As a consequence, if $M_{\text{SM}} \geq 1$ we kill the ejection with $u_{\theta_{\text{SAD}}} = 0$ but if $M_{\text{SM}} < 1$ we let $u_{\theta_{\text{SAD}}}$ free (as an outflow condition). In practice, the flow is sub-SM on almost the whole SAD. Note that this condition is not consistent with $\vec{u}_p \parallel \vec{B}_p$. This is not redhibitory as field lines anchored in the outer region remain sub-Alfvénic, thus the outflowing atmosphere cannot reach a steady state anyway.

The vertical magnetic field is identical to that on the JED:

\begin{equation} \label{eq:BthetaTronque}
    B_\theta = -B_{\text{d}} \left( \frac{R}{r_{\text{d}}}\right)^{\alpha - 2} - B_{\text{ext}}
\end{equation}

However here it becomes dominant in the outer regions ($r > 10^2$) where the power law drops off. It was introduced to provide additional collimation of the jet-atmosphere interface.

%%%%%%%%
\begin{figure}
    \centering
\includegraphics[width=\linewidth]{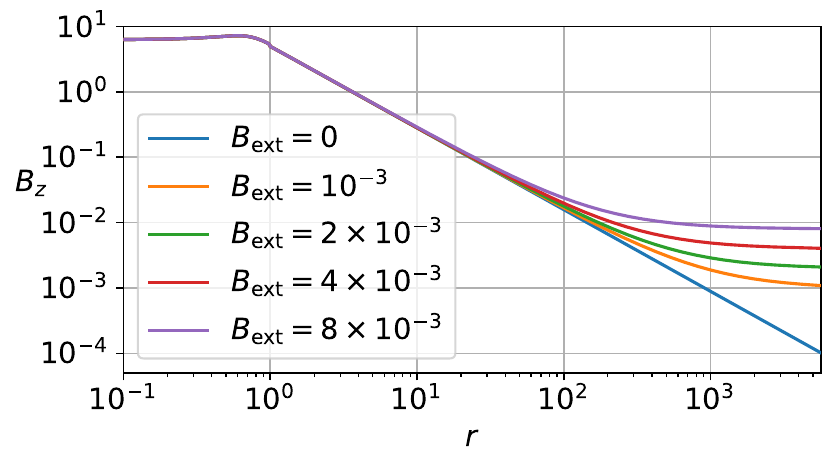}
    \caption{Radial distributions of the vertical magnetic field $B_z$ at the ejection boundary for different values of the external magnetic field $B_{\text{ext}}$ (note that in code units $B_{\text{d}} = 10$). The JED follows a Blandford \& Payne distribution $r^{-5/4}$ and is established from $r =1$ to $r=10$, with a SAD beyond $r =12$ (see text and Fig.~\ref{fig:FrontiereOmegaTronque}). The reference simulation O1 is the green one, with $B_{\text{ext}}=2\times 10^{-3}$ (see Table~\ref{tab:ParametresSimusTronquees}).}
    \label{fig:FrontiereBextTronque}
\end{figure}
%%%%%%%

On Fig. \ref{fig:FrontiereBextTronque} we show the boundary conditions on $B_z$. We do so for the five simulations exploring the influence of $B_{\text{ext}}$, from $B_{\text{ext}} = 0$ (simulation B1) to $B_{\text{ext}} = 8 \times 10^{-3}$ (simulation B5). The simulation with $B_{\text{ext}} = 2 \times 10^{-3}$ is the fiducial simulation, later called O1. Here and everywhere below, the magnetic field is given in code units, with $B_{\text{d}} = 10$. In order to reduce rotation in the non-ejecting region as much as possible, we limit the toroidal magnetic field. However, this cannot be done by simply setting $B_{\upvarphi_{\text{SAD}}} = 0$, as the appropriate number of disc ejection conditions are already fixed: five if the flow is sub-SM, six if it is super-SM. To avoid setting overconstrained boundary conditions, we gradually limit the toroidal magnetic field with

\begin{equation} \label{eq:ReductionBphi}
    B_\upvarphi (r,t+dt) = B_\upvarphi (r,t) e^{-\frac{dt}{\mathcal{T} T_K (r)}}
\end{equation}

where $T_K(r) = \sqrt{r^3/(GM)}$ is the local Keplerian timescale at radius $r$ and $\mathcal{T} = 10^{-6}$ is an adimensioned number quantifying how fast the toroidal magnetic field decreases with time. While simulations with $B_{\upvarphi_{\text{SAD}}} = 0$ did not converge, certainly because of causality issues, this workaround enabled the production of fast-converging simulations. To avoid introducing excessive discontinuities, this condition is enforced for $r > r_s$, where $r_s > r_{\text{j}}$.

\subsubsection{Transition between the JED and the SAD}

\hfill\break

The injected magnetic field $B_\theta$ only has one profile on both the JED and the SAD (equation \ref{eq:BthetaTronque}). For $B_R$ and $B_\upvarphi$ we use outflow boundary conditions, with an additional vanishing of $B_\upvarphi$ on the SAD (equation \ref{eq:ReductionBphi}).

For the other quantities ($\rho$, $P$, $u_R$, $u_\theta$ or $u_\upvarphi$), the transition between their JED and SAD profiles occurs for $r \in [r_{\text{j}}; r_s$]. In the steady-state simulations presented in this paper, $r_{\text{j}} = 10$ and $r_s = 12$. The transition is made via the spline function 

\begin{equation}
\begin{aligned}
    f_n \colon  x \loongrightarrow &  \left( 1 - 3 x^2 +2x^3 \right)^{\text{n}}\\
      [0;1] \mapsto & [0;1] 
\end{aligned}
\end{equation}

\begin{equation*}
\begin{aligned}
    \text{where $n \in \mathds{N}$ and } x(r) \equiv
    \begin{cases}
    0 & \text{if $r_{\text{d}} \leq r < r_{\text{j}}$}\\
    \frac{r - r_{\text{j}}}{r_S - r_{\text{j}}} & \text{if $r_{\text{j}} \leq r \leq r_S$}\\
    1 & \text{if $r_S < r \leq r_{\text{ext}}$}
    \end{cases}
\end{aligned} 
\end{equation*}

The profiles of density, pressure and velocities are then

\begin{equation}
    \mathcal{U}(r) = \mathcal{U}_{\text{JED}}(r) f_n \left( x(r) \right) + \mathcal{U}_{\text{SAD}}(r) \left[1-f_n \left( x(r) \right)\right]\\
\end{equation}

where $\mathcal{U}$ is $\rho$, $P$, $u_R$, $u_\theta$ or $u_\upvarphi$. Density and pressure are linked by the steep spline function $f_{20}$ while the velocities are linked by the smoother function $f_1$. As the transition region $[r_{\text{j}} = 10; r_s = 12]$ is small, this does not change much.

\subsection{Normalization} \label{sec:Normalization}

The MHD simulation results will be presented in dimensionless units. Unless otherwise specified, lengths are given in units of $r_{\text{d}}$, velocities in units of $V_{K_{\text{d}}}= \sqrt{GM/r_{\text{d}}}$, time in units of $T_{d}= r_{\text{d}}/V_{K_{\text{d}}}$, densities in units of $\rho_{\text{a}}$, magnetic fields in units of $B_0= V_{K_{\text{d}}}\sqrt{\mu_0 \rho_{\text{a}}} = B_{\text{d}} / 10$, mass fluxes in units of $\dot M_{\text{d}}= \rho_{\text{a}} r_{\text{d}}^2 V_{K_{\text{d}}}$ and powers in units of $P_{\text{d}} = \rho_{\text{a}} r_{\text{d}}^2 V_{K_{\text{d}}}^3$.

\begin{figure*}
\centering
\begin{subfigure}
  \centering
  \includegraphics[trim=0 10 0 5,clip,width=.58\linewidth]{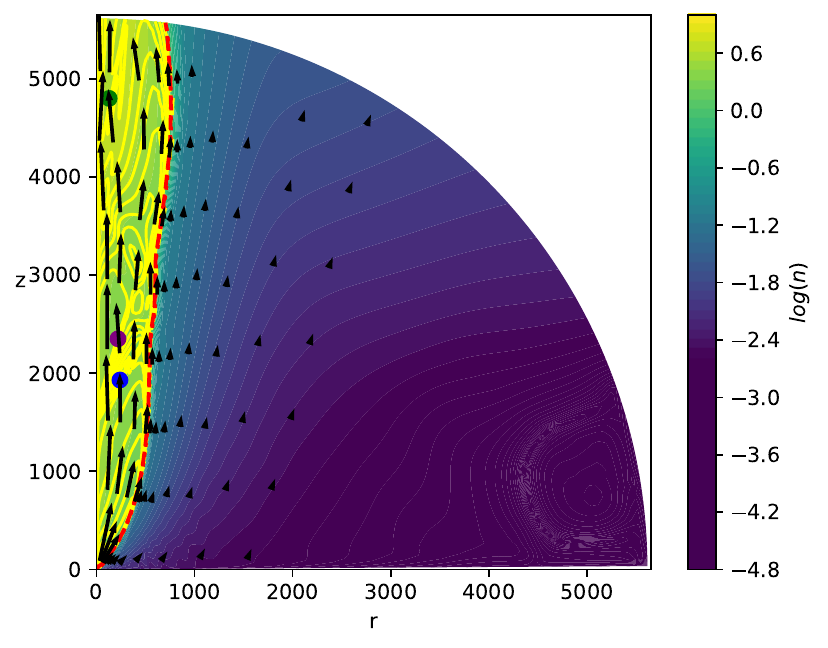}
\end{subfigure}%
\begin{subfigure}
  \centering
  \includegraphics[trim=0 10 0 5,clip,width=.41\linewidth]{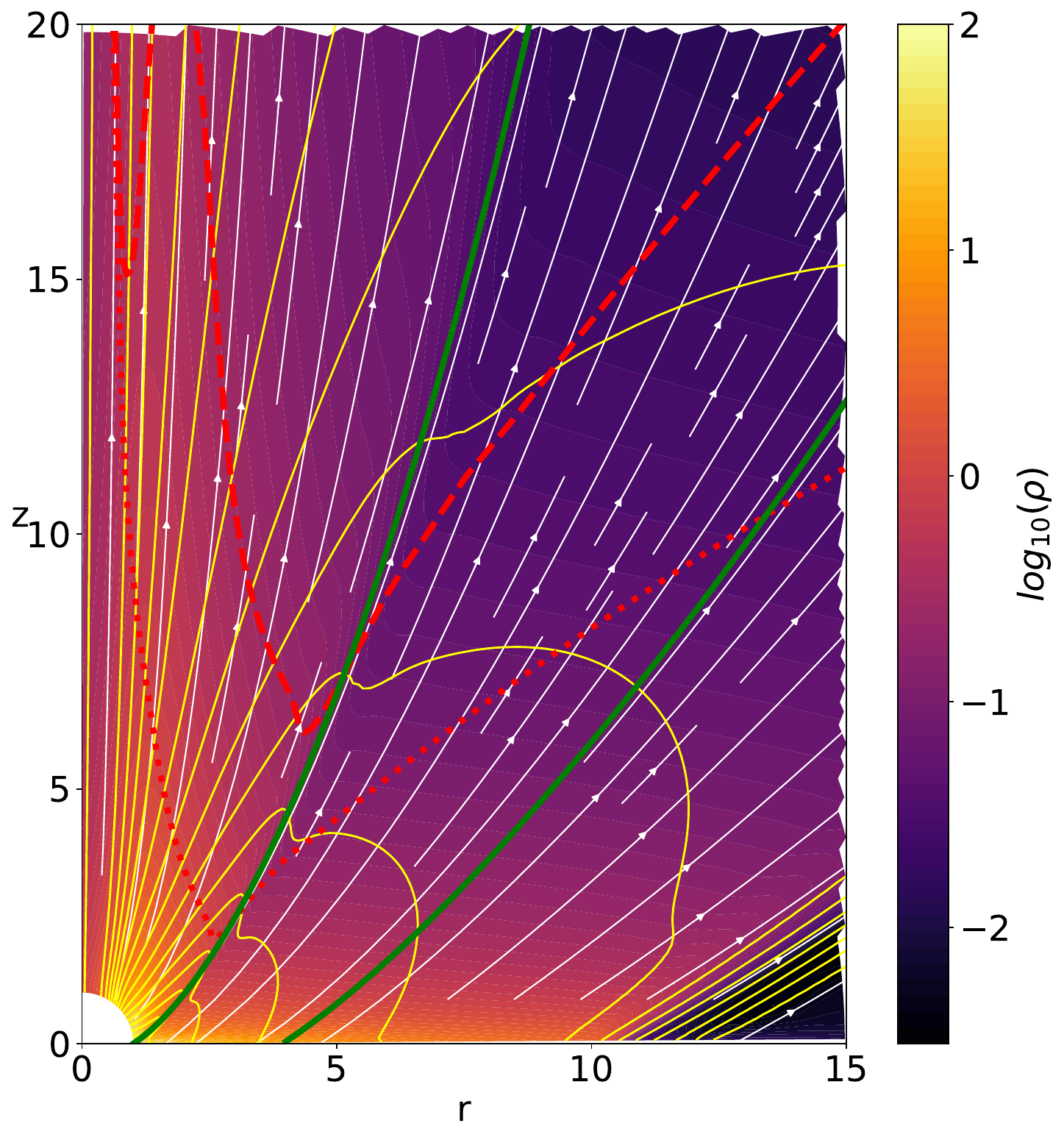}
\end{subfigure}
\caption{Snapshots of the reference simulation O1 at $t_{\text{end}}$. 
{\bf Left: } Global view showing the logarithm of the FM mach number $n$ in background color,  poloidal velocity vectors (black arrows) on field lines  anchored at different radii ($r_0= 2; 4; 7; 15; 40; 80; 160; 320; 600; 1000; 1500$), isocontours (yellow lines) of the poloidal electric current and the FM critical surface (red dashed line). The colored points (blue, purple and green) show where the field line anchored at $r_0 = 4$ meets the three standing recollimation shocks. 
{\bf Right: } Zoomed in view with the logarithm of the plasma density in background color. The white and green lines are poloidal magnetic field lines, the green ones being anchored on the disc at $r_0 \in [1.02;4]$. The red lines are the  Alfv\'en (dotted) and FM (dashed) critical surfaces. The yellow lines show the poloidal electric circuit, flowing out of disc, from the JED ($z=0$, $r \in \left[ 1;10 \right]$) and outer transition zone ($z=0$, $r \in \left[ 10;12 \right]$) and closing into the central object ($R = \sqrt{r^2 + z^2} = 1$). }
\label{fig:ReferenceSimulationTruncated}
\end{figure*}

For the case of a young star, assuming a  star of one solar mass with an innermost disc radius $r_{\text{d}}= 0.1$au and a density on the axis $\rho_{\text{a}} = 10^{-15} \text{g.cm}^{-3}$, this gives

\begin{equation}
\begin{split}    
        V_{K_{\text{d}}}&= 94.3 \left( \frac{M}{\text{M}_\odot} \right)^{1/2} \left( \frac{r_{\text{d}}}{0.1 \text{au}} \right)^{-1/2} \, \mbox{km.s}^{-1} \nonumber \\  
        \dot M_{\text{d}}&= 3.35 \times 10^{-10} \left( \frac{\rho_{\text{a}}}{10^{-15} \text{g.cm}^{-3}} \right) \left( \frac{M}{\text{M}_\odot} \right)^{1/2} \left( \frac{r_{\text{d}}}{0.1 \text{au}} \right)^{3/2} \, \text{M}_\odot.\mbox{yr}^{-1} \nonumber \\
        P_{\text{jet}}&= 1.26 \times 10^{12} \left( \frac{\rho_{\text{a}}}{10^{-15} \text{g.cm}^{-3}} \right) \left( \frac{M}{\text{M}_\odot} \right)^{3/2} \left( \frac{r_{\text{d}}}{0.1 \text{au}} \right)^{1/2} \, \text{ergs.s}^{-1} \nonumber \\
        B_0&= 10.6 \left( \frac{\rho_{\text{a}}}{10^{-15} \text{g.cm}^{-3}} \right)^{1/2} \left( \frac{M}{\text{M}_\odot} \right)^{1/2} \left( \frac{r_{\text{d}}}{0.1 \text{au}} \right)^{-1/2} \, \mbox{G} \nonumber \\
        T_{\text{d}}&= 1.8 \left( \frac{M}{\text{M}_\odot} \right)^{-1/2} \left( \frac{r_{\text{d}}}{0.1 \text{au}} \right)^{3/2} \, \mbox{days.} 
\end{split}
\end{equation}

%%%%%%%%%%%%%%%%%%%%%%%%%%%%%%%%%%%%%%%%%%%%%%%%%%%%%%%%%%%%%%%%%%
\section{The fiducial case} \label{sec:ReferenceStationaryCase}
%%%%%%%%%%%%%%%%%%%%%%%%%%%%%%%%%%%%%%%%%%%%%%%%%%%%%%%%%%%%%%%%%%

\subsection{General description} 
%%%%%%%%%%%%%%%%%%

Several simulations have been performed using the above setup and have all reached a steady-state. They are listed in Table~\ref{tab:ParametresSimusTronquees}. In this section we present the reference simulation O1, but they all share the same basic properties. Simulation O1 was done with no rotation at the axis ($\Omega_{*_{\text{a}}} = 0$) so as to exemplify the role of the disc as the only electromotive force (e.m.f.) present in the system, and with a moderate external magnetic field $B_{\text{ext}} = 2 \times 10^{-3}$. 

%% electric circuit
Fig.~\ref{fig:ReferenceSimulationTruncated} shows the final outcome at $t_{\text{end}} = 2.68 \times 10^{7}$ of O1. The global view on the left clearly shows a super-FM spine and jet of finite radius, surrounded by a sub-Alfv\'enic outflowing atmosphere. The poloidal electric current, which defines the global electromagnetic system described here as a super-FM outflow, is mostly flowing within a zone of a finite radial size around 650 at $z=4000$. This is in strong contrast with the `self-similar' situation obtained in \citetalias{jannaud2023} and is a direct consequence of a launching zone with a finite radial extent. Note however that both the jet transverse equilibrium and acceleration are self-consistently determined. The right panel displays a zoom of the launching region, showing the typical butterfly shape of the electric current circuit: the collimating current flows down along the axis towards the central engine (outside our domain inside $R=1$) and goes up along the outflow emitted from disc up to $r_0=12$, closing progressively radially inside the jet along with plasma acceleration. This situation, namely this radial electric current stratification and its progressive decline with altitude, is perfectly consistent with the expectations from a stationary accretion-ejection structure.  

\begin{figure}
\centering
    \includegraphics[trim=0 10 0 5,clip,width=\linewidth]{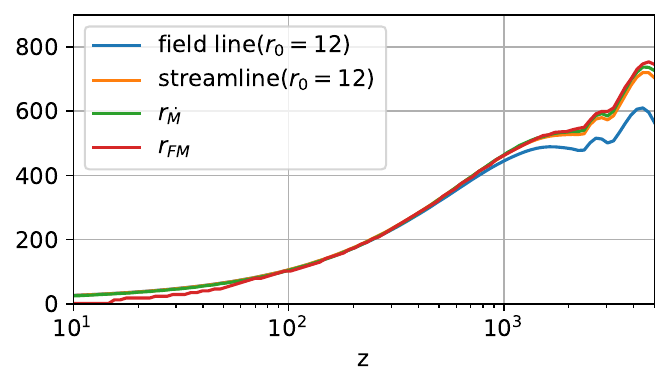}
\caption{Evolution of various cylindrical radii with altitude $z$ along the jet for simulation O1. In blue and orange are respectively the magnetic field lines and streamlines anchored on the disc $r_0 = 12$. In red is the radius $r_{\text{FM}}$ of the FM surface. In green is the radius $r_{\dot{M}_{\text{j}}}$ inside which the total ejected mass flux is located. For a YSO jet, this equates to a jet collimated down to a radius of about 70 au at 500 au from the disc.}
\label{fig:RadiiEvolution}
\end{figure}

%%% radii and definitions
The major goal of this work is to relate the initial launching conditions (e.m.f., mass flux from the JED) to the jet asymptotic state, which requires addressing extremely large spatial and time scales. While the latter has been nicely accommodated with a specially designed temporal acceleration scheme (see details in \citetalias{jannaud2023}), it was prohibitive to maintain a perfect spatial resolution in spherical coordinates up to very large distances. 
Fig.~\ref{fig:RadiiEvolution} illustrates this point. This graph shows the evolution of several radii with altitude at the final state of the simulation. In red is the FM radius $r_{\text{FM}}$, which is also the radius of the jet interface with the ambient medium (outflowing atmosphere) in Fig.~\ref{fig:ReferenceSimulationTruncated}. The radius of the last JED magnetic surface (anchored at $r_0 = r_{\text{j}} = 10$) is shown in blue. While the two are superimposed almost up to $z=10^3$, a clear deviation appears beyond: at $z = 5000$, $r(r_0 = 10) \simeq 360 \ll 650 \simeq r_{\text{FM}}$. If, starting from the outer boundary, we were to trace back to the disc the anchoring radius $r_0$ of the last {\em field line}\footnote{We define a field line as a curve tangent to $\vec{B}_{\text{p}}$ everywhere.} that becomes super-FM, we would find $r_0 =55$. But this makes no sense since there is no mass being steadily accelerated beyond $r_0=12$. This is because that ejected material is actually drifting outwardly because of unavoidable numerical diffusion at large distances. 

We checked this in two independent ways. First, we computed the {\em streamline}\footnote{We define a streamline as a curve tangent to $\vec{u}_{\text{p}}$ everywhere.} emitted from the anchoring radius $r_0=12$ (there is no MHD Poynting flux beyond this point). The radius $r(z)$ of this streamline is shown in orange in Fig.~\ref{fig:RadiiEvolution} and it follows closely (to numerical uncertainties) $r_{\text{FM}}$, as expected. As a second test, we computed the radius $r_{\dot{M}_{\text{j}}}(z)$ where all the ejected mass (in the spine and from the disc up to $r_0=12$) is actually recovered at each altitude $z$. Its precise definition is 
\begin{equation}\label{eq:DefRMdotJ}
    \dot{M}_{\text{spine}} + \dot{M}_{\text{jet}} = 2 \pi \int_{r=0}^{r=r_{\dot{M}}(z)} \rho(r,z) u_z (r,z) r dr \, 
\end{equation}
\noindent where the mass fluxes on the left-hand side are fixed and imposed by the boundary conditions (section \ref{sec:BoundaryConditions}). The radius  $r_{\dot{M}_{\text{j}}}(z)$ is shown in green and, downstream of $z\simeq 60$, follows $r_{\text{FM}}$ as expected. Thus, for all practical means, we will hereafter refer to the `jet' as the super-FM outflow emitted from the JED and transition zone ($r_\text{0}\in [1;12]$). The `spine' refers to the super-FM outflow flowing along the axis and corresponds to the mass ejected from $r_\text{0}\in [0;1]$. Indeed, mass diffusion is almost negligible along the axis (as shown in Sect~3.4). Note that the spine itself is not visible in the left panel of Fig.~\ref{fig:ReferenceSimulationTruncated}, as the innermost field line (probed by its velocity vector) is anchored at $r_0 = 4$.  But ideal MHD does not hold perfectly for the outer streamlines beyond a distance that is several hundred times the inner radius. This is a caveat of the numerical procedure used, but it does not impact our results or our main conclusions.     

%%%%%%%%%
%%%%%%%%%%%

%% valeurs num, profils 
The values of the mass flux and power carried in the jet and in the spine are shown in Table~\ref{tab:ParametresSimusTronquees}. For our chosen density contrast $\delta= \rho_d/\rho_a=10^2$, the mass flux carried by spine is roughly a third of the total mass flux ejected at super-FM speeds, while its power is about half of the total power. This ratio could easily be changed (up and down), mostly by changing  $\delta$. However, it sounds reasonable to us to expect much larger values for $\delta$ in real astrophysical systems. As for the outer outflowing atmosphere, the outflow rate (computed for all radii beyond $r_0=12$) remains less than 1\% of the JED mass outflow rate. Note that this holds for all the simulations done in this paper. 

\begin{figure}
\centering
    \includegraphics[trim=0 10 0 5,clip,width=\linewidth]{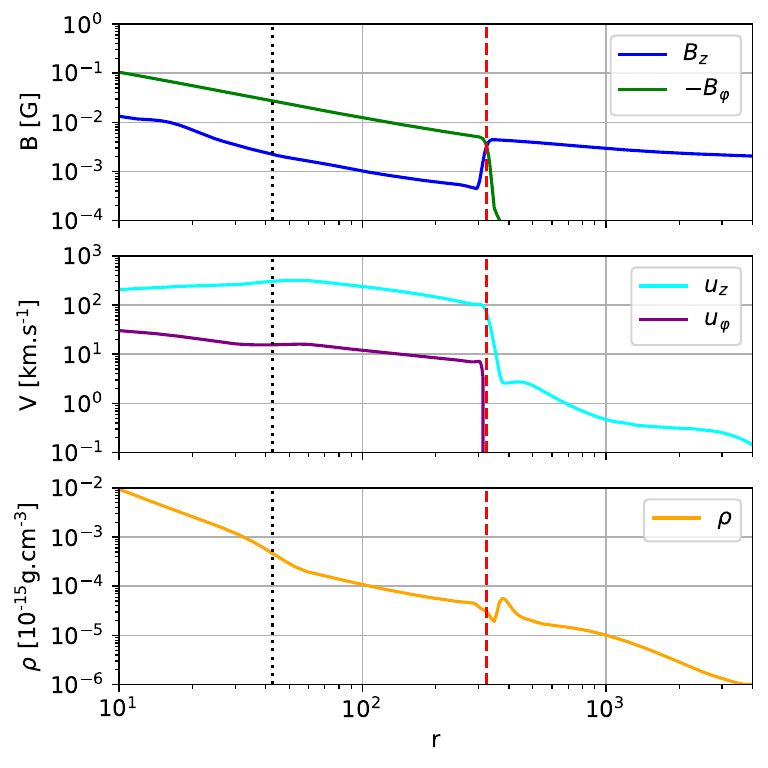}
\caption{Radial (cylindrical) distributions of several jet quantities (in physical units, see section \ref{sec:Normalization}) along the horizontal $z=500$ for simulation O1. The black dotted line marks the outer bound of the spine (field line anchored at $r_0=1$), while the red dashed line marks the interface between the jet and the ambient medium (the FM critical surface). Beyond this interface resides a sub-Alfv\'enic (hence unsteady) atmosphere.}   
\label{fig:Z500Profiles}
\end{figure}

Fig.~\ref{fig:Z500Profiles} shows the radial (cylindrical) distributions of several relevant quantities at an intermediate altitude $z=500$. The red dashed line marks the position of the FM critical surface and the outer edge of the jet, whereas the black dotted line is the spine/jet interface. The ratio $|B_\varphi/B_z | \sim 10$ is quite high throughout all the super-FM outflow, in agreement with the usual assumptions in super-FM outflows (e.g. \citealt{Pelletier1992}). The toroidal field vanishes nevertheless at the axis and decreases sharply at the jet/atmosphere interface. Since our outflow is cold, the total pressure is dominated by the magnetic pressure in the whole computational domain. As a consequence, what confines the jet, namely what determines its radius $r_{\text{FM}}$, is a force balance between the outwardly directed toroidal magnetic pressure with the outer, inwardly directed, poloidal magnetic pressure.     

The central panel in Fig.~\ref{fig:Z500Profiles}  displays the radial profiles of the two dominant velocity components. With no surprise, rotation is negligible with respect to the vertical velocity, whose distribution appears rather flat with a slight peak at the spine/jet interface. However, since the density profile is much steeper, (close to $r^{-2}$), the jet thrust is dominated by the inner regions.

Let us consider the normalisations used in section \ref{sec:Normalization}. The jet is collimated down to a radius of 70 au at z=500 au from its source (see Fig. \ref{fig:RadiiEvolution}). It has a density weighted jet velocity of $169 \, \text{km.s}^{-1}$, a total (jet + spine) one-sided mass loss rate of $1.85\, 10^{-8} \, \text{M}_\odot.\mbox{yr}^{-1}$, a total one-sided power of $3.7 \, 10^{14} \, \text{ergs.s}^{-1}$ and a vertical field of $10^{-3} B_\text{d} = 0.1 \, \text{G}$ at $z=50 \, \text{au}$ on the axis.

%%%%%%%%
%%%%%%%%

\subsection{Standing recollimation shocks}
%%%%%%%%%%%%%%%%%%%%%%%

%%%%%%%%
\begin{figure}
\centering
  \includegraphics[width=\linewidth]{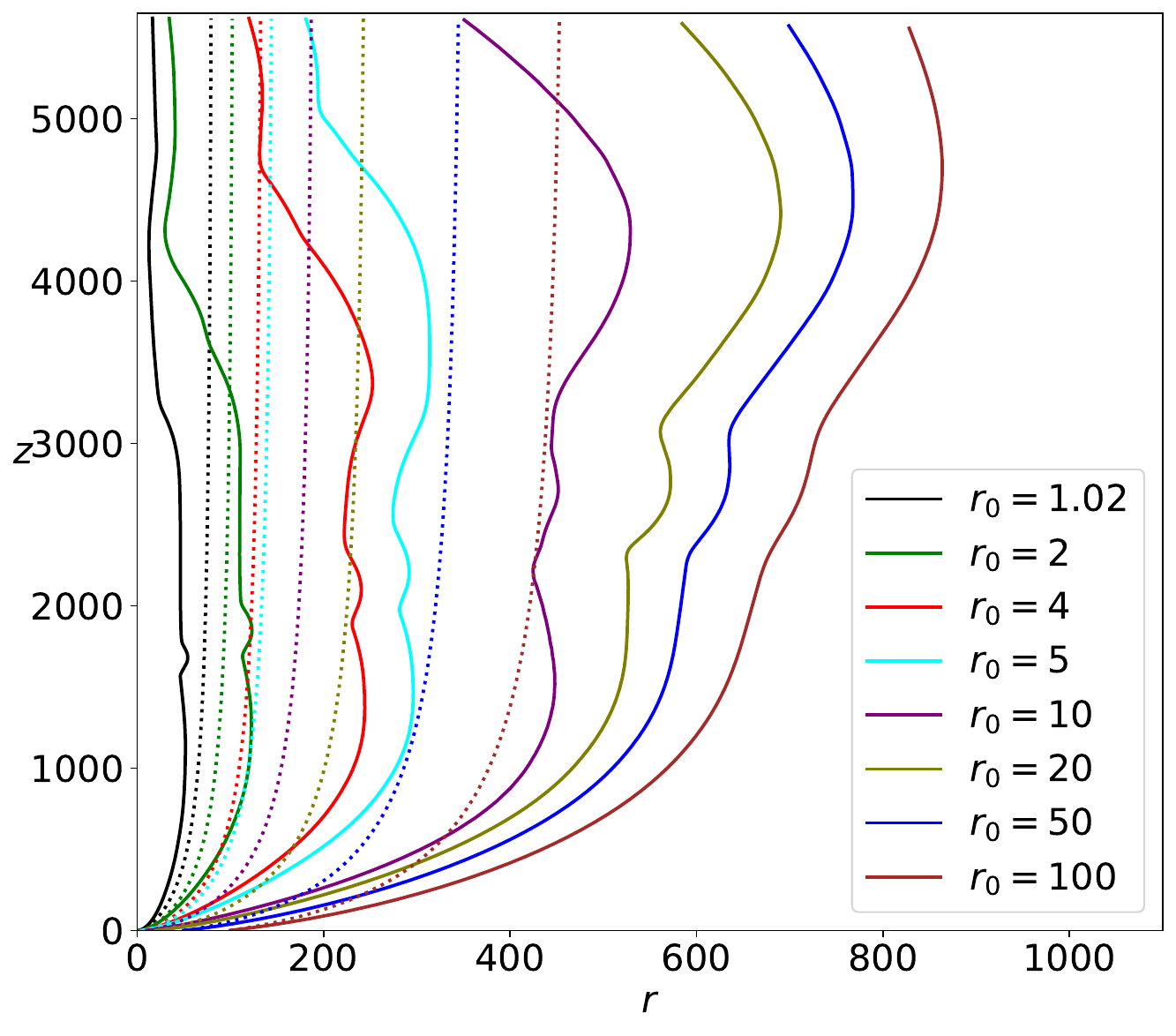}
\caption{Shape of magnetic surfaces anchored at the different anchoring radii $r_0$ (see inlet) for simulation O1. The initial magnetic configuration is potential and shown in dotted lines. The final (at  $t_{\text{end}}$) full MHD configuration is shown with the same color in solid lines. The wiggles seen in final magnetic surfaces reveal the position of several shocks. Note that only surfaces anchored up to $r_0=10$ are stationary.}
\label{fig:EvolutionFieldLinesTronquees}
\end{figure}
%%%%%%

The intersections of these standing shocks with the field line anchored on the disc at $r_0 = 4$ are shown by colored dots on Fig.~\ref{fig:ReferenceSimulationTruncated}. The first two (blue and purple) leave the polar axis at $Z_{\text{shock}} = 1600$, and extend until the jet/medium interface. The last one (green) leaves the polar axis at $z \sim 4200$; this shock is harder to study, partly because it is not fully enclosed in the simulation domain, and partly because it is located in a place where the grid cells are large ($\Delta R \propto R$). For this reason, we will mostly focus on the other two (blue and purple). These recollimation shocks are similar to those observed in the simulations of \citetalias{jannaud2023}, with different opening angles while merging at the same point near the axis. 

As first proposed in \citet{Pelletier1992} and later confirmed and analyzed by \citet{ferreira1997} and \citetalias{jannaud2023}, standing recollimation shocks are an intrinsic property of MHD self-collimation. The reason invoked was the pinching force arising from the toroidal magnetic field component that dominates in the asymptotic regime. This refocusing always occurs once the flow is super-FM so that only an oblique shock can deviate it along the axis. \citetalias{jannaud2023} have shown that these are weak shocks that induce a small density compression, a decrease of the toroidal velocity (hence an increase of the post-shock electric current) and a refraction of the magnetic surfaces. The post-shock material is then again reaccelerated with a reopening of the magnetic surface, becomes super-FM again and may thereby undergo a new recollimation towards the axis farther up (and a new shock). As such, each shock acts as a local e.m.f. that drives a new poloidal electric circuit, mostly disconnected from the source (central engine and accretion disc). The presence of these shocks in jets emitted from a JED of finite radial extent, hence of a finite MHD Poynting flux, is quite a strong result. It shows that their formation is not a bias of self-similarity, but a consequence of the underlying disc ejection conditions. This has important consequence that will be discussed in section \ref{sec:Discussions}. 

Nevertheless the presence of these standing shocks, with a corresponding wiggling of the outer jet interface (seen for instance at $z \simeq 2200 $ and $z \simeq 3050$ in Fig.~\ref{fig:ReferenceSimulationTruncated}), may have observational counterparts since the jet radius is non-monotonous with the distance. This is best seen in Fig.~\ref{fig:EvolutionFieldLinesTronquees}, where several magnetic surfaces anchored at radii going from $r_0=1$ to 100 are shown. These shocks follow the MHD characteristics (see e.g. fig.~3 in \citealt{Ferreira2004}) and are propagating radially inside-out, affecting the surrounding unsteady atmosphere (field lines anchored beyond $r_0=10$).

\subsection{MHD self-collimation and confinement}

Fig. \ref{fig:EvolutionFieldLinesTronquees} is also particularly illustrative of the self-collimation process inside the jet. For magnetic field lines anchored in the disc at various radii $r_0$, the initial state (potential field) is plotted in dotted lines and the final state (full MHD) in solid lines. Comparing the two is striking because of the non-monotonous behavior of the full MHD solution. However, looking at the largest distance from the source, we find that only the magnetic surfaces anchored below $r_0=5$ have reached a smaller radius than their initial potential configuration. All surfaces anchored between $r_0=5$ and 10 have been pushed against the outer medium, whose pressure allowed a transverse equilibrium.

Thus while we define the jet as the super-FM material ejected from $r_0 \in [1;12]$, only its inner region is self-collimated. It is customary to assume that the electric current\footnote{Defining $I=- 2 \pi rB_\varphi/\mu_o >0$ as the total current carried by a magnetic surface of radius $r$ at a given altitude $z$, then the MHD Poynting flux writes $\vec{ S}=\Omega_* I \vec{B}_p/2\pi$.} is entirely carried in by the jet so that if we define $r_I$ as the radius where the current vanishes ($B_\varphi=0$), one expects $r_I \lesssim r_{\text{FM}}$ (most works on jet asymptotics enforce an equality). Magnetic self-collimation refers to the Z-pinch that is provided when the hoop-stress ($-B_\varphi^2/r$, always directed towards the axis) dominates the outwardly directed toroidal magnetic pressure term $-\partial (B_\varphi^2/2)/\partial r$. This occurs when $J_zB_\varphi >0$, requiring a negative electric current density $J_z$. Since the electric current must be closed within the same outflowing structure ($\nabla \cdot \vec J = 0$), the outer parts of the jet must carry a positive electric current density $J_z>0$, leading to $J_zB_\varphi <0$. Within the jet there is therefore a radius $r_m$ where $J_z=0$ and the electric current achieves its maximum value\footnote{fig. 6 of \cite{Rezgui2025} is a clear illustration of that.}. Only the jet region below $r_m$ is magnetically self-collimated, whereas the jet radius  $r_{\text{FM}}$ is determined by the outer pressure $P_{\text{ext}}$. How close $r_m$ is to $r_{\text{FM}}$, or what fraction of the jet is actually self-collimated, depends on the subtle radial (transverse) distribution of the electric current density. For our simulation O1 and at the outer edge of the box, the self-collimated region appears to be $r_0 \lesssim 5$ or $r \lesssim 200$.

\subsection{Time evolution}
%%%%%%%%%%%%%%%%%%

\begin{figure}
\centering
    \includegraphics[trim=0 10 0 5,clip,width=\linewidth]{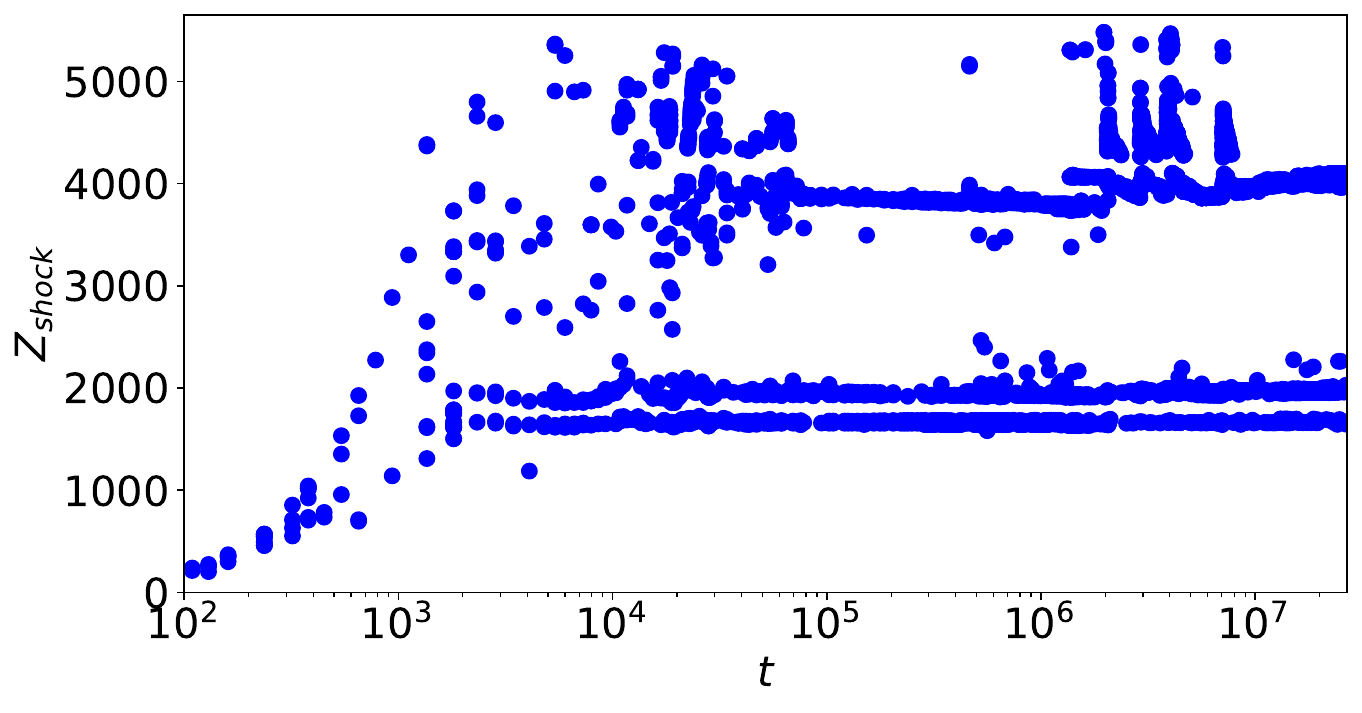}
\caption{Evolution of the shock altitudes over time detected on the field line anchored at $r_0 = 2$ for simulation O1.}
\label{fig:ShockAltitudeEvolution}
\end{figure}

We described in \citetalias{jannaud2023} the way the global poloidal electric circuit develops itself inside the jet. As it propagates through the ambient medium, the current follows a series of sudden adjustments over very long time scales. This is shown in Fig.~\ref{fig:ShockAltitudeEvolution}, where we plot the altitude $Z_{\text{shock}}$ of all shocks detected in the simulation over time, for the same  magnetic field line anchored on the disc at $r_0 = 2$. This value was chosen small to pick up shock altitudes near the axis, but not too small to limit boundary effects on the axis.

The first few $10^3 \, T_d$ correspond to the time it took the jet to reach the outer boundary, namely approximately $t_{\text{ext}}(r_0) = Z_{\text{ext}}(r_0)/u_z(r_0)$, with $v_z(r_0)$ a few $V_K(r_0)$ so that $t_{\text{ext}} \propto \sqrt{r_0}$ within the JED. Shortly after, the two main shocks appear (blue and purple dots on Fig.~\ref{fig:ReferenceSimulationTruncated}). They are located at altitudes $z \simeq 1600$ and $z \simeq 1800$. The positions of these two shocks remain relatively unchanged until the final state of the simulation, at $t_{\text{end}} = 2.68 \times 10^{7}$. This illustrates the fact that these first recollimation shocks, associated with the first acceleration zone of the plasma from its source, have indeed reached a steady-state. These shocks appear much faster than in the simulations of \citetalias{jannaud2023}, because jet arises here from a much smaller region. It takes therefore much less time for a global radial (cross-field) steady-state adjustment of the jet.   

A third shock is detected at $z \geq 4000$, which corresponds to the green dot on Fig. \ref{fig:ReferenceSimulationTruncated}. As it is located closer to the outer boundary, its detection is rather unreliable: it is a much weaker shock, located in a region of particularly low resolution. That is why it takes longer to be detected, and probably also why its altitude appears to change beyond $t \sim 10^6$. Nevertheless, it has little influence upstream, and in particular on the lower shock system. This phenomenon is rather interesting though and would require further investigations (beyond the scope of this paper). We note however that it is reminiscent of a transitional period that preceded the formation of an additional shock in the reference simulation of \citetalias{jannaud2023}, right before it reached a stationary state.

%However, this shock informs us of an interesting event happening between $t \simeq 3 \times 10^6$ and $t \simeq 8 \times 10^6$: quasi-periodic wobblings of the jet. On Figure \ref{fig:ShockAltitudeEvolution} it is evidenced by oscillations in the altitude of the highest shock, as well as detections of pseudo-shocks downstream. We also see that this wobbling has little influence upstream, and in particular on the lower shock system. This phenomenon, although transitional, is rather interesting and will be the subject of a following paper. It is very reminiscent of the transitional period that preceded the formation of an additional shock in the reference simulation of \citetalias{jannaud2023}, right before it reached a stationary state.

Also, while \citetalias{jannaud2023} had five shocks, only three were detected here. One reason may be the smaller resolution used here, as these shocks are rather weak (see section 5.2.1 of \citealt{janna2023} for a more thorough description of the impact of resolution on shock properties). Although recollimation is intrinsic to MHD collimation, the position and henceforth number of shocks appearing within the computational domain depend also on the outer confining medium. So this may be another reason for this difference. 

%%%%%%%
%%%%%%%%

\subsection{Comparison to MHD jet theory}
%%%%%%%%%%%%%%%%%%%%%%%%%%

%%%%%%%%
\begin{figure}
\centering
    \includegraphics[trim=0 10 0 5,clip,width=\linewidth]{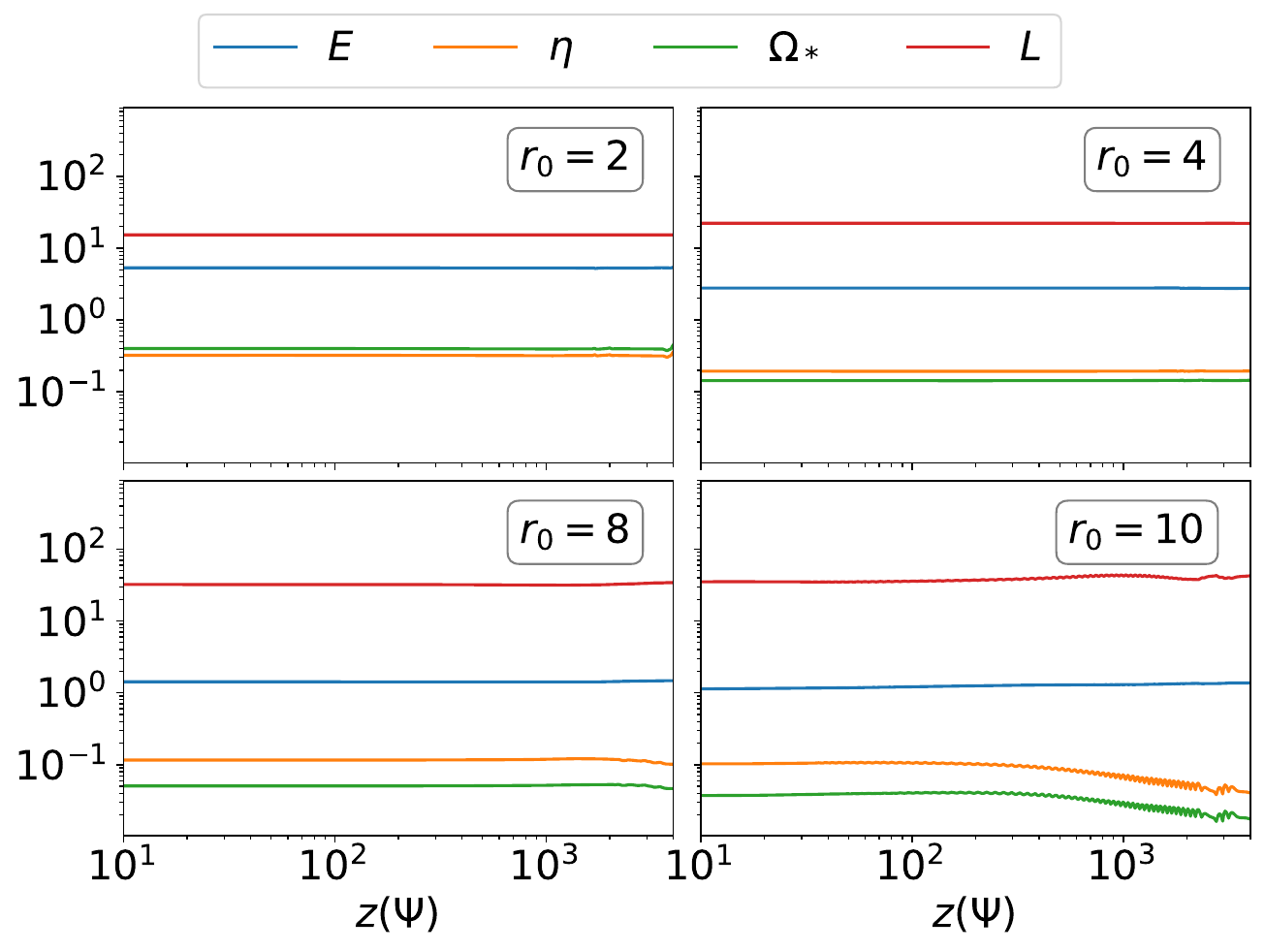}
\caption{Evolution of the 4 MHD invariants with altitude along various field lines anchored in the Jet-Emitting Disc at $t_{\text{end}}$, for simulation O1.}
\label{fig:Invariants}
\end{figure}
%%%%%%%%

Let us now come back to the final steady-state of the jet and verify how the magnetic surfaces behave compared to what is expected from steady-state, ideal MHD, axisymmetric jet theory. In the case of polytropic jets of index $\Gamma$, each magnetic surface should possess five MHD invariants, namely
\begin{itemize}
    \item the mass to magnetic flux ratio $\eta = \mu_0 \rho u_p / B_p $,
    \item the rotation rate of the magnetic surface $\Omega_* = \Omega - \eta B_\upvarphi / (\mu_0 \rho r)$
    \item the total specific angular momentum $L = \Omega r^2 - r B_\upvarphi / \eta$,
    \item the Bernoulli invariant $E = u^2 / 2+H+\Phi_G-\Omega_* r B_\upvarphi / \eta$,
    \item the specific entropy $S = P / \rho^\Gamma$,
\end{itemize}
\noindent where $\Omega = u_\upvarphi / r$ and $H = \frac{\Gamma}{\Gamma - 1} \frac{P}{\rho}$ is the enthalpy. These invariants are only a function of the magnetic flux $\Psi$ and should thus remain constant along magnetic field lines.

\subsubsection{MHD acceleration}

On Fig.~\ref{fig:Invariants} we represent the evolution of four\footnote{The specific entropy is a constant by definition in our setup.} of these quantities along several magnetic field lines anchored in the JED. We see very little variation, confirming the good agreement between our simulations and steady-state ideal MHD theory. The simulated jets here, despite being launched from a finite radial zone, exhibit the same relations and properties (such as $z_A\sim r_A$) than the classical self-similar cold solutions \citep{ferreira1997}. However, we notice that for $r_0 = 8$ and even worse fo $r_0=10$, the invariants $\eta$ and $\Omega_*$ (very sensitive to $\eta$) decrease for $z \gtrsim 500$. As discussed previously, this is caused by a (numerical) outward mass diffusion. This is unavoidable here given the huge spatial scales involved.  

%%%%%%%%
\begin{figure}
\centering
    \includegraphics[trim=0 10 0 5,clip,width=\linewidth]{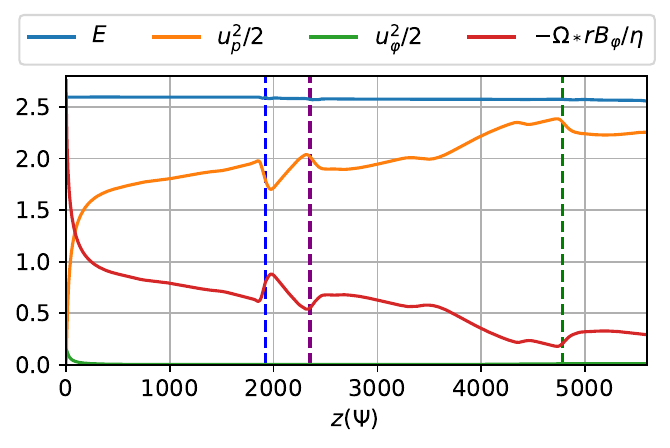}
\caption{Bernoulli invariant $E$ and evolution of its components (see inlet) along a magnetic surface of anchoring radius $r_{0} = 4$ at $t_{\text{end}}$, for the simulation O1. Gravity and enthalpy (negligible here) have been omitted for the sake of clarity. The Alfv\'en point ($m=1$) is located at $z_A= 10$ and the FM point ($n=1$) is at $z_{\text{FM}}= 33$. The vertical dashed lines correspond to the altitudes of the colored dots on Fig.~\ref{fig:ReferenceSimulationTruncated} where the surface encounters  the shocks: $z=1920$ (blue), $z=2350$ (purple) and $z=4780$ (green).}
\label{fig:EnergyTruncatedSimulation}
\end{figure}
%%%%%%%%

Fig.~\ref{fig:EnergyTruncatedSimulation} details the contributions to the Bernoulli invariant $E$ along the magnetic surface anchored at $r_0 = 4$. That surface is the second innermost one on Fig.~\ref{fig:ReferenceSimulationTruncated}, represented by arrows on the left panel and in green on the right panel. Enthalpy is negligible for our cold jets and gravity becomes also rapidly negligible with distance so they have both been omitted for the sake of clarity. This behavior is in line with what is expected from steady-state theory, and with what was seen in the simulations of \citetalias{jannaud2023}. Along this field line, the MHD invariants vary by less than 2\%, even through the various shocks (all invariants are conserved in polytropic MHD shocks). 
The presence of the three shocks can be seen at altitudes $z \simeq 1920, 2350, 4780$, evidenced by the blue, purple and green vertical lines\footnote{Some wiggles, seen at $z\simeq 3500, 4200$ in the kinetic and magnetic energy curves, are probably due to even weaker shocks that our post-processing procedure has missed. A look at Fig.~\ref{fig:EvolutionFieldLinesTronquees} shows that the shape of the $r_0=4$ magnetic surface is (almost) unaffected.}. They always induce a decrease of poloidal and toroidal plasma velocity and a localized increase of the toroidal magnetic field. This post-shock increase of toroidal field is associated with a new post-shock poloidal electric circuit: the shock acts as a new e.m.f.  

The total energy is conserved and, far from the source ($z>1000$), it is composed of two major components. The magnetic term is dominant near the disc and, as the magnetic surface opens up and the flow is accelerated, the poloidal kinetic term rises composing the vast majority of the stored energy at the outer boundary. At large distances, the poloidal Alfv\'enic Mach number $m= u_p/V_{Ap} \sim 40$ becomes much larger than unity and the magnetic surface radius achieves $r(z)\sim 150 > r_A$ with $r_A\sim 14$ being the Alfv\'en radius (cylindrical radius where $m=1$). As a consequence, the usual analytical approximations are rather valid for this magnetic surface for $z>1000$. Under the limits $m^2 \gg 1, r \gg r_A$,  one gets $u_\varphi \simeq \Omega_* r_A^2/r$ which is much smaller than the poloidal speed $u_p$.

%%%%%%%%
\begin{figure}
\centering
  \includegraphics[trim=0 8 0 5,clip,width=\linewidth]{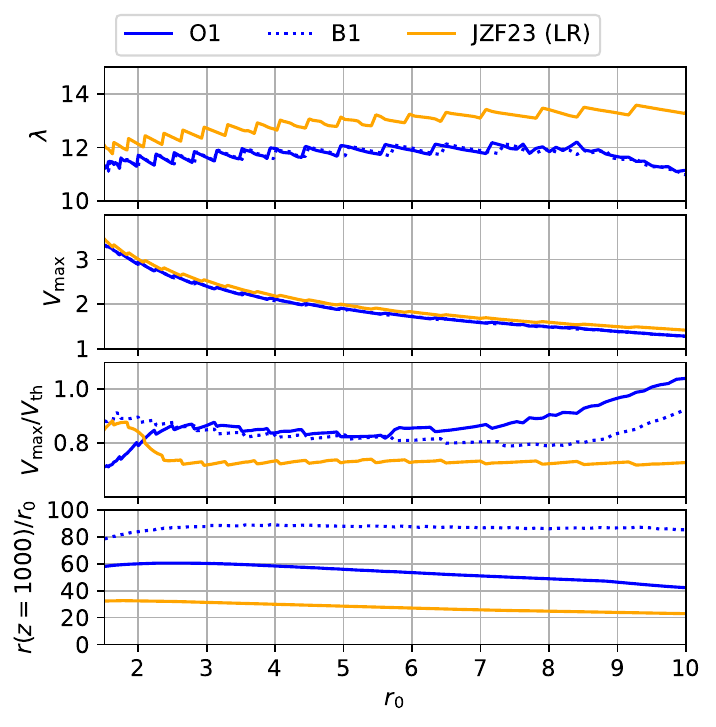}
\caption{Transverse distributions of several jet quantities as function of the anchoring radius $r_0$ within the JED (from top to bottom): magnetic lever arm $\lambda$, asymptotic speed $V_{\text{max}}$ (in code units) achieved at $R_{\text{ext}}$, ratio $V_{\text{max}}/V_\infty$ and radius $r(z=1000)/r_0$ at $z=1000$ of the magnetic surface, normalized to its anchoring radius $r_0$.  In blue are shown two simulations from this work: the fiducial simulation O1 in solid lines, and the simulation B1 without external magnetic field in dotted lines. Orange solid lines display the low-resolution reference simulation of \citetalias{jannaud2023}.}
\label{fig:EvolutionLambda}
\end{figure}

\subsubsection{Asymptotic behavior}

It is thus relevant to compare the jet speeds attained in our simulations to steady-state asymptotic jet theory. In that ansatz, the critical parameter is the magnetic arm lever $\lambda$, defined by \cite{blandford1982} as $\lambda (r_0) \equiv (\Omega_* r_{\text{A}}^2)/(\Omega_{K_0} r_0^2)$, for each magnetic surface. Along each surface, the asymptotic theoretical speed would be $ V_{\text{th}} \equiv V_K(r_0) \sqrt{2 \lambda -3}$, when all available initial magnetic energy has been finally converted into poloidal kinetic energy. Fig.~\ref{fig:EvolutionLambda} shows the transverse distribution of the magnetic lever arm $\lambda$ as function of the anchoring radius $r_0$ within the JED (top), as well as the field line radius $r_{\text{FM}}$ normalized to its anchoring radius $r_0$ (bottom) for two simulations. Our reference simulation O1 is shown in blue solid line and a comparable (done at low resolution) `self-similar' simulation of \citetalias{jannaud2023}, done with the same JED parameters (orange solid line). We are thus probing here the effect of truncating the JED size.
   
The magnetic lever arm $\lambda$ remains roughly constant (around 12) throughout the whole JED. In all cases, the jet nearly reaches its theoretical asymptotic speed. The second panel shows $V_{\text{max}}$, poloidal speed reached on the magnetic surface at the outer edge of the domain. The third panel shows that speed being very close to $ V_{\text{th}} $, indicating that the jet is (close to) its asymptotic state. We note however a slight discrepancy with the `self-similar' simulation, which provides systematically larger  $\lambda$ (by 10\%), hence larger ratios $r_A/r_0$. But this does not mean that the asymptotic jet radius is also larger, as evidenced in the bottom panel. In fact, the jet radius (as measured at this altitude) is actually smaller in the `self-similar' case than in the finite size case, not larger (as one would naively extrapolate from the behavior of $r_A$). In the `self-similar' case, the outer (beyond $r_0=10$) pressure, due to an outflowing wind, is greater than the pressure due to our initial (static) atmosphere  in the O1 case. This naturally changes how the jet is collimated, and thus accelerated. The effect of the external pressure on jet dynamics must therefore be investigated.

%%%%%%%%%%%%%%%%%%%%%%%%%%%%%%%%%%%%%%%%%%%%%%%%%%%%%%%%%%%%%%%%%%
\section{Steady-state jets: Parametric studies} \label{sec:ParametricStudies}
%%%%%%%%%%%%%%%%%%%%%%%%%%%%%%%%%%%%%%%%%%%%%%%%%%%%%%%%%%%%%%%%%%

\begin{figure*}
\includegraphics[trim=0 0 0 0, clip, height=.5\linewidth]{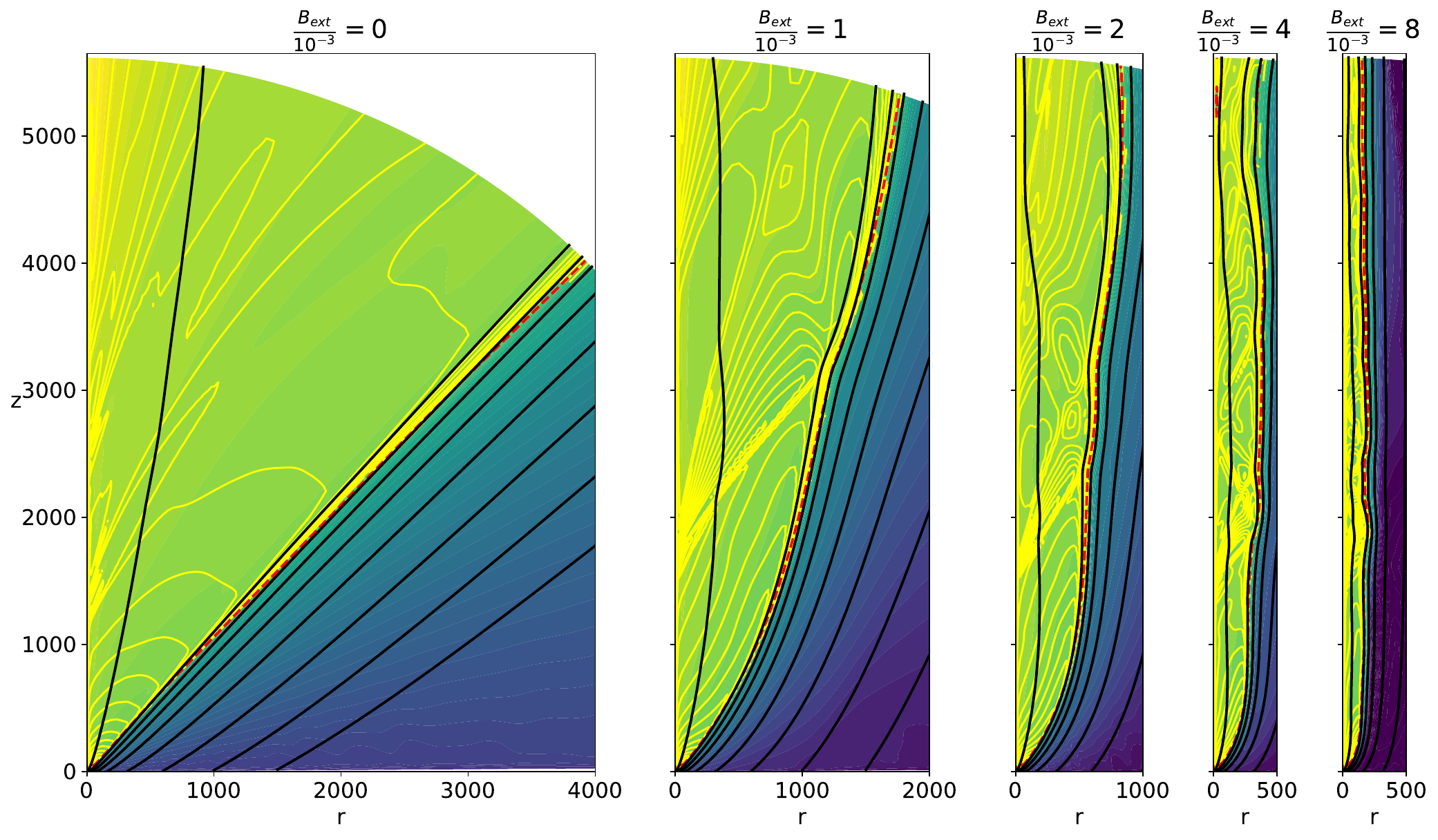}
\caption{Influence of the external magnetic field $B_{\text{ext}}$ on the final stage of jets obtained with $\Omega_{*_{\text{a}}} = 0$ (from left to right: B1, B2, O1, B3, B4). The background color provides the logarithm of the FM mach number $n$. The black solid lines are the poloidal magnetic surfaces anchored in the disc at $r_0= 3; 15; 40; 80; 160; 320; 600; 1000; 1500$. The yellow solid lines are isocontours of the poloidal electrical current and red dashed lines the critical FM surfaces.}
\label{fig:SimulationsBext}
\end{figure*}

\begin{figure*}
      \includegraphics[trim=0 0 0 0, clip, height=.63\linewidth]{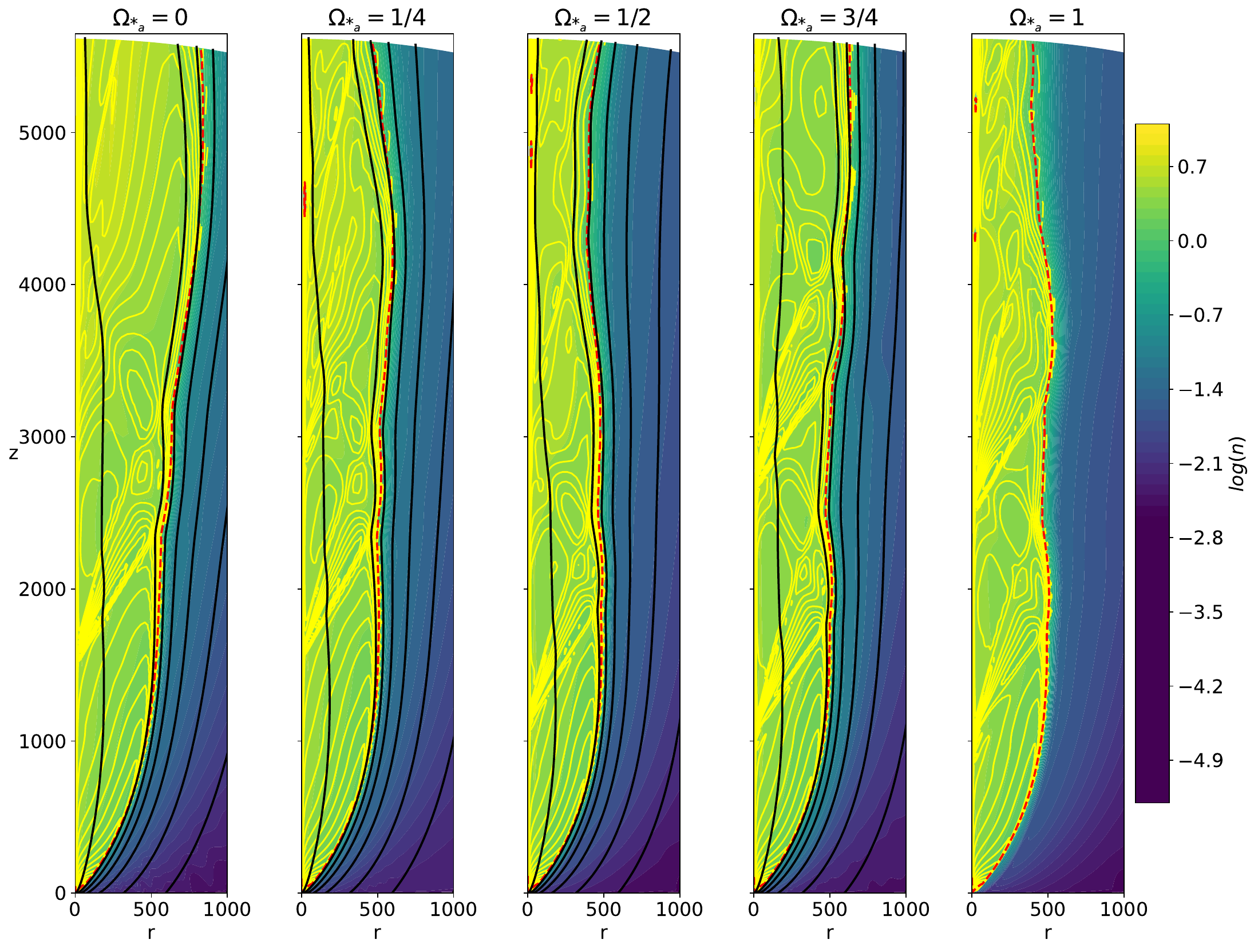}
      \caption{Influence of the axis rotation $\Omega_{*_{\text{a}}}$ on the final stage of jets obtained with $B_{\text{ext}}=2 \times 10^{-3}$ (from left to right: O1 to O5). The background color displays the logarithm of the FM mach number $n$. The black solid lines are the poloidal magnetic surfaces anchored in the disc at $r_0= 3; 15; 40; 80; 160; 320; 600; 1000; 1500$. The yellow solid lines are isocontours of the poloidal electrical current and the red lines the critical FM surfaces.}
\label{fig:SimulationsOmegaEtoile}
\end{figure*}

The platform simulation described above is the first to showcase stationary recollimation shocks in a jet carrying a finite Poynting flux, emitted from finite radial extent of a keplerian disc. To further understand the impact of jet launching conditions on collimation and shock behavior, we now explore the influence of two parameters:

\begin{itemize}
    \item The external vertical magnetic field on the disc $B_{\text{ext}}$.
    \item The rotation of magnetic surface on the axis $\Omega_{*_{\text{a}}}$.
    %\item The magnetic flux profile via $\alpha$.
\end{itemize}

All simulations have become stationary at a final time $t_{\text{end}}$ and are presented in Table~\ref{tab:ParametresSimusTronquees} and shown in Figs. \ref{fig:SimulationsBext} and \ref{fig:SimulationsOmegaEtoile}. Apart from $\Omega_{*_{\text{a}}}$ or $B_{\text{ext}}$, the setups are unchanged from one simulation to another. The first two columns describe the modified parameters, and the others show various properties of the global super-FM outflow (in code units) measured at $t_{\text{end}}$. These are the altitude $Z_\text{shock}$ of the first standing shock close to the axis, the mass loss rates in the spine $\dot{M}_{\text{spine}}$ and in the jet $\dot{M}_{\text{jet}}$ (super-FM outflow emitted from the JED) as defined in equation~\ref{eq:DefRMdotJ} and their respective total power $P_{\text{spine}}$ and $P_{\text{jet}}$. 
The total power carried by one component $\mathcal{X}$ is computed as  
\begin{equation}
P_\mathcal{X} =  \int_{\mathcal{X}}  \rho \vec{u} E \cdot \vec{dS} 
\end{equation}

Unless some specific heating mechanism would lead to the selection of particular magnetic surfaces, hence to specific speeds (such as, e.g. dissipative interaction with the ambient medium selecting only the outer streamlines), it seems reasonable to define as `the' jet speed the density-weighted velocity. And since the spine is also super-FM, there is no reason here to differentiate the spine from the jet emitted from the JED. The last two columns of Table~\ref{tab:ParametresSimusTronquees} thus provide the density-weighted jet speed $V_\text{j}$, 
\begin{equation}
   V_\text{j} =\left( \int_{\mathcal{S}}  \rho \vec{u} \cdot \vec{dS} \right) / \left( \int_{\mathcal{S}}  \rho dS \right)
\end{equation}
where $\mathcal{S}$ is a surface that depends on the distance considered. For the average velocity $V_\text{j}^{z=500}$, the surface $\mathcal{S}$ is a disc centered on the z axis at $z=500$ and of radius the last radius $r_{\text FM}$ where the jet is super-FM. For the average velocity $V_\text{j}^{R_\text{ext}}$, the surface $\mathcal{S}$ is the spherical cap at the boundary $R=R_\text{ext}$ which encompasses the super-FM jet. Clearly, MHD acceleration decays with distance but it takes a while to fully vanish. Emitted at the disc surface with a super-SM speed of $0.1 V_{K_0}$, the flow gets accelerated and reaches a velocity already close to $V_{K_0}$ at the Alfv\'en critical point \citep{ferreira1997}. This factor 10 in velocity, from the disc surface to the Alfv\'en point, is done at the expense of the MHD Poynting flux, as can be seen below $z=200$ in Fig.~\ref{fig:EnergyTruncatedSimulation}. Beyond the Alfv\'en point, acceleration is still present, although less efficient and depending on the jet confinement. It allows the plasma to possibly reach the maximum allowed speed $V_{\text{th}}$, namely to gain a factor $\sqrt{2\lambda-3}\sim 4$ in poloidal speed. The difference between the average speed  $V_{\text{j}}^{z=500}$ measured at $z=500$ and the average speed $V_\text{j}^{R_\text{ext}}$ measured at the boundary is therefore informative of this second stage of jet acceleration.  

Not surprisingly, since all platform simulations have been done with the same injection parameters and a steady-state has been reached, the mass outflow rates remain unchanged. The total power is also almost the same. This implies that, for all these cold simulations, the MHD Poynting flux has not been affected much by playing around with $B_{\text{ext}}$ or $\Omega_{*_{\text{a}}}$. Since the available MHD Poynting flux along a given magnetic surface is determined by the Alfv\'en critical point, this tells us that the influence of these parameters (if any) will only be seen farther away, affecting therefore jet confinement and collimation. Another generic result can be derived from Table~\ref{tab:ParametresSimusTronquees}: all simulations done here display standing recollimation shocks. This is an important result that points to the fact that these shocks are due to the internal MHD dynamics and are only weakly dependent on the outer conditions (at seen with our computational domain).    
   
%%%%%%%%%%%%%%%%%%%
\renewcommand{\arraystretch}{0.5}
\begin{table}
\centering
\begin{tabular}{ |p{0.3cm}||p{0.2cm}|p{0.2cm}||p{0.4cm}|p{0.4cm}|p{0.4cm}|p{0.4cm}|p{0.3cm}|p{0.5cm}|p{0.4cm}|p{0.5cm}|p{0.6cm}|p{0.6cm}|}
\hline
    Name & $\Omega_{*_{\text{a}}}$ & $\frac{B_{\text{ext}}}{10^{-3}}$ & $\frac{t_{\text{end}}}{10^5}$ & $Z_{\text{shock}}$ & $\dot{M}_{\text{jet}}$ &  $\frac{\dot{M}_{\text{spine}}}{\dot{M}_{\text{jet}}}$ & $P_{\text{jet}}$ & $\frac{{P}_{\text{spine}}}{{P}_{\text{jet}}}$ & $V_{\text{j}}^{z=500}$ & $V_{\text{j}}^{R_{\text{ext}}}$\\
\hline
\hline
    O1 & 0 & 2 & 268 & 1600  & 36.8 & 0.52 & 146 & 1.07 & 1.80 & 3.54\\
\hline
\hline
    O2 & 1/4 & 2 & 99.2 & 1600 & 36.8 & 0.52 & 147 & 1.09 & 1.88 & 3.86\\
\hline
    O3 & 1/2 & 2 & 22.7 & 1150 & 36.8 & 0.52 & 147 & 1.13 & 1.94 & 4.06\\
\hline
    O4 & 3/4 & 2 & 104 & 1050 & 36.8 & 0.52 & 146 & 1.16 & 2.01 & 2.65\\
\hline
    O5 & 1 & 2 & 22.6 & 1000 & 36.8 & 0.52 & 147 & 1.19 & 2.05 & 3.99\\
\hline
\hline
    B1 & 0 & 0 & 22.9 & 1250 & 36.6 & 0.53 & 146 & 1.07 & 1.82 & 4.93\\
\hline
    B2 & 0 & 1 & 25.5 & 1500 & 36.7 & 0.53 & 146 & 1.07 & 1.82 & 3.81\\
\hline
    B3 & 0 & 4 & 99.8 & 1450 & 37.0 & 0.52 & 147 & 1.06 & 1.80 & 3.06\\
\hline
    B4 & 0 & 8 & 29.3 & 1550 & 37.3 & 0.52 & 148 & 1.05 & 1.75  & 2.92\\
\hline
\end{tabular}
\begin{comment}
\hline
    AT1 & 0 & 2 & 15/16 & 66.4 & 1500 & 0.075 & 0.075 & 58.6 & 0.26 & 172 & 0.65 & 1.16 \\
\hline
    AT2 & 0 & 2 & 1 & 0.012 & DNE & DNE & DNE & 69 & 0.54 & 172 & 0.02 & 6.3 $\times$ 10$^{-6}$ \\
\hline
    AT3 & 0 & 2 & 17/16 & 0.014 & DNE & DNE & DNE & 82 & 0.29 & 187 & 0.02 & 7.4 $\times$ 10$^{-7}$ \\
\hline
    AT4 & 0 & 2 & 18/16 & 0.016 & DNE & DNE & DNE & 99 & 0.22 & 203 & 0.03 & 2.1 $\times$ 10$^{-6}$ \\
\hline
\end{comment}
\normalsize
\caption{Simulations of this paper with the Blandford \& Payne radial exponent $\alpha=3/4$ and run until a final time $t_{\text{end}}$. Our reference simulation is O1. The variable input parameters, $\Omega_{*_{\text{a}}}$ and $B_{\text{ext}}$, are defined in section \ref{sec:BoundaryConditions}. $Z_{\text{shock}}$ is the altitude near the axis of the lowest shock found in the simulation. $\dot{M}_{\text{jet}}$ and $P_{\text{jet}}$ are the mass flux and power emitted in the jet only. $\dot{M}_{\text{spine}}$ and $P_{\text{spine}}$ are the mass flux and power emitted in the spine only. $V_{\text{j}}$ is the density-weighted speed of the jet, either computed at $z=500$ or at the outer boundary $R_\text{ext}$. All quantities are in code units.}
\label{tab:ParametresSimusTronquees}
\end{table}

\subsection{Outer magnetic pressure}\label{sec:ExternalBfield}%%%%%%%%%%%%%%%%%%%%%%%%%%%

%% Objectives and method
Five simulations varying the constant external magnetic field $B_{\text{ext}}$ were run: B1 to B4 and O1. They are shown on Fig. \ref{fig:SimulationsBext} Because the Blandford \& Payne power law of the radial distribution of the vertical magnetic field is quite steep ($B_z  \propto r^{-5/4}$) and the grid is large, only a small variation of  $B_{\text{ext}}$ was needed to see an impact on the initial (potential) magnetic field, as illustrated in Fig.~\ref{fig:FrontiereBextTronque}. For $B_{\text{ext}}=0$ (simulation B1) the vertical magnetic field keeps its Blandford \& Payne radial  exponent all the way in the outer SAD region, whereas for $B_{\text{ext}}=8 \times 10^{-3} = 8 \times 10^{-4} B_\text{d}$ (simulation B4), the potential field levels off at a radius $\sim 300$ and becomes constant. Thus, while the initial launching conditions remain unchanged ($B_{\text{ext}} \ll B_\text{d}$) within the spine and JED, we expect significant modification of the outer jet ambient (external) pressure.     

% This limited effect on the inner ejecting disc can be seen in the variations of $\dot{M}_{\text{jet}}$ on Table \ref{tab:ParametresSimusTronquees}. As in the JED the poloidal velocity is set such that $\vec{u}_p \parallel \vec{B}_p$, increasing the vertical magnetic field increases $u_z$ and with it the jet mass loss rate. However this variation is extremely marginal, with a relative mass flux rate variation of only one percent between the two extremes. Naturally, the effect is the same on jet power.

%%%%%%%
\begin{figure}
    \centering
\includegraphics[trim=0 10 0 4, clip,width=\linewidth]{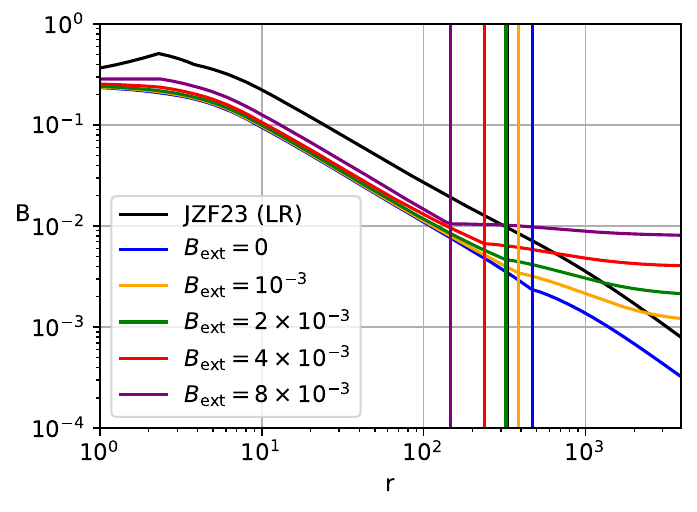}
    \caption{Radial (cylindrical) distributions of the total magnetic field $B(r)$ at $z=500$ (in code units). Each color corresponds to a simulation with increasing  $B_{\text{ext}}$ (respectively B1, B2, O1, B3, B4),. The black line is the low-resolution `self-similar' simulation of \citetalias{jannaud2023}. The vertical lines mark the positions of the FM surface at $z=500$.}
    \label{fig:B500Tronque}
\end{figure}
%%%%%%%

%%% results: confinement and collimation

The jet radius, as measured by $r_{\text{FM}}$, is strongly affected by $B_{\text{ext}}$. In Fig.~\ref{fig:B500Tronque} the vertical lines mark the position of $r_{\text{FM}}$ for the various simulations. At this altitude, it goes from $\sim 150$ to $\sim 500$, namely a factor $\sim 5$. This is quite striking since both power mass loss rates and jet power remain unaffected by $B_{\text{ext}}$. As shown in Fig. \ref{fig:SimulationsBext}, the effect is stronger at larger altitudes.

Fig.~\ref{fig:B500Tronque} also shows that, in the `self-similar' simulation of \citetalias{jannaud2023}, where a JED is established over the whole domain, the outer disc wind provides a confining pressure on the inner regions that is even stronger than in our simulation B4. As a consequence, the outflow emitted from $r_0=1$ to $r_0=10$ in this simulation has a radius even smaller (see Fig.~\ref{fig:EvolutionLambda}), although its magnetic lever arm is actually larger than B4: a larger magnetic lever arm does not necessarily give rise to a larger jet radius.  

%Indeed, a fit $z \propto r_{\text{FM}}^\omega$, done below $z=500$ (namely before the shocks come into play) provides $\omega$ increasing monotonously from 1.2 for B1\footnote{The values given here are just informative, as the shape of the jet is later modified by recollimation. For B1 for instance, since recollimation is not yet seen in $r_{\text{FM}}$ inside our computational domain, a fit up to $z=1000$ provides $\omega=1$, namely a conical jet.}  to 3.2 for B4. {\bf Check VALUES}  

\begin{figure}
\center \includegraphics[trim=0 9 0 5,clip,width=\linewidth]{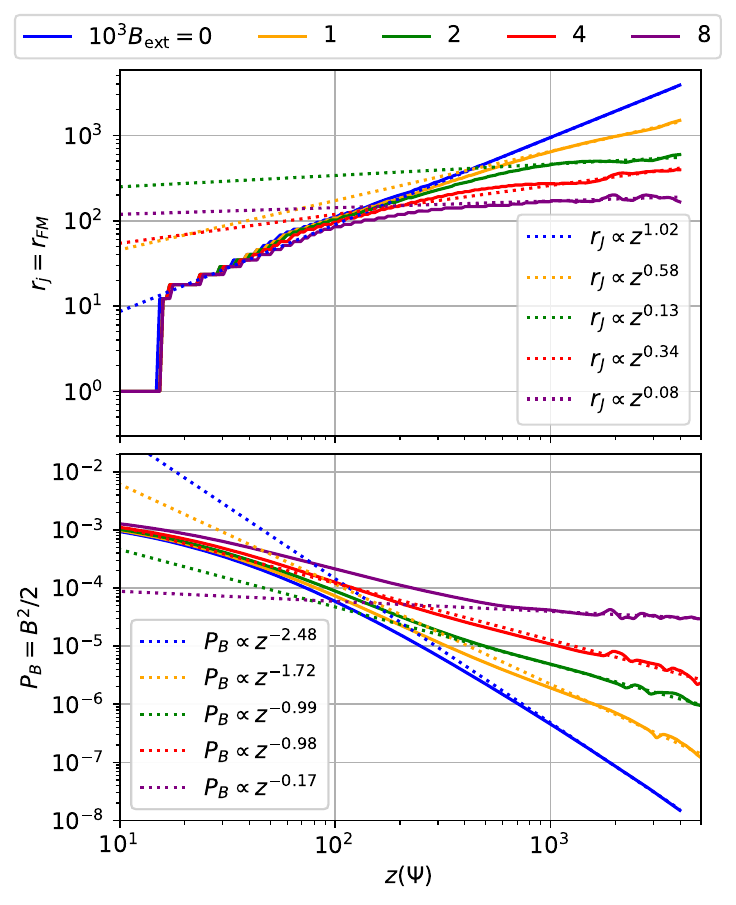}
\caption{Linear fits of the jet radius $r_{\text{FM}} \propto z^\omega$ (top) and the external pressure $P_\text{B} \propto z^{-k}$ (bottom) for  different simulations at $t_\text{end}$. All fits are made for $z \geq 10^3$. Each color is associated with the external magnetic field $B_{\text{ext}}$ shown in the inlet. The dotted lines are the linear fits made at large distances. The external (magnetic) pressure is measured along the magnetic surface anchored in the SAD region at $r_0=60$.}
\label{fig:SimulationsBextFits}
\end{figure}

These results illustrate the fact that the jet radius is determined by the external pressure. The top row of Fig.~\ref{fig:SimulationsBextFits} shows power-law fits of the jet radius $r_{\text{FM}} \propto z^\omega$ at large distances for B1 as function of $P_{\text{ext}}$. As $B_{\text{ext}}$ increases, $\omega$ decreases accordingly with values $\omega= 1.01; 0.74; 0.47; 0.18; 0.20$. The magnetic surfaces become more paraboloidal, almost cylindrical for the largest $B_{\text{ext}}$ (see also Fig. \ref{fig:SimulationsBext}). Since the jet radius is determined by the outer confining pressure, one should find a correspondance between the jet profile $r_{\text{FM}}(z)$ and how the pressure $P_{\text{ext}}(z)$ decays with the distance. Note that the values for $\omega$ are just informative, as jets recollimate further out.

The bottom row of  Fig.~\ref{fig:SimulationsBextFits} displays power-law fits $P_{\text{ext}} \propto z^{-k}$ in the same conditions. We choose to follow the total magnetic pressure $P_\text{B}$  along the field line anchored at $r_0=60$, far enough from the JED to avoid contamination but close enough to ensure temporal convergence. The values obtained are respectively $k=2.48; 1.72; 0.99; 0.98; 0.17$ for increasing $B_{\text{ext}}$. 
We recover qualitatively the results obtained by \citet{komissarov2007} and \citet{komissarov2009}. They found that for $k>2$ the jet becomes conical ($\omega=1$), whereas for $k<2$ their jet is of paraboloidal shape with $\omega <1$, which is quite exactly what we found. In our case however, neither the jet shape nor the external pressure are imposed a priori and both emerge self-consistently. The simulation is initiated with a potential field which is the sum of a steeply decreasing power-law potentiel field and a constant $B_{\text{ext}}$ field. How the total (here magnetic only) external pressure ends up scaling with the distance from the source emerges from the simulation, as the interplay between the confining external pressure and the outward push due to the jet itself. The final outer pressure stratification is thus different from the initial one, as the initial exponents were respectively $k=2.40; 0.64; 0.39; 0.22; 0.12$ (with increasing $B_{\text{ext}}$).  

\begin{figure}
\includegraphics[trim=0 7 0 5, clip, width=\linewidth]{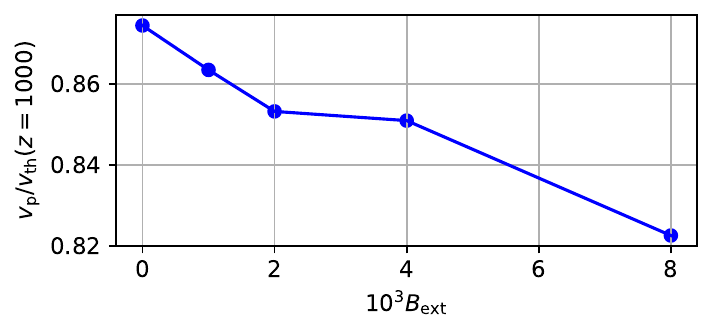}
\caption{Ratio of the jet poloidal speed $V_{\text{p}}$ achieved at $Z=1000$ and at $t_\text{end}$ for the field line anchored at $r_0=4$ to the maximum theoretical jet speed $V_{\text{th}}= V_{K_0}\sqrt{2\lambda -3}$ for cold outflows, as function of the external magnetic field $B_{\text{ext}}$ (confining pressure).}
\label{fig:Vp-Vth}
\end{figure}

Although the jet power remains the same regardless of $B_{\text{ext}}$, a lower external pressure allows for a larger jet widening and thereby to a more efficient plasma acceleration. This can be seen in the last two columns of Table~\ref{tab:ParametresSimusTronquees}, which display the average (density-weighted) super-FM outflow velocity at $z=500$ and at the outer boundary. A clear monotonous trend emerges for $V_{\text{j}}^{R_{\text{ext}}}$, which keeps on decreasing as the jet becomes narrower ($B_{\text{ext}}$ increases). This shows that the magnetic structure keeps more and more energy, stored in the form of MHD Poynting flux (thereby electric current), as the jet is prevented from opening up because of an increasing outer pressure. 
Assuming a cold outflow and a negligible rotation velocity wrt to the poloidal one (both assumptions verified here), the jet poloidal velocity writes
\begin{equation}  
V_{\text{p}}(z) = \frac{V_{\text{th}}}{\sqrt{1 + \sigma(z)}}
\end{equation}
where $V_{\text{th}}= V_{K_0}\sqrt{2\lambda -3}$ is the maximum theoretical speed, achieved only when the ratio $\sigma(z)$ of the MHD Poynting flux to the kinetic energy flux becomes much less than unity. The ratio $V_{\text{p}}/V_{\text{th}}$ at $z=1000$ is shown in Fig.~\ref{fig:Vp-Vth} for the five simulations. Clearly, the stronger the outer pressure, the less efficient the outflow. By less efficient, we mean here that the magnetic structure keeps a significant fraction of the initial energy so that $\sigma$ remains large. For instance, B1 keeps stored roughly only 19\% of the initial magnetic energy ($\sigma=0.23$) whereas B4 keeps nearly 36\% of the initial energy with $\sigma=0.56$.   

%%%%%%%%
%%%%%%%

%%%%%%%%
%%%%%%%

%The impact of the external magnetic field on collimation is also visible on Figure \ref{fig:B500Tronque}. The full lines represent the total magnetic field $B = \sqrt{\vec{B} \cdot \vec{B}}$ along the horizontal $z=500$ for all five simulations. The vertical line, $r_{\text{FM}} (z=500)$, tracks the  interface between the inner super-FM jet and the outer sub-FM atmosphere. It gets closer to the axis with increasing values of the external magnetic field. We see that this interface is associated with a change of regime, from the self-similar profile in the jet to a flatter one in the atmosphere. This behavior is very natural, and is expected from analytical works (see e.g. Figure 3 of \cite{Beskin2009}). In addition, the value of the magnetic field on the axis is seen to be increasing with the external magnetic field. This is not a direct consequence of the boundary conditions, as the external field $B_{\text{ext}}$ is negligible compared to that in the innermost disc regions ($ \lvert B_r / B_z \rvert \ll 1$). It is due to the growing collimation. As along a magnetic field line the poloidal field varies in $B_p \propto 1/r^2$, the magnetic field on the axis is less diluted for more collimated jets, that is for higher values of $B_{\text{ext}}$. Likewise, the jet of \citetalias{jannaud2023} being more collimated than those of this work, its magnetic field is larger. The toroidal field is not considered here because it vanishes on the axis.

%%% results: shocks
The global behavior discussed above, namely the impact of the confining outer pressure on the jet radius, must however be quantitatively revised because of the existence of standing recollimation shocks. These can be seen in the electric current isocontours and in the sudden outward refractions of the magnetic surfaces (left panel of Fig.~\ref{fig:ReferenceSimulationTruncated} or Fig. \ref{fig:SimulationsBext}). Because of their presence, the jet radius, defined here as $r_\text{FM}(z)$, cannot be simply represented as a monotonous power law of the distance. Launched from the disc surface, the jet undergoes a first lateral expansion ($B_r>0$) leading to a super-FM speed until a maximum turning radius $r_t$ is reached and a recollimation towards the axis is triggered ($B_r<0$, see e.g. fig.~12 in \citealt{ferreira1997}). The converging super-FM flow is then refracted outwardly in the recollimation shock, accelerated again to a super-FM speed and the whole process (expansion, recollimation, shock) can repeat itself. 

As discussed in \citetalias{jannaud2023}, these shocks are intrinsic to the MHD self-collimation process and seem to occur (for all simulations but B1) regardless of the outer confining pressure. The position of the first shock is distinct only for the $B_{\text{ext}}=0$ B1 simulation. For this simulation, $Z_\text{shock}=1250$ is significantly smaller than in the four other simulations, where $Z_\text{shock}$ varies between 1450-1600. The conical simulation B1 corresponds also to the most efficient acceleration and the electric current decreases therefore much faster than in the other cases. We will come back to this later in section~\ref{sec:CurrentClosure}.

\subsection{Rotation of the inner spine}\label{sec:RotationSpine}%%%%%%%%%%%%%%%%%%%%%%%%%%%

%% Objectives and method
With $B_{\text{ext}}$ we explored the effect of the outer pressure on the collimation properties of a jet emitted from a keplerian accretion disc. The launching conditions were those of a typical near-equipartition JED, with $B_z \propto r^{-5/4}$, settled between $r_0=1$ and $r_0=10$, around a central object whose rotation $\Omega_{*_{\text{a}}}$ was vanishingly small wrt to the inner rotation of the JED. Our reference simulation O1 assumed $\Omega_{*_{\text{a}}}=0$, in order to diminish as much as possible the impact of the central spine on the jet propagation and collimation. In this section, we vary $\Omega_{*_{\text{a}}}$ from 0 to 1\footnote{A value $\Omega_{*_{\text{a}}}>1$ would correspond to a situation where the central engine is rotating faster than the innermost disc radius $r_0=1$. This is something that could be realized, but probably not following the magnetic field distribution used here. If the central object has its own magnetosphere for instance, it would be in a propeller regime and the whole system would not be steady. If it is a maximally rotating Kerr black hole, then a plunging region should also be considered. Both situations are beyond the scope of this work.}(simulations O1 to O5), leading to the radial distributions $\Omega_*(r)$ of the magnetic surfaces shown in Fig.~\ref{fig:FrontiereOmegaTronque}. 

As discussed previously, the jet power can be simply related to the total e.m.fthat drives the electric current flowing inside the jet. This e.m.f. $\mathcal{E}$ is provided by Faraday's induction law and depends therefore on both the rotation of the central engine and the vertical magnetic field threading it. In MHD, it writes $\mathcal{E} \simeq \int \Omega r B_z dr$, where $\Omega$ is the rotation of the plasma ($\sim \Omega_*$) and the integration should be done from the axis to $r_0=12$. Our boundary conditions at $R=1$ are such that the e.m.f. related to the spine is in series with that of the disc, so that there is a unique electric circuit fed by one total e.m.f. $\mathcal{E}= \mathcal{E}_\text{spine} + \mathcal{E}_\text{disc}$ (see right panel in Fig.~\ref{fig:ReferenceSimulationTruncated}). So, by choosing $\Omega_{*_{\text{a}}}=0$, our reference simulation O1 strongly lowered the impact of the spine, with $\mathcal{E}\simeq \mathcal{E}_\text{disc}$ only. In contrast, simulation O5 with $\Omega_{*_{\text{a}}}=1$ has $\mathcal{E}_\text{spine} \simeq \mathcal{E}_\text{disc}$, leading therefore to an almost twice larger electric current. How such enhanced electric current impacts the jet collimation properties is the focus of this section.  
 
%Four simulations were run varying the rotation of the magnetic surfaces on the axis $\Omega_{*_{\text{a}}}$. Contrary to \citetalias{jannaud2023}, here the energy on the axis is not modified: $e_{\text{a}} = 4$ as in the reference simulation O1. The axis rotation is modified from $\Omega_{*_{\text{a}}} = 0$ (Schwarzchild black hole, or young star whose disc corotation radius with the star is cast to infinity) to $\Omega_{*_{\text{a}}} = 1$ (Kerr black hole, or young star with coinciding truncation and corotation radii). The case $\Omega_{*_{\text{a}}} > 1$ corresponds to highly rotating black holes that would develop strong relativistic \cite{blandford1977} jets, or to rapidly rotating stars in the "propeller regime" (see \cite{Romanova2009,Zanni2009}). Both are out of reach of this work.

%%%%%%%%
%%%%%%%%%

%%%%%%%%
\begin{figure}
    \centering
    \includegraphics[trim=6 0 6 0, clip,height=.7\linewidth]{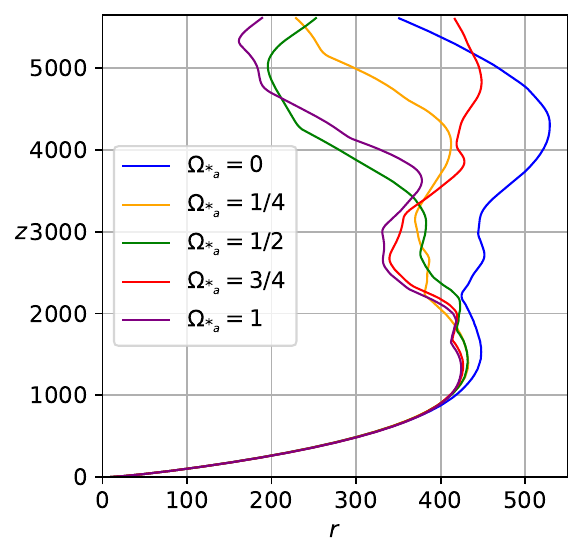}
    \includegraphics[trim=6 0 6 0, clip,height=.7\linewidth]{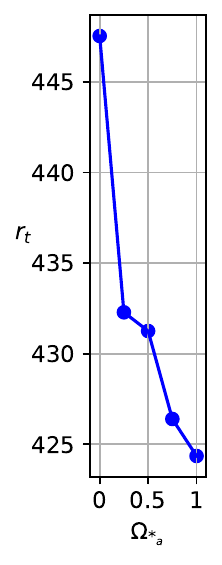}
    \caption{Influence of the axis rotation $\Omega_{*_{\text{a}}}$ on the magnetic surface anchored at $r_0 = 10$. Left: shape of the magnetic surfaces at the final stage of the simulations. Right: first turning radius $r_t$  as a function of $\Omega_{*_{\text{a}}}$.}
    \label{fig:OmegaTronqueCollimation}
\end{figure}
%%%%%%%

%%% results: influence on outer jet collimation and confinement 
The five simulations O1 to O5 are presented at their final time $t_{\text{end}}$ in Fig. \ref{fig:SimulationsOmegaEtoile} and some of their general properties are highlighted in Table~\ref{tab:ParametresSimusTronquees}. Like for $B_{\text{ext}}$, all simulations have reached a steady-state so that $\Omega_{*_{\text{a}}}$ has no impact as well on the mass loss rates (from both the spine and the JED) and neither on the power $P_\text{jet}$ carried away the jet emitted from the JED. This last part is more surprising as one would naively expect a stronger toroidal field at the disc surface, hence an increase of the jet power. 

But remember that our platform simulations do not take into account the feedback on the disc mass loss rate. This implies that the toroidal magnetic field is determined on each magnetic surface by the Alfv\'enic regularity condition associated with the mass load. In other words, whatever the spine e.m.f. $\mathcal{E}_\text{spine}$, the steady-state toroidal field at the disc surface will be mostly determined by the imposed mass loss rate. Thus, any increase in $\mathcal{E}_\text{spine}$ leads to an increase of the electric current flowing down the axis, namely to an extra electric circuit mostly related to the spine itself. This is why the power carried by the spine $P_\text{spine}$ increases from O1 to O5 (although by only $\sim$ 10\%), which leads to a small increase in the average (density-weighted) jet velocities, both $V_{\text{j}}^{z=500}$ and $V_{\text{j}}^{R_{\text{ext}}}$ (see Table~\ref{tab:ParametresSimusTronquees}).

%% transition vers shocks
At first sight, the influence of the axis rotation on the jet overall structure appears to be negligible. This is exemplified in Fig.~\ref{fig:OmegaTronqueCollimation}, which shows how the magnetic surface anchored at $r_0=10$ is modified with $\Omega_{*_{\text{a}}}$. The left panel shows the final stage at  $t_{\text{end}}$ of that surface: only the lower regions have a monotonous behavior with the distance to the disc. Beyond $z\simeq 1000$, the existence of shocks, occurring at different altitudes for different simulations, introduces huge downstream differences in the jet radius. {\em We stress that all simulations are in steady-state and that this is only a consequence of the various shocks}. One way to measure quantitatively the impact of $\Omega_{*_{\text{a}}}$ on jet collimation is to evaluate the first cylindrical radius $r_t$ where the jet recollimates (hence $B_r=0)$  as function of  $\Omega_{*_{\text{a}}}$. This is shown in the right panel of Fig.~\ref{fig:OmegaTronqueCollimation}. As expected,  there is a monotonous decrease of $r_t$ as the spine e.m.f. is increased, due to a larger hoop-stress acting on the spine. However, this remains a marginal effect (the relative variation is around 5\%), because the spine itself occupies a tiny volume compared to the jet.          

%However, axis rotation still has a small impact on jet collimation below the shocks. Figure \ref{fig:OmegaTronqueCollimation} shows on the left panel the magnetic field line anchored at $r_0 = 10$ for all five simulations, and on the right panel the evolution of the turning radius $r_t$ with $\Omega_{*_{\text{a}}}$ for those field lines. It is defined as the maximal field line radius before recollimation, where $B_r = 0$. We see a small decrease in $r_t$ with the axis rotation. This is a direct consequence of a more collimated spine, that allows for higher collimation in the outer jet. This leads to a small increase in $u_{\text{jet}}^{z=500}$ (see Table \ref{tab:ParametresSimusTronquees}), the turning point happening for $z \sim 500$. The increase in collimation with the central object rotation is due to the growing emf $e_{obj} \simeq \int_{\theta = 0}^{\pi / 2} \Omega r B_R d \theta$. On the left panel of Figure \ref{fig:OmegaTronqueCollimation} we see that beyond the turning point, this order is disrupted by the presence of the shocks. As a consequence, there no clear trend on the influence of axis rotation on large-scale jet collimation.

%%%%%%%%
\begin{figure}
    \centering
\includegraphics[trim=5 5 0 5, clip,width=.91\linewidth]{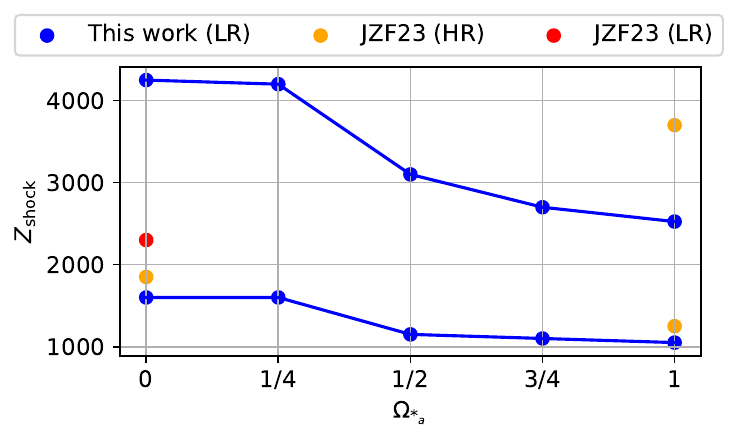}
    \caption{Evolution of the altitude of the shocks (measured at the axis) with the axis rotation $\Omega_{*_{\text{a}}}$.    Results from this work are shown in blue, results from \citetalias{jannaud2023} in orange (at high resolution) and in red (at low resolution).}
    \label{fig:OmegaTronqueChoc}
\end{figure}
%%%%%%%%

%%% results: shock 
The most striking effect of the axis rotation $\Omega_{*_{\text{a}}}$ is its impact on the altitude (measured at the axis) of the recollimation shocks, as shown in Fig.~\ref{fig:OmegaTronqueChoc}: as $\Omega_{*_{\text{a}}}$ increases, the shocks decrease to lower altitudes. This is consistent with an increasing Z-pinch, consequence of a stronger electric current flowing along the axis. Interestingly, as first shown and discussed in \citetalias{jannaud2023}, our computational domain allows to capture two sets of recollimation shocks (the two blue curves in Fig.~\ref{fig:OmegaTronqueChoc}). Not only both shocks decrease to lower altitudes as the axis rotation is increased, but also their separation gets smaller. 

For the sake of completeness, we put also the shock positions measured in the `self-similar' simulations of \citetalias{jannaud2023}. In the low-resolution case, done with $\Omega_{*_{\text{a}}}=0$, only one set of recollimation shocks was observed, at an altitude higher than in our finite JED case. We see no obvious reason for that. The `self-similar' simulation displays a stronger confinement but, as shown in Table~\ref{tab:ParametresSimusTronquees}, confinement does not seem to have a direct influence on $Z_\text{shock}$. However, the `self-similar' simulations done at higher resolution not only provide shock altitudes closer to the finite JED case, but also show the same trend with $\Omega_{*_{\text{a}}}$, namely a decrease of the shock altitude. A second set of shocks is even appearing at $\Omega_{*_{\text{a}}}=1$.  We therefore conclude that this trend is rather generic, although it may be blurred by numerical effects.

%The shocks are still present, with a structure similar those in the reference O1. Shock altitude is lowered by an increased rotation due to a stronger Z-pinch. Figure \ref{fig:OmegaTronqueChoc} compares this decrease to the simulations of \citetalias{jannaud2023}. We see that passing from no-rotation to solid-body rotation has the same impact on both cases: the reduction of shock altitude by a factor of roughly one third. A second set of shocks appears at higher altitudes, its altitude being roughly twice that of the first set of shocks. This set of shocks is present in all simulations, and renders drawing conclusions on the influence of $\Omega_{*_{\text{a}}}$ on global jet collimation difficult.

%%%%%%%%%%%%%%%%%%%%%%%%%%%%%%%%%%%%%%%%%%%%%%%%%%%%%%%%%%%%%%%%%%
\section{Discussion}\label{sec:Discussions}
%%%%%%%%%%%%%%%%%%%%%%%%%%%%%%%%%%%%%%%%%%%%%%%%%%%%%%%%%%%%%%%%%%

\subsection{Caveats}\label{sec:Caveats}
%%%%%%%%%%%%%%%%%%%%

%When interpreting any numerical work, one needs to be very careful with the assumptions made. Below, we list some of the limitations of this setup. They were all already present in \citetalias{jannaud2023} except the last one, the unphysical nature of the SAD-like boundary condition ($r_0 \geq 10$).

As usual with complex numerical simulations, our experiments suffer from several limitations. Most of them have been already extensively discussed in \citetalias{jannaud2023}, but we briefly recall them here for the sake of completeness. 

%% non-relativistic and polytropic (cold) jets
%Second, the simulations presented in this paper are non-relativistic. In the relativistic case the inertia is greater $(mv/c)$ and there is an additional term in the equation of motion due to the electric field. These effects are expected to weaken the confinement induced by the Z-pinch, especially at high Lorentz factors. As a consequence, in similar GRMHD simulations, it is not clear that the recollimation shocks would still be present. And if they do appear, they should be located at higher altitudes. Thus, comparisons between the results of this paper and AGN or X-ray binary jets can only be qualitative.

Our numerical experiments address only non-relativistic, axisymmetric and cold polytropic jets. Since we are not focusing on jet observational signatures but mostly on their collimation and confinement properties, we believe these simplifying assumptions introduce no relevant limitation. This is also suggested by observations which estimate radiative losses to be much smaller than jet kinetic power \citep[see e.g.][and references therein]{Ghisellini2014}. As \citet{heyvaerts2003b} showed analytically that relativistic jets share the same asymptotic properties as non-relativistic ones, we conclude that our results (namely the existence of standing recollimation shocks) remain generic and can be safely extrapolated to relativistic jets. However, because of the presence of a strong electric field in the relativistic case, the physical scales involved in relativistic jets are expected to be much larger than in non-relativistic ones. This is also the reason why no self-consistent MHD simulation has yet connected the central engine to the asymptotic jet regime (see e.g. \citealt{komissarov2021} and references therein). This, as well as numerical difficulties (see below), may be the reason why no standing recollimation shocks have been detected so far in GRMHD simulations (see however \citealt{chatterjee2019}).

% axisymmetry assumption
The axisymmetry assumption used in our simulations is more problematic as it severely limits the validity of our analysis. Clearly, axisymmetry facilitates the appearance of recollimation shocks inside the simulation domain and it is unclear if such structures would survive in 3D. Moreover, 3D jets are prone to several instabilities such as, for instance, the kink instability \citep{bromberg2016,tchekhovskoy2016,barniolduran2017,Mizuno2014} or the Kelvin-Helmholtz shear instability at the interface between the jet and the outer medium \citep{Bodo1994,Baty2002,Baty2006}. In any case, the analysis of these 3D instabilities requires first to precisely determine the underlying 2.5D (stratified) jet structure to be perturbed. Our stratified 2D outflow is physically consistent with a low density spine surrounded by a Blandford \& Payne wind launched by a JED of finite radial extent. Thus, we believe that our work presents the most realistic rotating finite MHD jet configuration to date. Unfortunately, making 3D simulations of such a highly stratified outflow remains numerically challenging. 
%% Sausage ?
Noteworthily, given the wavy, almost sausage-like, appearance of our jets one could suspect the presence of the axisymmetric sausage instability. But we checked that the local Suydam criterion \citep{suydam1958} is verified nowhere in our jets. Although recollimation is intrinsic to MHD collimation it is not the result of an instability. As it will be argued below, it may be instead a way for the jet structure to self-adjust radially so as to maintain a coherent poloidal electric current configuration. 
 
% Numerical difficulties
We would like to stress that our numerical experiments have been carefully tailored to capture the shocks. In most of the computational domain, we use a HLLD solver \citep{miyoshi2005} as well as a second order in space piecewise reconstruction with various slope limiters: Monotonized Center for the density \citep{VanLeer1977}, Van Leer for the velocity and the magnetic field \citep{VanLeer1974} and Minmod  for the pressure \citep{Roe1986}. Such a reconstruction scheme (including the slope limiters) is the standard in PLUTO, and is well-suited to capture shocks. However, to avoid excessive computational costs we had to employ a flat reconstruction scheme and a HLL  solver \citep{harten1983} in the cells closest to the axis, as well as in those of very low density and high Alfv\'en speed. Moreover, the simulation results depend on their spatial resolution. At higher resolution the shocks altitude and strength are roughly the same, but their structure is more complex. One shock system at low resolution appears to be composed, when looked at higher resolution, of many smaller shocks (see for instance Fig. \ref{fig:OmegaTronqueChoc} above, section 3.4.2. of \citetalias{jannaud2023} or section 5.2.1 of \citealt{janna2023}). 

The numerical effort required to correctly capture MHD jet physics is tremendous. Not only are the physical scales gigantic, with simulations going ideally from $z=0$ to several $10^4 r_d$, but resolution should be kept good enough so that ideal MHD is maintained all the way. Despite our efforts, ideal MHD is violated at large distances for field lines anchored beyond $r_0=10$ (Fig.~\ref{fig:Invariants}). This is an unavoidable computational limitation, shared by all similar jet simulations (see for instance fig.~12 in \citealt{ramsey2019} and discussion in \citealt{komissarov2007}). Besides, the lack of resolution at large distances may forbid the identification of possible recollimation shocks. On the other hand, it is unclear until which distance from the source astrophysical jets can still be accurately described within the ideal MHD framework. In the context of protostellar jets for instance, the lack of ionisation of the jet material may eventually cause a drift between neutrals and ionized species, leading to the existence of a resistivity (ambipolar diffusion) and the breaking up of the ideal MHD approximation (see for instance \citealt{Frank1999}). Thus, although our results are numerically biased in that aspect, we believe the outcome remains physically viable, namely with plasma streamlines progressively less confined than their magnetic counterpart. 

% Physical conditions: spine, JED, SAD
A last word of caution must be said about the conditions used at the injection boundary. In our simulations, only the JED is realistic when compared to models of discs launching outflows, although valid only for cold outflows with $M_S = 10$. As a consequence, the radial distribution of the vertical magnetic field has been chosen to follow the \citet{blandford1982} distribution $B_z \propto R^{\alpha - 2}$ with $\alpha=3/4$, in agreement with a tiny disc ejection efficiency and fast outflows (see \citealt{Zimniak2024} and references therein). Changing the value of $\alpha$ introduces no significant modification to the overall picture described in this paper as long as $\alpha <1$ (see \citetalias{jannaud2023}). We were unable to obtain steady-state jets with values of $\alpha \geq 1$, corresponding to massive and slow winds. As shown in \citetalias{jannaud2023}, such a magnetic field distribution induces a poloidal electric current that enters the disc at its surface and requires therefore, in order to close the electric circuit, to come out at larger distances. But this electric circuit configuration is impossible to achieve within our simplified JED-SAD framework. Dealing with such a situation would require to compute the disc structure itself, which is outside the scope of this paper.   

Below $r_0 \leq 1$, namely inside the region related to the central object and generating the axial spine, and beyond $r_0 \geq 10$, namely in the SAD-like region, the injection conditions have been drastically simplified. The spine ejection conditions were taken to limit its influence on the fast jet dynamics, although as we saw on section~\ref{sec:RotationSpine} it does have an impact on the shock altitude. The spine conditions were set to smoothly go from those wanted at the inner edge of the disc to those wanted on the axis. But, as such, they do not properly model a stellar wind nor a \citet{blandford1977} jet of course. Moreover, massive unsteady outflows such as the magnetospheric ejecta located between the inner stellar wind and the outer disc wind \citep{zanni2013}, have been voluntarily discarded. In the same vein, the outer region surrounding the fast jet from the disc has been chosen with no outflow, as in the conventional JED-SAD picture. The easiest way to do that with a platform simulation was to impose $\Omega_{*_{\text{SAD}}} = 0$ beyond  $r_0 = 10$. This is of course unrealistic. Indeed, whenever the disc is threaded by a large-scale magnetic field, a wind is expected to be launched and the smaller the disc magnetization, the more massive the wind \citep{zhu2018,jacquemin-ide2019,jacquemin-ide2021}.       

In both cases, namely in the presence of a strong inner spine and of an outer massive disc wind, modifications of the final outcome of the fast outflow emitted from the JED (in terms of acceleration, collimation and confinement) are therefore expected. But the aim of this paper is precisely to provide the analysis of this outcome when these external conditions are not significant and so, they have been minimized. Nevertheless, their impact is most certainly important and requires future investigations.

\subsection{Comparisons to previous simulations}\label{ref:ComparisonsStateOfTheArt}
%%%%%%%%%%%%%%%%%%%%%%%%%%%%%

In this section, we argue that the simulations presented in this work are unique, although some aspects have been already identified and discussed previously (see \citetalias{jannaud2023} for more general references on jets). In the following, we are only interested in numerical simulations of jets that (1) are launched from a zone of finite extent and (2) are connected to their source. This last condition requires that jet material is launched at the bottom of the computational domain with a speed smaller than the escape velocity and reaches, at the outer boundary, a super-FM (Fast Magnetosonic) speed. In terms of poloidal electric circuit, this implies that the first accelerating current circuit is fully enclosed within the computational domain.  

% celles contenant disque : 3D
For obvious technical limitations, 3D simulations computing both the turbulent accretion disc and its jets have provided so far seldom information on the jets themselves. This holds for both relativistic (GRMHD, e.g. \citealt{porth2019}) and non-relativistic situations (e.g. \citealt{zhu2018}). Even if physical scales may be quite large, sometimes up to $z\sim 10^4 r_{\text{in}}$ (in units of the innermost radius $r_{\text{in}}$ of the simulation), images and jet quantities are restricted to a much smaller region (usually below 20-50$r_{\text{in}}$, up to $z\sim 180 r_{\text{in}}$ in \citealt{jacquemin-ide2021}). Following turbulence is time consuming and jets are probably computed farther out at a lower resolution, challenging thereby the trust on their outcome (see discussion in \citealt{komissarov2021}). 

% celles contenant disque : 2D
An alternative is to use ad-hoc prescriptions for the turbulence \citep{Shakura1973}, allowing to compute 2D (axisymmetric) accretion-ejection structures \citep{casse2002,casse2004,zanni2007,tzeferacos2009,murphy2010,sheikhnezami2012,stepanovs2016}. While earlier works used a rather small box ($z$ from 50 to 180) on a rather short duration ($t_{\text{end}}\sim 40-60$), latter studies reached larger distances ($z<840$ in \citealt{murphy2010}, $z<1500$ in \citealt{stepanovs2016}) and much longer times ($t_{\text{end}}=953$ and 1500, respectively). However, these space and time scales are still slightly too small to allow a comparison with our own work. Moreover, the physical conditions are too different. \citet{stepanovs2016} start with a self-similar accretion disc and the width of the jet keeps on increasing in time, following the size of the launching region (as in \citetalias{jannaud2023}). On the contrary, \citet{murphy2010} do have a jet emitted from a zone of finite size, but this was due to their initial radially (very steeply) decreasing disc magnetisation\footnote{The 2D simulations of a non turbulent, ambipolar diffusion dominated, protostellar accretion disc of \citet{martel2022} also obtained a hybrid disc configuration, with an inner region launching a fast jet. However, the domain size is far  too small ($z < 50$) to look for any sign of standing recollimation shocks.}. They do not report any recollimation within their domain. However, since these launching conditions are very different from those used here, a direct comparison would be anyway difficult. 

% platforms
A much better alternative to study the outcome of MHD jets is to use `platform' simulations,  where ideal MHD jets are computed above the turbulent disc (and the central object), by using freely specified boundary conditions at the bottom of the computational box.  This is the approach followed here and by many other authors \citep{ustyugova1995, ustyugova1999, ouyed1997a, ouyed1999, krasnopolsky2003, anderson2005, pudritz2006, fendt2006, porth2010, ramsey2011, ramsey2019,meskini2024}. 

Of these earlier works, only \citet{krasnopolsky2003, anderson2005} computed jets from a finite launching surface, from $r=1$ to $r=10$. However, in strong contrast to our work, the last magnetic field line has been chosen to lie exactly on the equatorial plane with $u_z=0$. With this boundary condition, the field lines are forced to open up and to fill in the whole space (as in the fan shaped-winds of \citealt{ferreira2013}). This leads to a profound modification of the outflow emitted below $r=10$, despite their use of \citet{blandford1982} boundary conditions. Nevertheless, these earlier studies were computationally limited. Both space and time scales are too small to allow direct comparisons (e.g. $z<100$ in \citealt{krasnopolsky2003, anderson2005}, $z<80$ in \citealt{pudritz2006}, $z<200$ in \citealt{fendt2006}). It is also worth mentioning that many of these parametric works used arbitrarily free radial dependences for the magnetic field and density distributions. Although such freedom is mathematically allowed, it is unclear if some of the choices made would be achieved in astrophysical systems (most of them are actually inconsistent with JED scalings).

%%%%%%% Porth & Fendt 2010, Ramsey & Clarke 2011, 2019
It is illustrative to compare our work with the axisymmetric simulations of \citet{porth2010} and \citet{ramsey2011, ramsey2019}. The former were relativistic whereas the others were done for YSO jets but, as argued previously, the general outcome of jets is the same for relativistic and non-relativistic jets. In both cases, the \citet{blandford1982} mass loading parameter $\kappa$ (defined in equation~\ref{eq:DefintionKappa}) is roughly $\kappa \sim v_{\text{inj}}\beta$, where $v_{\text{inj}}=u_z/V_K$ controls the injection speed and $\beta = P/(B_z^2/2\mu_0) = \beta_d (r/r_d)^{-1/2}$ is the fixed plasma beta at the jet basis. Note that while $\kappa$ measures the role of inertia in jet dynamics, $\beta$ is a measure of the relative importance of the jet thermal pressure with respect to the magnetic forces. Thus, cold jets with $\beta_d \ll1$ are mostly determined by magnetic forces, whereas outflows with $\beta_d \geq 1$ are expected to be strongly affected by the gas pressure gradient. A stark difference between these works and our own is the range of $\beta_d$ used. Indeed, \citet{porth2010} used $\beta_d=2, 1, 0.2$ whereas \citet{ramsey2019} did a parametric survey, from a so-called strong-field case (their simulations A to D) with $\beta_d$ ranging from 0.1 to 2.5, to a weak field case (their simulations G and H) with $\beta_d=160, 640$. In our case however, $\beta_d = 2 \epsilon^2/\gamma \mu^2 \sim \epsilon^2 \sim 10^{-4}$: our jets are therefore much colder than those computed in these two works (with $v_{\text{inj}}\sim M_S \epsilon =0.1$). Moreover, because of the different radial exponents used (again, inconsistent with JED calculations), these two works lead to a much less steep decrease of the magnetic pressure ($B_z \propto r^{-1}$ instead of $r^{-5/4}$), as well as to an outwardly increase of the influence of the magnetic field  $\beta \propto r^{-1/2}$ instead of constant). Altogether, these differences tend to enforce a stronger collimation of the outflows in their case.   

A direct comparison with \citet{porth2010} is impossible to make since the only images provided are for the $\beta_d=0.2$ case. Moreover, since the material is injected at sub-magnetosonic speeds, $v_{\text{inj}}$ is self-consistently determined and the values of $\kappa$ in these simulations remain unknown. However, they report the existence of simulations done with $v_{\text{inj}}>0.4$ (their simulations I04-I12), with some close to our $\kappa = 0.1$. They report that increasing the mass flux further enhances collimation. However, these observations were done for simulations in a grid of $z < 200 r_{\text{in}}$ only, and a duration of 500 inner orbits, leaving open any further jet evolution. Their fig.~6 showing the poloidal electric circuit is nevertheless highly instructive: as expected for  $\alpha=1$, the current enters the inner disc at its surface and leaves it in the outer parts. As a consequence, current closure is strongly affected by the radial boundary conditions, in particular coming from the unsteady sub-Alfv\'enic outer region. 

The simulations of \citet{ramsey2011, ramsey2019} have been done in a much larger grid, up to $z=81400  r_{\text{in}}$ (with 9 levels of adaptive mesh refinement), and on a much longer duration ($t_{\text{end}}\sim 10^4$ inner orbits). In these huge simulations, jets never reach the end of the computational domain and are therefore never strictly steady. The outflow is injected with $v_{\text{inj}}=10^{-3}$ so that $\kappa \sim 10^{-3}\beta$ is also tiny. Their simulations A and B, with respectively $\beta_d=0.1$ and 0.4 may represent the same situation as those shown in \citet{porth2010}. However, signs of shocks can be seen for instance at distances much larger than $z=200 r_{\text{in}}$ (see their fig.~9), leaving open the issue of stationarity. Only their simulations F and G, with respectively $\beta_d=40$ and $160$, are comparable to ours, with similar magnetic field and density at $r_d$ and thus $\kappa \sim 0.1$. However, they do not observe steady or recollimation shocks. Only their `medium field' simulation F shows periodic knots that are travelling along the jet.
The magnetic lever arm $\lambda$ seems to be much smaller in their simulation: we estimate it to be around 2.8 for the field line anchored at $r_0=1$ au, whereas we obtain $\lambda \simeq 11$. This is in agreement with a less confined outflow in our case. Besides the previously discussed differences in prescribed radial exponents, which would naturally enhance collimation in their case, this may also be due to the continuous mass injection at all radii. Indeed, that leads to a continuous increase of the size of the launching region, hence to an increase of the outer pressure.  
 
It turns out that the freedom in radial distributions allowed by platform simulations is an important limitation. It is also the reason why we only consider here axisymmetric (2D) simulations. Indeed, 3D jet simulations focus on the non-linear outcome of instabilities triggered in stratified 2D outflows. But as long as these 2D flows have not been physically justified, we believe that going to 3D remains premature, since no general nor definite conclusion on astrophysical objects can be drawn from it.

%
%%%%%%% Stute+08, Matsakos09,  etc..
There is yet another interesting way to use platform simulations. Instead of prescribing ad-hoc conditions at the inner (injection) boundary and starting with a static atmosphere and a potential magnetic field, the whole computational domain is filled in with a stationary self-similar MHD wind solution. It can be either a pure disc wind \citep{gracia2006}, a pure stellar wind \citep{matsakos2008}, a two-component wind composed of two disc winds \citep{stute2008}, or an inner stellar wind surrounded by an outer disc wind \citep{matsakos2009}. The obvious advantage of this approach is in the use of initial (and boundary) conditions that are consistent with analytical models of outflows. However, the results of the axisymmetric simulations remain quite hard to interpret. Indeed, not only the physical scales involved are too small to allow a full development of the outflow (distances up to $z < 200$ only), but the interplay between the two jet components involves the ad-hoc transition zone that has been obtained by mixing the two analytical models. It is therefore unclear if the computed interplay is generic or if it is a consequence of the mixing procedure used. We note that none of these approaches provided a jet of finite size inside their computational domain. 

%
%%%%%%% Komissarov+07,09 
To conclude, the only previous simulations addressing the acceleration of astrophysical jets of a finite size are those of \citet{komissarov2007, komissarov2009}. These relativistic jet simulations were done with a specifically designed grid extension technique (with elliptical coordinates) that allowed them to cover six to nine orders of magnitude in spatial scales (comparable to self-similar solutions, \citealt{ferreira1997,vlahakis2003a}). They obtained stationary jets of finite size by computing 2D flows inside a funnel with a rigid wall of prescribed shape $z\propto r^a$, with $a$ ranging from 1 to 3. The poloidal magnetic field is taken uniform within the jet, the flow is cold and injected at a super-slow magnetosonic speed with a constant density. As the study was focused on relativistic jets from black holes, rotation was assumed uniform (although an ad-hoc differential rotation profile was also analyzed). We already discussed in section~\ref{sec:ExternalBfield} the similarities with our work, namely the correspondance between the vertical profiles of the jet radius $r_{\text{FM}} \propto z^{\omega}$ and the external pressure $P_{\text{ext}} \propto z^{-k}$, as well as their impact on the jet acceleration efficiency, as measured by the ratio $\sigma$ of the MHD Poynting flux to the kinetic energy flux at the outer boundary. Our work not only confirms this generic behavior of MHD winds (valid for both relativistic and non-relativistic flows) but it extends also their results since, in our case, the jet profile emerged self-consistently and was not imposed. 

A simple argument based on a Bennett relation for jet asymptotics allows to recover the existence of the limiting case $k=2$ between paraboloidal and conical wind. Assuming almost vertical surfaces and neglecting any curvature effect, the transverse balance amounts to a simple screw-pinch equilibrium where the local current $I(r)$ must verify $I^2 = 2 \mu_0 \left( \int_0^r 2rP_{\text{tot}} dr - r^2P_{\text{tot}}(r) \right)$, where $P_{\text{tot}}$ is the sum of thermal and magnetic pressures \citep[see e.g. equation 4.25 in][]{Freidberg1982}. Evaluated at the jet edge $r_{\text{FM}}$ where $P_{\text{tot}}= P_{\text{ext}}$ and assuming a steep decrease with the radius of $P_{\text{tot}}$ within the jet, this relation leads at infinity to  
\begin{equation}
\frac{I_\infty^2}{2\mu_0} = r_c^2 P_{\text{tot,c}} - r_{\text{FM}}^2P_{\text{ext}}(z)
\end{equation} 
where $r_c$ and $P_{\text{tot,c}}$ are respectively the radius and total pressure of some innermost core (possibly related to the spine). With profiles chosen as $P_{\text{ext}} \propto z^{-k}$ and $ r_{\text{FM}} \propto z^\omega$, it can be readily seen that if $k> 2\omega$, a finite asymptotic current $I_\infty \neq 0$ is possible, whereas if $k< 2\omega$, the current must vanish \citep{heyvaerts2003c}.   

Note finally that magnetic surfaces with a non negligible $\sigma$ at the outer domain boundary must therefore carry a non-vanishing electric current $I$. As a consequence, the current closure is done outside the computational domain. As argued below, this is an issue and a bias of the approaches done so far.

\subsection{Current closure issue and jet asymptotics}\label{sec:CurrentClosure}
%%%%%%%%%%%%%%%%%%%%%%%%%%%%

\subsubsection{Current and collimation in classical jets}

An MHD jet is usually seen as outflowing plasma carrying magnetic fields and electric currents. And, usually, the electric current density is seen as just the rotational of the magnetic field, almost a consequence of it (Barlow wheel effect). We favour thinking of the magnetic field as generated by the electric currents, the latter being driven by the electromotive forces in the plasma. Since a large-scale vertical magnetic field is usually assumed to thread the rotating source (in our case, the disc), we are mostly interested in the poloidal electric current (related to the toroidal magnetic component). In steady-state, there is no charge accumulation nor rarefaction and the poloidal electric circuit must be closed. So, MHD jets can also be seen as electric circuits growing in size as the plasma propagates outwardly. Now, maintaining $\nabla \cdot \vec J = 0$ whatever the extent (radial and vertical) of the jet is an issue (see discussion in e.g. \citealt{heyvaerts1989,okamoto2001,heyvaerts2003a,heyvaerts2003b,heyvaerts2003c}).     

When projected along and perpendicular to a poloidal magnetic surface, the MHD force writes \citep{ferreira1997}  
\begin{eqnarray}
F_{\varphi} & = & - \frac{B_p}{2\pi r} \nabla_{\parallel} I \nonumber \\
F_{\parallel} & = & \frac{B_{\varphi}}{2\pi r}\nabla_{\parallel} I
\label{eq:force}\\
F_{\perp} & = & B_pJ_{\varphi} + \frac{B_{\varphi}}{2\pi r} \nabla_{\perp} I
\nonumber
\end{eqnarray}
\noindent where $I = - 2 \pi r B_{\varphi}/\mu_0 > 0 $ is the total current flowing within this magnetic surface, $\nabla_{\parallel}\equiv (\vec{B}_p \cdot \nabla)/B_p$ and $\nabla_{\perp}\equiv (\nabla \Psi \cdot\nabla)/|\nabla \Psi |$. MHD acceleration is basically a conversion of electric current into organized plasma motion\footnote{From this point of view, the distinction commonly made in the literature between `magneto-centrifugally driven' jets \citep{blandford1982} and `magnetic tower' jets \citep{LyndenBell1996,LyndenBell2003} is meaningless.}. 
 
\begin{figure}
\centering
\includegraphics[width=\linewidth,trim=0 9 0 5, clip]{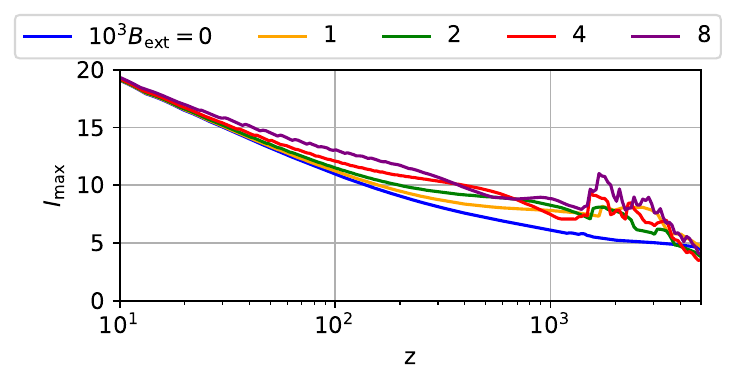}
\caption{Evolution along the altitude $z$ of the maximum absolute value $I_{\text{max}}$ (code units) of the poloidal electric current within the jet at $t_{\text{end}}$, for increasing values of $B_{\text{ext}}$.}
\label{fig:Imax}
\end{figure}

Making a cut in the jet at an altitude $z$, the transverse (radial) profile of the electric current $I(r)$ starts from zero at the axis, achieves its maximum absolute value $| I_{\text{max}}| $ at a radius $r_m$ (where $J_z=0$) and goes back to zero at a radius $r_I$. Jet acceleration corresponds therefore to a decrease of $I_{\text{max}}(z)$ as function of the distance $z$. This is shown in Fig.~\ref{fig:Imax} for different values of $B_{\text{ext}}$. In agreement with our previous remark on jet acceleration efficiency, the stronger $B_{\text{ext}}$ the slower the decay of current with distance. However, as analytically shown by \citet{heyvaerts2003c} and confirmed here, the current decreases very slowly with $z$, as $1/\ln (z)$. Moreover, Fig.~\ref{fig:Imax} shows that in all simulations but B1 the presence of recollimation shocks give rise to a sudden increase of the available current $I_{\text{max}}$. 

\begin{comment}
    It is goes from $0.9 \, r_{\text{FM}}$ at $z=500$ to $0.55 \, r_{\text{FM}}$ at $z=3000$ to $0.05 \, r_{\text{FM}}$ at $z=5000$
\end{comment}

In the classical picture, as the jet propagates further up and gets accelerated as a whole while opening up, there is a radial adjustment of the poloidal electric circuit: the radius $r_m(z)$ where $J_z=0$ decreases. If, at lower altitudes, the positive current density $J_z$ was mostly localized near the jet outer edge, the fraction of the jet area occupied by a positive $J_z$ increases with the distance. This is expected in a medium with a decreasing external pressure. Indeed, since only the $r<r_m$ zone is self-collimated, whenever the outer pressure hence the confinement decreases, the outer parts expand laterally. This dilution of the outer current density is consistent with a more efficient jet acceleration in these outer regions.

\citet{komissarov2007} and \citet{komissarov2009} proposed that this situation would lead to the formation of outer zones no longer carrying a toroidal field, thereby forcing the poloidal current to close inside the jet body. This situation would arise whenever the lateral jet expansion (due to the external pressure decay) would occur faster than the jet is able to radially adjust, leading thereby to a loss of radial connectivity between the axis and these outer regions. According to \citet{heyvaerts2003c} this would correspond to a conical jet profile inside which cylindrical surfaces (carrying a non vanishing current $I_\infty$) could exist. However, these authors argue that $I_\infty \neq 0$ is inconsistent with a super-FM outflow everywhere, so that $I_\infty$ must vanish to zero in any case (albeit on a logarithmic scale).

%%%%%%%%
%%%%%%%%%

\subsubsection{Recollimation shocks and causal connectivity}

\begin{figure}
\centering
\includegraphics[width=\linewidth,trim=0 9 0 5, clip]{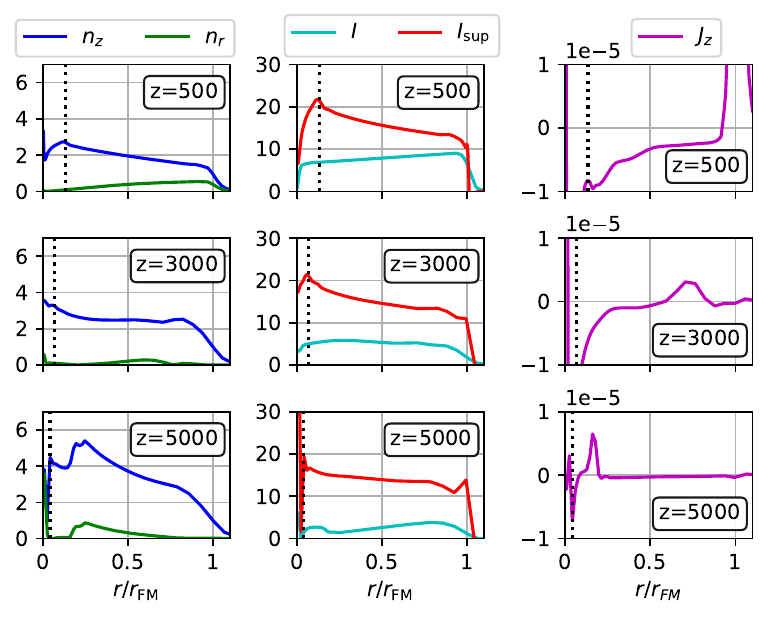}
\caption{Radial distributions of several quantities at different altitudes $z$, for reference simulation O1 at $t_{\text{end}}$. Left: vertical $n_z$ and radial $n_r$ FM Mach numbers. Middle: poloidal electric current $I$ and its analytical upper limit $I_{\text{sup}}$ (in code units). Right: Vertical current density $J_z$ (in code units). Vertical dashed lines display the position of the magnetic surface anchored at $r_0=1$. The cylindrical radius $r$ has been normalized to the jet radius $r_{\text{FM}}(z)$.}
\label{fig:3RadialProfiles}
\end{figure}

Our simulations display a quite different behavior. Fig.~\ref{fig:3RadialProfiles} shows the radial profiles of several quantities at three different altitudes for our reference simulation O1. To ease comparisons, the cylindrical radius $r$ has been normalized to the jet radius $r_{\text{FM}}(z)$. As expected for super-FM flows, the current $I < I_{\text{sup}}(\Psi) = \frac{4\pi}{3\mu_0}\frac{\eta E}{\Omega_*}$ for all magnetic surfaces \citep{heyvaerts2003c}. Before causal connectivity is lost between the jet axis and its outer edge, the jet undergoes a recollimation towards the axis. Such a non-monotonous behavior of the jet profile was not envisioned in the studies mentioned above. As shown in Fig.~\ref{fig:3RadialProfiles}, the FM Mach number $n=u_p/V_{\text{FM},p}\simeq n_z= u_z/V_{\text{FM},z}$ is indeed always larger than unity but, thanks to recollimation, the radial FM Mach number $n_r= u_r/V_{\text{FM},r}$ remains always smaller than unity, ensuring causal connectivity.

%% mecanisme de fermeture, analogue a un court circuit ?
Several systems of shocks are eventually necessary to maintain this connectivity all along the flow. As the flow expands (due to the decrease of the external pressure), it seems like the poloidal electric circuit chooses to reconnect to the axis along a transversal thin shock layer (standing recollimation shock). This prevents any loss of causal connectivity, with jet outer layers propagating ballistically with no more $B_\varphi$. From this point of view, each standing shock corresponds to some short-circuit, allowing to separate one current cell zone from the other. The force responsible for jet recollimation is the magnetic pinching force $F_{\perp}<0$. It corresponds to the dominant hoop stress in most of the jet transverse structure where $J_z<0$ \citep{Pelletier1992,ferreira1997}. However, in the outer region where $J_z>0$ then $I$ vanishes, the collimating force is instead $F_{\perp}\simeq B_pJ_\varphi < 0$, caused by the poloidal magnetic pressure (see top panel in Fig.~\ref{fig:Z500Profiles}).

%%% cellules de courant
On the right column of Fig. \ref{fig:3RadialProfiles} we see that the recollimation shocks also modify the profile of the vertical current density. We still see two regions of negative and positive $J_z$. But here, they do not take up the whole jet: they are localised around the shock, that acts as a current sheet. The left  panel of Fig.~\ref{fig:SchemaCollimationAsymptotique} provides a sketch of the mechanism at play in our simulations, while the right panel show relevant circuits in our fiducial simulation. Plasma ejected from the disc is accelerated up to super-Alfv\'enic ($m=u_p/V_{A}$>1) and then super-FM  ($n=u_p/V_{\text{FM}}>1$) speeds in the first region of maximal current $I_0$. This first acceleration region connects the disc (JED) and its e.m.f. to the first recollimation shock (shown as a simple conical red line). In the next region or cell, the shock itself serves as an e.m.f for the downstream MHD flow. The subsequent regions play a similar role, carrying their own electric current $I_n$ and allowing an interplay between the kinetic and magnetic energies.
%%%%%%%%%%

\subsubsection{The vanishing of the asymptotic current}

The real asymptotic regime, valid for all magnetic surfaces within the jet, corresponds to $ \nabla_{\parallel} I_\infty =0$. It can be achieved in two ways: (a)  $I_\infty=0$ for 
kinetic dominated jets with $\sigma_\infty=0$ (hence $B_\varphi (z \rightarrow \infty) =0$), or (b) current-carrying force-free jets with $I_\infty \neq 0$. While case (a) corresponds mostly to a ballistic motion, case (b) raises the issue of current closure. A negative current density $J_z$ flowing within the jet cannot be compensated by a positive $J_z$ flowing along the jet surface without violating the $ \nabla_{\parallel} I_\infty =0$ condition (${\bf J}_p$ must be parallel to ${\bf B}_p$). In numerical simulations, this is however acceptable once the jet leaves the computational domain, since the current closure is no longer a constraint. But astrophysical jets have no such boundary and current closure must be done for steady-state to be achieved.

%% Evolution selon z 
In steady-state, MHD invariants are conserved so that each standing recollimation shock acts as a new e.m.f for the next region. So, as long as no energy is dissipated within the shocks, this pattern of successive shocks could in principle be repeating itself indefinitely.  Under these circumstances, a MHD jet could maintain a non vanishing current on scales much larger than those envisioned by \citet{heyvaerts2003c}. This is reminiscent of the early jet model of \citet{chan1980} showing quasi periodic  oscillations of the radius, although in their model no connection with the source was taken into account. Besides energy dissipation within the shocks, the other possibility to diminish the current available in each cell is of course through the external pressure. A smaller pressure leads to an opening of the jet radius, leading to a further conversion of electromagnetic energy into kinetic energy. 

\begin{figure}
\centering
\includegraphics[width=.8\linewidth,trim=0 0 0 200, clip]{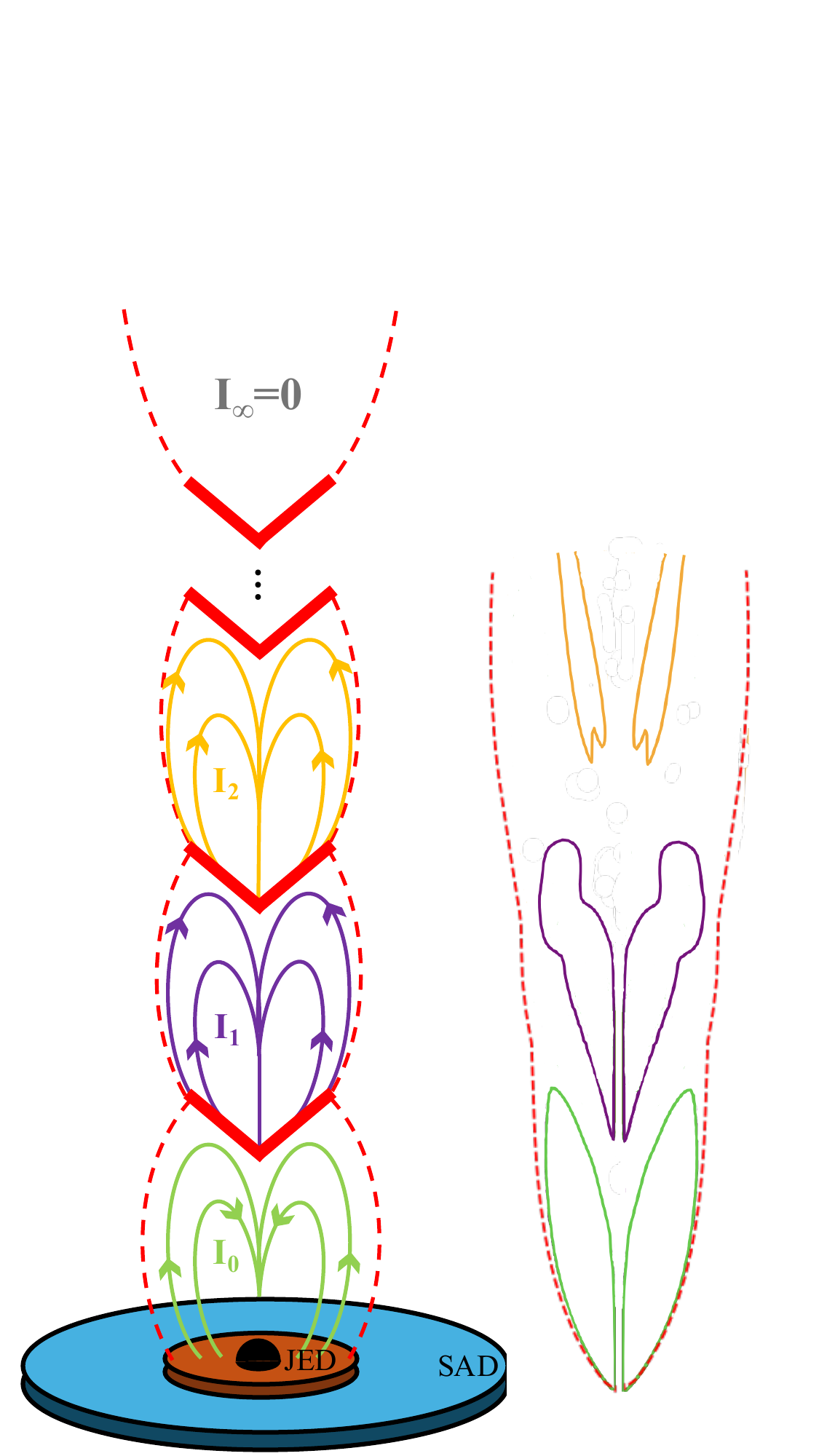}
\caption{Left: Sketch of an axisymmetric MHD jet of finite size. The accretion disc is threaded by a large-scale vertical field but only the innermost region, the Jet-Emitting Disc in orange, is assumed to launch a super-FM jet (the FM critical surface is represented as the dashed red surface). The jet is first accelerated and collimated at the expense of the poloidal electric current $I$, whereas its radius is determined by the external pressure (not represented here). Jet acceleration corresponds to the butterfly shape of the electric current lines. The first electric circuit (current $I_0$ in green) leads plasma from the disc to the first recollimation shock (thick red line). Successive shocks $k=1,2...$ define regions of (almost self-contained) poloidal electric current $I_k$, leading to jet re-acceleration and jet widening until refocusing to the next shock. If some dissipation occurs in shocks, this sausage-like configuration could end up in a final nozzle point, beyond which a kinetic dominated jet with $I_\infty=0$ is refracted and propagates with a paraboloidal shape.\\
Right: Poloidal electric circuits for the reference simulation O1 at $t_{\text{end}}$, using the same color code.}
\label{fig:SchemaCollimationAsymptotique}
\end{figure}

In our simulations, even though the shocks lead to a local jump in current, it still decreases with altitude ($I_0 > I_1 > I_2>...$) as seen on Figs. \ref{fig:Imax} and \ref{fig:3RadialProfiles}. This might eventually lead to a vanishing electric current ($I_\infty = 0$), as illustrated in Fig. \ref{fig:SchemaCollimationAsymptotique}.

\subsubsection{Information propagation in the outer sheath}
    
Since our simulations are stationary, it is a proof that our outflows are stable to axisymmetric perturbations. One could even argue that recollimation, with the subsequent sausage-like jet profile, is the means for our jets to maintain transverse causal connectivity and stay thereby stable. But this holds only for 2D instabilities. As discussed above, 3D simulations should be done in order to assess the complete stability of these jets. However, no instability is expected below the Alfv\'en surface due to the stabilizing effect of the dominant magnetic tension. In the subsequent super-Alfv\'enic zone where the jet undergoes its lateral expansion, it has been shown that the growth rate of Kelvin-Helmholtz and current-driven instabilities is actually dramatically reduced by this expansion \citep{rosen2000,moll2008,porth2015,kim2016,kim2018}. So, one would expect conventional 3D instabilities to occur only once the jet has achieved its asymptotic state. However, our simulations undergo a series of lateral expansion and recollimation and a state with no curvature effect is never achieved. 

There is nevertheless an intrinsic 3D instability, so-called recollimation or centrifugal instability, that may play an important role once the jet starts to recollimate \citep{matsumoto2013,gourgouliatos2018a,gourgouliatos2018b}. Although the magnetic field has been claimed to possibly inhibit this instability \citep{komissarov2019, matsumoto2021}, the physical conditions of the studied cases were quite different from the situation described in this paper. Such an instability remains thereby an actual possibility that should be investigated. Note however that recollimation instability does not necessarily lead to jet destruction but could instead drive a jet wobbling. 

%%%%%%%%%% XRB jets and QPOs
Interestingly, our simulations of jets of finite radius $r_{\text{FM}}(z)$ show that the total current $I(z)$ is not entirely conveyed inside the super-FM jet. Indeed, the radius $r_I(z)$ where the total current vanishes is actually slightly larger than $r_{\text{FM}}(z)$ (see for instance Figs.~\ref{fig:Z500Profiles}, \ref{fig:3RadialProfiles} and \ref{fig:SchemaCollimationAsymptotique}). This means that the current sheath surrounding the super-FM jet is (partly) sub-FM. This is not surprising since our outflow has been launched in a medium filled in with an initial potential field. In fact, this situation, namely an outflow propagating inside a magnetized medium, should be rather generic in astrophysics. As a consequence, any strong perturbation (e.g. instability, wobbling, interaction with the ambient medium) occurring at the jet interface will propagate as a FM wave downstream, but also upstream along the jet interface and down to the disc. If the jet itself can be seen as a supersonic flow, forbidding thereby the upstream propagation of any perturbation, its magnetic sheath remains subsonic and allows information to flow back towards the source. It has been proposed that this generic situation could explain the low frequency quasi-periodic oscillations seen in X-ray binaries whenever jets are present \citep{ferreira2022}.

\subsection{Constraints from observed astrophysical jets}
\label{sec:CompObs}
%%%%%%%%%%%%%%%%%

We mentioned that jets are expected around most accreting objects. Around AGNs and young stars in particular, collimated outflows are routinely observed. It would be tempting to directly link their shape to the collimation of their magnetic surfaces. However, it is unclear what mechanisms lead them to shine (and the recollimation shocks described here do certainly offer a plausible one). Indeed, unless jets are resolved spatially in their transverse direction, observational signatures could either probe only the axis/spine or the (emissivity averaged) body volume, or only the jet/ambient medium interface. If it is the latter ($r_{\text{obs}} \sim r_{\text{FM}}$), then the confining external pressure would set the jet shape. If it is the former ($r_{\text{obs}} \ll r_{\text{FM}}$), it would be forced by the self-collimating hoop stress. Thereby concluding on jet dynamics without knowing which jet region is actually shining is a difficult task, and usually requires to \textit{choose} a paradigm. The implications drawn from our MHD jet dynamics will hopefully provide valuable assistance.

\subsubsection{Jets around active galactic nuclei}\label{sec:ComparisonAGNObservations}

Several moving but also standing knots are observed along AGN jets (see e.g. \citealt{lister2009,lister2013,mertens2016,doi2018} or \citealt{boccardi2017} for a review). In the M87 jet, stationary bright features are observed near the HST-1 complex. These standing knots could be caused by particle acceleration occurring at standing recollimation shocks. Indeed, the distances at which recollimation shocks appear in our simulations are on the same scale as the closest standing knots in M87 (see e.g. fig.~2 of \citealt{asada2012}). We also expect standing shocks to appear at much larger scales. First, the current in our jets is still far from zero when it reaches the outer boundary of the box (see Fig. \ref{fig:Imax}). Because it decreases as $1 / \ln{z}$, the current should become negligible much further away. Moreover, one should keep in mind that our simulations are non-relativistic. In the relativistic case, the increased inertia and the electric field should weaken the jet confinement. It is not sure that recollimation shocks would still appear in relativistic simulations with a similar setup\footnote{Note however that \citet{chatterjee2019} found time dependent radial variations of the jet radius in their 2D relativistic simulations, whose origin remains unidentified.}, but if they do, they should be at greater altitudes.

A transition from parabolic to conical shape is commonly observed in AGN jets, such as M87 \citep{asada2012}, NGC 6251 \citep{Tseng2016}, NGC 4261 \citep{nakahara2018}, NGC 315 \citep{boccardi2021} or 3C273 \citep{akiyama2018, Okino2022}. While jets such as Cygnus A \citep{boccardi2016, nakahara2019} or 3C84 \citep{Giovannini2018} have different behaviors, the parabolic-to-conical profile seems rather generic among both Fanaroff-Riley jet types \citep{Algaba2017,Pushkarev2017,Kovalev2020}. One should not be too eager to associate this transition to the $I_\infty = 0$ asymptotic state. First, because \cite{heyvaerts1989} showed that the case $I_\infty = 0$ corresponds to asymptotic magnetic surfaces that are paraboloidal, not conical. Then, because the observed jet width $r_{\text{obs}}$ would not necessarily follow any magnetic field line. Certainly, if $r_{\text{obs}} \ll r_{\text{FM}}$, then a drop of current would lead to a decrease in collimation. But it is not sure what jet shape this would result in.

Shocks are thought to be the cause of changes in polarisation in jets \citep[see e.g.][]{Hodge2018}. The MHD recollimation shock is a natural candidate as it provides a sudden increase in $B_\varphi$. Steady recollimation shocks can also power flares \citep[see e.g.][and references therein]{fichetdeclairfontaine2021,fichetdeclairfontaine2022}. But when the jet has consumed all its current (or $B_\varphi$), then it does not recollimate anymore. In the asymptotic state $I_\infty = 0$, there are no recollimation shocks anymore. The knots, polarisation changes and flares that are caused by those shocks should stop appearing. Naturally recollimation shocks will not be the sole cause of these observables, necessating a specific care when interpreting the observations.

\subsubsection{Jets around protostars}\label{sec:ComparisonYSOObservations}
%%%%%%%%%%%%%%%%%%%%%%%%%%%%%

Making comparisons between our simulations and YSO jets is not an easy task, due to their complexity. In HH30 for instance, we can see both an inner fast and highly collimated atomic jet in the optical \citep{burrows1996}, as well as an outer, slower and less collimated molecular wind in radio \citep{louvet2018}. Similar structures are found in many other objects, such as DG Tau for instance \citep[see e.g.][]{agra-amboage2011,Agra-Amboage2014,devalon2020,delabrosse2024}. In fact, it has been argued that YSO outflows have several nested components \citep{Ferreira2006b}: an inner spine made of a hot, tenuous stellar wind, surrounded by time-dependent star-disc massive magnetospheric ejecta \citep{zanni2013} confined by a self-collimated jet emitted from the inner disc \citep{combet2008}, itself confined by a more massive disc wind \citep{bai2016,lesur2021}. It seems therefore difficult to make a direct correspondance between our simplified jet simulations and these highly structured outflows.  

%Protostellar jets present lots of interesting features such as Herbig-Haro objects. They show quasi-periodic features along the flow (see \citealt{lee2017} for HH212 or \citealt{louvet2018} for HH30) that are often interpreted as bow shocks. As mentioned in \citetalias{jannaud2023}, this is often interpreted as the result of jet instabilities, or a time-dependent jet production mechanism (see \citealt{tabone2018} and references therein for HH212). On the contrary, we think that those structures can be the result of the successive recollimation shocks along the jet. After a first recollimation shock has refracted the jet away from the axis, it is then collimated again by Z-pinch until forming a new recollimation shock, and so forth.

For instance, YSO jets present lots of interesting features such as Herbig-Haro objects, which are quasi-periodic knots present along the outflow axis (e.g. \citealt{lee2017} for HH212 or \citealt{louvet2018} for HH30). They are often interpreted as the result of jet instabilities, or due to a time-dependent jet production mechanism (e.g. \citealt{raga1990}, see \citealt{tabone2018} and references therein for HH212). And, indeed, moving knots that have a bow-like shape do advocate for the last interpretation. In DG Tau for instance, a standing recollimation shock has been however detected \citep{white2014}, also seen in X-rays \citep{guedel2008}. In strong contrast with AGN jets, the position of this standing shock is between 30 and 50 au from the central star, namely only 300 to 500 times the innermost disc radius, hence much closer to the source. While several observational arguments do favor the MHD refocusing model (see discussion in \citealt{white2014}), obtaining a first recollimation shock so close to the source is quite challenging. A look at our Fig.~\ref{fig:OmegaTronqueChoc} shows for instance that an extra e.m.f due to a fast rotating core ($\Omega_{*_{\text{a}}}=1$) would be required. Such a situation could actually be achieved if the disc truncation is near the co-rotation radius for that particular object. More observations are needed to assess whether this is a common trend in YSOs or particular to DG Tau.

Finally, our simulations indicate that the magnetic pressure collimating the jet should come from an outer MHD wind. In our simulation B4, $r_\text{FM}$ is collimated down to 15 au at an altitude of 50 au. If $r_{\text{obs}} \sim r_{\text{FM}}$, then it is the scale as some collimated protostellar jets \citep[see e.g.][]{lee2017,Lee2025}. At 50 au, we provide a magnetic field of 10 mG. This is two to three orders of magnitude higher than the organised field measured in dense prestellar cores \citep{Lee2017a}. As shown by \cite{cabrit2007}, this should mean that the external field is provided by outer MHD wind.

\section{Conclusion}\label{sec:Conclusion}
%%%%%%%%%%%%%%%%%%%%%%%%%%%%%%%%%%%%%%%%%%%%%%%%%%%%%%%%%%%%%%%%%%%

We performed 2.5D ideal MHD simulations of jets emerging from a Keplerian JED of finite size, propagating in a surrounding potential field. Those are the first large-scale platform simulations of steady-state jets. We recover most previous results on jets, in particular the link between the jet profile and the external pressure profile, and its impact on jet efficiency.

All simulations display standing recollimation shocks at large distances from the disc, as in the extended jet simulations of \citetalias{jannaud2023}. As the simulations of this work stray further away from self-similarity, this further proves that such shocks are intrinsic to self-collimated jets, rather than a bias of the self-similar approach.

We argue that recollimation is a natural outcome of the MHD collimation process, in order to maintain transverse causal connectivity. Over most of the jet body, recollimation is caused by the hoop-stress. The jet self-organizes as a series of successive cells, carrying their own poloidal electric circuit, separated by a recollimation shock. Due to this insulation effect, the poloidal electric current decreases less rapidly than the $1/\ln z$ scaling found by \citet{heyvaerts2003c}.

Our super-FM jets are always surrounded by a sub-FM magnetic sheath. This opens the way for information to propagate upstream alongside and may explain low frequency Quasi Periodic Oscillations (QPOs) as proposed in \cite{ferreira2022}. 3D simulations should be carried out to assess whether the recollimation instability is triggered in these circumstances. If so, it would be a generic feature of MHD jets, possibly leading to jet wobbling.

The shape of observed jets could depend both on the MHD self-collimation and on the confining external pressure, depending on which part of the jet is actually radiating. This prevents making direct comparisons between our simulations and the observations of collimated AGN and protostellar jets. However, the presence of standing shocks is expected to contribute to their luminosity. We propose that the standing knots observed in some of those jets are tracers of our steady recollimation shocks. We also confirm that any external pressure confining protostellar atomic jets must be greater than the pressure found in the unshocked interstellar medium.

The rotation of the central object was found to have a small influence on shock altitude. More refined modelling of the ejection properties inside the disc inner radius is required for a complete understanding of jet self-collimation. The magnetic field distribution ($\alpha$ in $B_z \propto r^{\alpha - 2}$) has been fixed to $\alpha = 3/4$ in all simulations of this work. It is a crucial parameter, as it sets not only the collimation of the jet at launch, but also how the electric currents close: on the central object for $\alpha < 1$ or on both the central object and the disk for $\alpha > 1$. As a consequence, the configurations with $\alpha > 1$ appear unstable. The consequences of this dichotomy on jet launching and collimation deserves some investigation.

\section*{Acknowledgements}

TJ wishes to thank Sylvie Cabrit, Catherine Dougados and Serguei Komissarov for fruitful scientific discussions. JF acknowledges financial support from the CNES French space agency and the ATPEM program of French CNRS. All the computations presented in this paper were performed using the GRICAD infrastructure (\url{https://gricad.univ-grenoble-alpes.fr}), which is supported by Grenoble research communities.

%%%%%%%%%%%%%%%%%%%%%%%%%%%%%%%%%%%%%%%%%%%%%%%%%%
\section*{Data Availability}

Simulation data will be made available upon reasonable request to
the corresponding author.



\bibliographystyle{mnras}
\bibliography{references5} % if your bibtex file is called example.bib

@ARTICLE{delabrosse2024,
       author = {{Delabrosse}, V. and {Dougados}, C. and {Cabrit}, S. and {Tabone}, B. and {Tychoniec}, L. and {Ray}, T. and {Podio}, L. and {McClure}, M.},
        title = "{JWST study of the DG Tau B disk-wind candidate. I. Overview and nested H$_{2}$-CO outflows}",
      journal = {\aap},
         year = 2024,
        month = aug,
       volume = {688},
          eid = {A173},
        pages = {A173},
          doi = {10.1051/0004-6361/202449176},
archivePrefix = {arXiv},
       eprint = {2403.19400},
 primaryClass = {astro-ph.SR},
       adsurl = {https://ui.adsabs.harvard.edu/abs/2024A&A...688A.173D}
}

@ARTICLE{LyndenBell2003,
       author = {{Lynden-Bell}, D.},
        title = "{On why discs generate magnetic towers and collimate jets}",
      journal = {\mnras},
         year = 2003,
        month = jun,
       volume = {341},
       number = {4},
        pages = {1360-1372},
          doi = {10.1046/j.1365-8711.2003.06506.x},
archivePrefix = {arXiv},
       eprint = {astro-ph/0208388},
 primaryClass = {astro-ph},
       adsurl = {https://ui.adsabs.harvard.edu/abs/2003MNRAS.341.1360L}
}

@ARTICLE{LyndenBell1996,
       author = {{Lynden-Bell}, D.},
        title = "{Magnetic collimation by accretion discs of quasars and stars}",
      journal = {\mnras},
         year = 1996,
        month = mar,
       volume = {279},
       number = {2},
        pages = {389-401},
          doi = {10.1093/mnras/279.2.389},
       adsurl = {https://ui.adsabs.harvard.edu/abs/1996MNRAS.279..389L}
}

@ARTICLE{Lee2025,
       author = {{Lee}, Chin-Fei and {Jhan}, Kai-Syun and {Moraghan}, Anthony},
        title = "{A magnetized protostellar jet launched from the innermost disk at the truncation radius}",
      journal = {Scientific Reports},
         year = 2025,
        month = aug,
       volume = {15},
       number = {1},
          eid = {29702},
        pages = {29702},
          doi = {10.1038/s41598-025-11602-w},
       adsurl = {https://ui.adsabs.harvard.edu/abs/2025NatSR..1529702L}
}

@ARTICLE{Hodge2018,
       author = {{Hodge}, M.~A. and {Lister}, M.~L. and {Aller}, M.~F. and {Aller}, H.~D. and {Kovalev}, Y.~Y. and {Pushkarev}, A.~B. and {Savolainen}, T.},
        title = "{MOJAVE XVI: Multiepoch Linear Polarization Properties of Parsec-scale AGN Jet Cores}",
      journal = {\apj},
         year = 2018,
        month = aug,
       volume = {862},
       number = {2},
          eid = {151},
        pages = {151},
          doi = {10.3847/1538-4357/aacb2f},
archivePrefix = {arXiv},
       eprint = {1806.07312},
 primaryClass = {astro-ph.GA},
       adsurl = {https://ui.adsabs.harvard.edu/abs/2018ApJ...862..151H}
}

@ARTICLE{white2014,
       author = {{White}, M.~C. and {McGregor}, P.~J. and {Bicknell}, G.~V. and {Salmeron}, R. and {Beck}, T.~L.},
        title = "{Multi-epoch sub-arcsecond [Fe II] spectroimaging of the DG Tau outflows with NIFS - I. First data epoch}",
      journal = {\mnras},
         year = 2014,
        month = jun,
       volume = {441},
       number = {2},
        pages = {1681-1707},
          doi = {10.1093/mnras/stu654},
archivePrefix = {arXiv},
       eprint = {1404.0728},
 primaryClass = {astro-ph.SR},
       adsurl = {https://ui.adsabs.harvard.edu/abs/2014MNRAS.441.1681W}
}

@ARTICLE{pelletier1992,
       author = {{Pelletier}, Guy and {Pudritz}, Ralph E.},
        title = "{Hydromagnetic Disk Winds in Young Stellar Objects and Active Galactic Nuclei}",
      journal = {\apj},
         year = 1992,
        month = jul,
       volume = {394},
        pages = {117},
          doi = {10.1086/171565},
       adsurl = {https://ui.adsabs.harvard.edu/abs/1992ApJ...394..117P}
}

@ARTICLE{heyvaerts2003a,
       author = {{Heyvaerts}, Jean and {Norman}, Colin},
        title = "{Global Asymptotic Solutions for Nonrelativistic Magnetohydrodynamic Jets and Winds}",
      journal = {\apj},
         year = 2003,
        month = oct,
       volume = {596},
       number = {2},
        pages = {1270-1294},
          doi = {10.1086/378220},
archivePrefix = {arXiv},
       eprint = {astro-ph/0309128},
 primaryClass = {astro-ph},
       adsurl = {https://ui.adsabs.harvard.edu/abs/2003ApJ...596.1270H}
}

@ARTICLE{heyvaerts2003b,
       author = {{Heyvaerts}, Jean and {Norman}, Colin},
        title = "{Global Asymptotic Solutions for Relativistic Magnetohydrodynamic Jets and Winds}",
      journal = {\apj},
         year = 2003,
        month = oct,
       volume = {596},
       number = {2},
        pages = {1240-1255},
          doi = {10.1086/378221},
archivePrefix = {arXiv},
       eprint = {astro-ph/0309132},
 primaryClass = {astro-ph},
       adsurl = {https://ui.adsabs.harvard.edu/abs/2003ApJ...596.1240H}
}

@ARTICLE{heyvaerts2003c,
       author = {{Heyvaerts}, Jean and {Norman}, Colin},
        title = "{Kinetic Energy Flux versus Poynting Flux in Magnetohydrodynamic Winds and Jets: The Intermediate Regime}",
      journal = {\apj},
         year = 2003,
        month = oct,
       volume = {596},
       number = {2},
        pages = {1256-1269},
          doi = {10.1086/378222},
archivePrefix = {arXiv},
       eprint = {astro-ph/0309143},
 primaryClass = {astro-ph},
       adsurl = {https://ui.adsabs.harvard.edu/abs/2003ApJ...596.1256H}
}

@ARTICLE{ferreira2013,
       author = {{Ferreira}, J. and {Casse}, F.},
        title = "{On fan-shaped cold MHD winds from Keplerian accretion discs}",
      journal = {\mnras},
         year = 2013,
        month = jan,
       volume = {428},
       number = {1},
        pages = {307-320},
          doi = {10.1093/mnras/sts012},
archivePrefix = {arXiv},
       eprint = {1209.3871},
 primaryClass = {astro-ph.SR},
       adsurl = {https://ui.adsabs.harvard.edu/abs/2013MNRAS.428..307F}
}

@ARTICLE{vlahakis2003a,
       author = {{Vlahakis}, Nektarios and {K{\"o}nigl}, Arieh},
        title = "{Relativistic Magnetohydrodynamics with Application to Gamma-Ray Burst Outflows. I. Theory and Semianalytic Trans-Alfv{\'e}nic Solutions}",
      journal = {\apj},
         year = 2003,
        month = oct,
       volume = {596},
       number = {2},
        pages = {1080-1103},
          doi = {10.1086/378226},
archivePrefix = {arXiv},
       eprint = {astro-ph/0303482},
 primaryClass = {astro-ph},
       adsurl = {https://ui.adsabs.harvard.edu/abs/2003ApJ...596.1080V}
}

@ARTICLE{Tzeferacos2009,
       author = {{Tzeferacos}, P. and {Ferrari}, A. and {Mignone}, A. and {Zanni}, C. and {Bodo}, G. and {Massaglia}, S.},
        title = "{On the magnetization of jet-launching discs}",
      journal = {\mnras},
         year = 2009,
        month = dec,
       volume = {400},
       number = {2},
        pages = {820-834},
          doi = {10.1111/j.1365-2966.2009.15502.x},
       adsurl = {https://ui.adsabs.harvard.edu/abs/2009MNRAS.400..820T}
}

@ARTICLE{Porth2019,
       author = {{Porth}, Oliver and {Chatterjee}, Koushik and {Narayan}, Ramesh and {Gammie}, Charles F. and {Mizuno}, Yosuke and {Anninos}, Peter and {Baker}, John G. and {Bugli}, Matteo and {Chan}, Chi-kwan and {Davelaar}, Jordy and {Del Zanna}, Luca and {Etienne}, Zachariah B. and {Fragile}, P. Chris and {Kelly}, Bernard J. and {Liska}, Matthew and {Markoff}, Sera and {McKinney}, Jonathan C. and {Mishra}, Bhupendra and {Noble}, Scott C. and {Olivares}, H{\'e}ctor and {Prather}, Ben and {Rezzolla}, Luciano and {Ryan}, Benjamin R. and {Stone}, James M. and {Tomei}, Niccol{\`o} and {White}, Christopher J. and {Younsi}, Ziri and {Akiyama}, Kazunori and {Alberdi}, Antxon and {Alef}, Walter and {Asada}, Keiichi and {Azulay}, Rebecca and {Baczko}, Anne-Kathrin and {Ball}, David and {Balokovi{\'c}}, Mislav and {Barrett}, John and {Bintley}, Dan and {Blackburn}, Lindy and {Boland}, Wilfred and {Bouman}, Katherine L. and {Bower}, Geoffrey C. and {Bremer}, Michael and {Brinkerink}, Christiaan D. and {Brissenden}, Roger and {Britzen}, Silke and {Broderick}, Avery E. and {Broguiere}, Dominique and {Bronzwaer}, Thomas and {Byun}, Do-Young and {Carlstrom}, John E. and {Chael}, Andrew and {Chatterjee}, Shami and {Chen}, Ming-Tang and {Chen}, Yongjun and {Cho}, Ilje and {Christian}, Pierre and {Conway}, John E. and {Cordes}, James M. and {Geoffrey} and {Crew}, B. and {Cui}, Yuzhu and {De Laurentis}, Mariafelicia and {Deane}, Roger and {Dempsey}, Jessica and {Desvignes}, Gregory and {Doeleman}, Sheperd S. and {Eatough}, Ralph P. and {Falcke}, Heino and {Fish}, Vincent L. and {Fomalont}, Ed and {Fraga-Encinas}, Raquel and {Freeman}, Bill and {Friberg}, Per and {Fromm}, Christian M. and {G{\'o}mez}, Jos{\'e} L. and {Galison}, Peter and {Garc{\'\i}a}, Roberto and {Gentaz}, Olivier and {Georgiev}, Boris and {Goddi}, Ciriaco and {Gold}, Roman and {Gu}, Minfeng and {Gurwell}, Mark and {Hada}, Kazuhiro and {Hecht}, Michael H. and {Hesper}, Ronald and {Ho}, Luis C. and {Ho}, Paul and {Honma}, Mareki and {Huang}, Chih-Wei L. and {Huang}, Lei and {Hughes}, David H. and {Ikeda}, Shiro and {Inoue}, Makoto and {Issaoun}, Sara and {James}, David J. and {Jannuzi}, Buell T. and {Janssen}, Michael and {Jeter}, Britton and {Jiang}, Wu and {Johnson}, Michael D. and {Jorstad}, Svetlana and {Jung}, Taehyun and {Karami}, Mansour and {Karuppusamy}, Ramesh and {Kawashima}, Tomohisa and {Keating}, Garrett K. and {Kettenis}, Mark and {Kim}, Jae-Young and {Kim}, Junhan and {Kim}, Jongsoo and {Kino}, Motoki and {Koay}, Jun Yi and {Patrick} and {Koch}, M. and {Koyama}, Shoko and {Kramer}, Michael and {Kramer}, Carsten and {Krichbaum}, Thomas P. and {Kuo}, Cheng-Yu and {Lauer}, Tod R. and {Lee}, Sang-Sung and {Li}, Yan-Rong and {Li}, Zhiyuan and {Lindqvist}, Michael and {Liu}, Kuo and {Liuzzo}, Elisabetta and {Lo}, Wen-Ping and {Lobanov}, Andrei P. and {Loinard}, Laurent and {Lonsdale}, Colin and {Lu}, Ru-Sen and {MacDonald}, Nicholas R. and {Mao}, Jirong and {Marrone}, Daniel P. and {Marscher}, Alan P. and {Mart{\'\i}-Vidal}, Iv{\'a}n and {Matsushita}, Satoki and {Matthews}, Lynn D. and {Medeiros}, Lia and {Menten}, Karl M. and {Mizuno}, Izumi and {Moran}, James M. and {Moriyama}, Kotaro and {Moscibrodzka}, Monika and {M{\"u}ller}, Cornelia and {Nagai}, Hiroshi and {Nagar}, Neil M. and {Nakamura}, Masanori and {Narayanan}, Gopal and {Natarajan}, Iniyan and {Neri}, Roberto and {Ni}, Chunchong and {Noutsos}, Aristeidis and {Okino}, Hiroki and {Oyama}, Tomoaki and {{\"O}zel}, Feryal and {Palumbo}, Daniel C.~M. and {Patel}, Nimesh and {Pen}, Ue-Li and {Pesce}, Dominic W. and {Pi{\'e}tu}, Vincent and {Plambeck}, Richard and {PopStefanija}, Aleksandar and {Preciado-L{\'o}pez}, Jorge A. and {Psaltis}, Dimitrios and {Pu}, Hung-Yi and {Ramakrishnan}, Venkatessh and {Rao}, Ramprasad and {Rawlings}, Mark G. and {Raymond}, Alexander W. and {Ripperda}, Bart and {Roelofs}, Freek and {Rogers}, Alan and {Ros}, Eduardo and {Rose}, Mel and {Roshanineshat}, Arash and {Rottmann}, Helge and {Roy}, Alan L. and {Ruszczyk}, Chet and {Rygl}, Kazi L.~J. and {S{\'a}nchez}, Salvador and {S{\'a}nchez-Arguelles}, David and {Sasada}, Mahito and {Savolainen}, Tuomas and {Schloerb}, F. Peter and {Schuster}, Karl-Friedrich and {Shao}, Lijing and {Shen}, Zhiqiang and {Small}, Des and {Sohn}, Bong Won and {SooHoo}, Jason and {Tazaki}, Fumie and {Tiede}, Paul and {Tilanus}, Remo P.~J. and {Titus}, Michael and {Toma}, Kenji and {Torne}, Pablo and {Trent}, Tyler and {Trippe}, Sascha},
        title = "{The Event Horizon General Relativistic Magnetohydrodynamic Code Comparison Project}",
      journal = {\apjs},
         year = 2019,
        month = aug,
       volume = {243},
       number = {2},
          eid = {26},
        pages = {26},
          doi = {10.3847/1538-4365/ab29fd},
archivePrefix = {arXiv},
       eprint = {1904.04923},
 primaryClass = {astro-ph.HE},
       adsurl = {https://ui.adsabs.harvard.edu/abs/2019ApJS..243...26P}
}

@ARTICLE{zhu2018,
       author = {{Zhu}, Zhaohuan and {Stone}, James M.},
        title = "{Global Evolution of an Accretion Disk with a Net Vertical Field: Coronal Accretion, Flux Transport, and Disk Winds}",
      journal = {\apj},
         year = 2018,
        month = apr,
       volume = {857},
       number = {1},
          eid = {34},
        pages = {34},
          doi = {10.3847/1538-4357/aaafc9},
archivePrefix = {arXiv},
       eprint = {1701.04627},
 primaryClass = {astro-ph.EP},
       adsurl = {https://ui.adsabs.harvard.edu/abs/2018ApJ...857...34Z}
}

@ARTICLE{zimniak2024,
       author = {{Zimniak}, N. and {Ferreira}, J. and {Jacquemin-Ide}, J.},
        title = "{Influence of the turbulent magnetic pressure on isothermal jet emitting disks}",
      journal = {\aap},
         year = 2024,
        month = dec,
       volume = {692},
          eid = {A99},
        pages = {A99},
          doi = {10.1051/0004-6361/202450501},
archivePrefix = {arXiv},
       eprint = {2412.06999},
 primaryClass = {astro-ph.HE},
       adsurl = {https://ui.adsabs.harvard.edu/abs/2024A&A...692A..99Z}
}

@ARTICLE{Frank1999,
       author = {{Frank}, Adam and {Gardiner}, Thomas A. and {Delemarter}, Guy and {Lery}, Thibaut and {Betti}, Riccardo},
        title = "{Ambipolar Diffusion in Young Stellar Object Jets}",
      journal = {\apj},
         year = 1999,
        month = oct,
       volume = {524},
       number = {2},
        pages = {947-951},
          doi = {10.1086/307840},
       adsurl = {https://ui.adsabs.harvard.edu/abs/1999ApJ...524..947F}
}

@ARTICLE{Komissarov2021,
       author = {{Komissarov}, Serguei and {Porth}, Oliver},
        title = "{Numerical simulations of jets}",
      journal = {\nar},
         year = 2021,
        month = jun,
       volume = {92},
          eid = {101610},
        pages = {101610},
          doi = {10.1016/j.newar.2021.101610},
       adsurl = {https://ui.adsabs.harvard.edu/abs/2021NewAR..9201610K}
}

@ARTICLE{chan1980,
       author = {{Chan}, K.~L. and {Henriksen}, R.~N.},
        title = "{On the supersonic dynamics of magnetized jets of thermal gas in radio galaxies}",
      journal = {\apj},
         year = 1980,
        month = oct,
       volume = {241},
        pages = {534-551},
          doi = {10.1086/158368},
       adsurl = {https://ui.adsabs.harvard.edu/abs/1980ApJ...241..534C}
}

@phdthesis{janna2023,
	author = {Jannaud, Thomas},
	note = {2023GRALY057},
	title = {Simulations Num{\'e}riques MHD de Jets Astrophysiques},
	url = {http://www.theses.fr/2023GRALY057/document},
	year = {2023}}

@ARTICLE{rosen2000,
       author = {{Rosen}, Alexander and {Hardee}, Philip E.},
        title = "{The Effect of Expansion on Mass Entrainment and Stability of Super-Alfv{\'e}nic Jets}",
      journal = {\apj},
         year = 2000,
        month = oct,
       volume = {542},
       number = {2},
        pages = {750-760},
          doi = {10.1086/317020},
archivePrefix = {arXiv},
       eprint = {astro-ph/0006102},
 primaryClass = {astro-ph},
       adsurl = {https://ui.adsabs.harvard.edu/abs/2000ApJ...542..750R}
}

@ARTICLE{moll2008,
       author = {{Moll}, R. and {Spruit}, H.~C. and {Obergaulinger}, M.},
        title = "{Kink instabilities in jets from rotating magnetic fields}",
      journal = {\aap},
         year = 2008,
        month = dec,
       volume = {492},
       number = {3},
        pages = {621-630},
          doi = {10.1051/0004-6361:200810523},
archivePrefix = {arXiv},
       eprint = {0809.3165},
 primaryClass = {astro-ph},
       adsurl = {https://ui.adsabs.harvard.edu/abs/2008A&A...492..621M}
}

@ARTICLE{Kim2016,
       author = {{Kim}, Jinho and {Balsara}, Dinshaw S. and {Lyutikov}, Maxim and {Komissarov}, Serguei S.},
        title = "{On the linear stability of sheared and magnetized jets without current sheets - non-relativistic case}",
      journal = {\mnras},
         year = 2016,
        month = sep,
       volume = {461},
       number = {1},
        pages = {728-741},
          doi = {10.1093/mnras/stw1051},
archivePrefix = {arXiv},
       eprint = {1603.00341},
 primaryClass = {astro-ph.HE},
       adsurl = {https://ui.adsabs.harvard.edu/abs/2016MNRAS.461..728K}
}

@ARTICLE{Kim2018,
       author = {{Kim}, Jinho and {Balsara}, Dinshaw S. and {Lyutikov}, Maxim and {Komissarov}, Serguei S.},
        title = "{On the linear stability of sheared and magnetized jets without current sheets - relativistic case}",
      journal = {\mnras},
         year = 2018,
        month = mar,
       volume = {474},
       number = {3},
        pages = {3954-3966},
          doi = {10.1093/mnras/stx3065},
       adsurl = {https://ui.adsabs.harvard.edu/abs/2018MNRAS.474.3954K}
}

@ARTICLE{matsumoto2013,
       author = {{Matsumoto}, Jin and {Masada}, Youhei},
        title = "{Two-dimensional Numerical Study for Rayleigh-Taylor and Richtmyer-Meshkov Instabilities in Relativistic Jets}",
      journal = {\apjl},
         year = 2013,
        month = jul,
       volume = {772},
       number = {1},
          eid = {L1},
        pages = {L1},
          doi = {10.1088/2041-8205/772/1/L1},
archivePrefix = {arXiv},
       eprint = {1306.1046},
 primaryClass = {astro-ph.HE},
       adsurl = {https://ui.adsabs.harvard.edu/abs/2013ApJ...772L...1M}
}

@ARTICLE{matsumoto2021,
       author = {{Matsumoto}, Jin and {Komissarov}, Serguei S. and {Gourgouliatos}, Konstantinos N.},
        title = "{Magnetic inhibition of the recollimation instability in relativistic jets}",
      journal = {\mnras},
         year = 2021,
        month = may,
       volume = {503},
       number = {4},
        pages = {4918-4929},
          doi = {10.1093/mnras/stab828},
archivePrefix = {arXiv},
       eprint = {2010.11012},
 primaryClass = {astro-ph.HE},
       adsurl = {https://ui.adsabs.harvard.edu/abs/2021MNRAS.503.4918M}
}

@ARTICLE{gourgouliatos2018a,
       author = {{Gourgouliatos}, Konstantinos N. and {Komissarov}, Serguei S.},
        title = "{Reconfinement and loss of stability in jets from active galactic nuclei}",
      journal = {Nature Astronomy},
         year = 2018,
        month = dec,
       volume = {2},
        pages = {167-171},
          doi = {10.1038/s41550-017-0338-3},
archivePrefix = {arXiv},
       eprint = {1806.05683},
 primaryClass = {astro-ph.HE},
       adsurl = {https://ui.adsabs.harvard.edu/abs/2018NatAs...2..167G}
}

@ARTICLE{gourgouliatos2018b,
       author = {{Gourgouliatos}, Konstantinos N. and {Komissarov}, Serguei S.},
        title = "{Relativistic centrifugal instability}",
      journal = {\mnras},
         year = 2018,
        month = mar,
       volume = {475},
       number = {1},
        pages = {L125-L129},
          doi = {10.1093/mnrasl/sly016},
archivePrefix = {arXiv},
       eprint = {1710.01345},
 primaryClass = {astro-ph.HE},
       adsurl = {https://ui.adsabs.harvard.edu/abs/2018MNRAS.475L.125G}
}

@ARTICLE{Komissarov2019,
       author = {{Komissarov}, Serguei S. and {Gourgouliatos}, Konstantinos N. and {Matsumoto}, Jin},
        title = "{Magnetic inhibition of centrifugal instability}",
      journal = {\mnras},
         year = 2019,
        month = sep,
       volume = {488},
       number = {3},
        pages = {4061-4073},
          doi = {10.1093/mnras/stz1973},
       adsurl = {https://ui.adsabs.harvard.edu/abs/2019MNRAS.488.4061K}
}

@ARTICLE{zanni2013,
       author = {{Zanni}, C. and {Ferreira}, J.},
        title = "{MHD simulations of accretion onto a dipolar magnetosphere. II. Magnetospheric ejections and stellar spin-down}",
      journal = {\aap},
         year = 2013,
        month = feb,
       volume = {550},
          eid = {A99},
        pages = {A99},
          doi = {10.1051/0004-6361/201220168},
archivePrefix = {arXiv},
       eprint = {1211.4844},
 primaryClass = {astro-ph.SR},
       adsurl = {https://ui.adsabs.harvard.edu/abs/2013A&A...550A..99Z}
}

@ARTICLE{akiyama2018,
       author = {{Akiyama}, Kazunori and {Asada}, Keiichi and {Fish}, Vincent L. and {Nakamura}, Masanori and {Hada}, Kazuhiro and {Nagai}, Hiroshi and {Lonsdale}, Colin J.},
        title = "{The Global Jet Structure of the Archetypical Quasar 3C 273}",
      journal = {Galaxies},
         year = 2018,
        month = jan,
       volume = {6},
       number = {1},
          eid = {15},
        pages = {15},
          doi = {10.3390/galaxies6010015},
       adsurl = {https://ui.adsabs.harvard.edu/abs/2018Galax...6...15A}
}

@ARTICLE{mertens2016,
       author = {{Mertens}, F. and {Lobanov}, A.~P. and {Walker}, R.~C. and {Hardee}, P.~E.},
        title = "{Kinematics of the jet in M 87 on scales of 100-1000 Schwarzschild radii}",
      journal = {\aap},
         year = 2016,
        month = oct,
       volume = {595},
          eid = {A54},
        pages = {A54},
          doi = {10.1051/0004-6361/201628829},
archivePrefix = {arXiv},
       eprint = {1608.05063},
 primaryClass = {astro-ph.HE},
       adsurl = {https://ui.adsabs.harvard.edu/abs/2016A&A...595A..54M}
}

@ARTICLE{nakahara2018,
       author = {{Nakahara}, Satomi and {Doi}, Akihiro and {Murata}, Yasuhiro and {Hada}, Kazuhiro and {Nakamura}, Masanori and {Asada}, Keiichi},
        title = "{Finding Transitions of Physical Condition in Jets from Observations over the Range of {}10$^{3}$-{}10$^{9}$ Schwarzschild Radii in Radio Galaxy NGC 4261}",
      journal = {\apj},
         year = 2018,
        month = feb,
       volume = {854},
       number = {2},
          eid = {148},
        pages = {148},
          doi = {10.3847/1538-4357/aaa45e},
       adsurl = {https://ui.adsabs.harvard.edu/abs/2018ApJ...854..148N}
}

@ARTICLE{nakahara2019,
       author = {{Nakahara}, Satomi and {Doi}, Akihiro and {Murata}, Yasuhiro and {Nakamura}, Masanori and {Hada}, Kazuhiro and {Asada}, Keiichi},
        title = "{The Cygnus A Jet: Parabolic Streamlines up to Kiloparsec Scales}",
      journal = {\apj},
         year = 2019,
        month = jun,
       volume = {878},
       number = {1},
          eid = {61},
        pages = {61},
          doi = {10.3847/1538-4357/ab1b0e},
       adsurl = {https://ui.adsabs.harvard.edu/abs/2019ApJ...878...61N}
}

@ARTICLE{Combet2008,
       author = {{Combet}, C. and {Ferreira}, J.},
        title = "{The radial structure of protostellar accretion disks: influence of jets}",
      journal = {\aap},
         year = 2008,
        month = feb,
       volume = {479},
       number = {2},
        pages = {481-491},
          doi = {10.1051/0004-6361:20078734},
archivePrefix = {arXiv},
       eprint = {0712.0913},
 primaryClass = {astro-ph},
       adsurl = {https://ui.adsabs.harvard.edu/abs/2008A&A...479..481C}
}

@ARTICLE{raga1990,
       author = {{Raga}, A.~C. and {Canto}, J. and {Binette}, L. and {Calvet}, N.},
        title = "{Stellar Jets with Intrinsically Variable Sources}",
      journal = {\apj},
         year = 1990,
        month = dec,
       volume = {364},
        pages = {601},
          doi = {10.1086/169443},
       adsurl = {https://ui.adsabs.harvard.edu/abs/1990ApJ...364..601R}
}

@article{bai2016,
	author = {{Bai}, X.-N. and {Ye}, J. and {Goodman}, J. and {Yuan}, F.},
	journal = {\apj},
	month = feb,
	pages = {152},
	title = {{Magneto-thermal Disk Winds from Protoplanetary Disks}},
	volume = 818,
	year = 2016}

@ARTICLE{lesur2021,
       author = {{Lesur}, Geoffroy R.~J.},
        title = "{Systematic description of wind-driven protoplanetary discs}",
      journal = {\aap},
         year = 2021,
        month = jun,
       volume = {650},
          eid = {A35},
        pages = {A35},
          doi = {10.1051/0004-6361/202040109},
archivePrefix = {arXiv},
       eprint = {2101.10349},
 primaryClass = {astro-ph.SR},
       adsurl = {https://ui.adsabs.harvard.edu/abs/2021A&A...650A..35L}
}

@ARTICLE{devalon2020,
       author = {{de Valon}, A. and {Dougados}, C. and {Cabrit}, S. and {Louvet}, F. and {Zapata}, L.~A. and {Mardones}, D.},
        title = "{ALMA reveals a large structured disk and nested rotating outflows in DG Tauri B}",
      journal = {\aap},
         year = 2020,
        month = feb,
       volume = {634},
          eid = {L12},
        pages = {L12},
          doi = {10.1051/0004-6361/201936950},
archivePrefix = {arXiv},
       eprint = {2001.09776},
 primaryClass = {astro-ph.SR},
       adsurl = {https://ui.adsabs.harvard.edu/abs/2020A&A...634L..12D}
}

@ARTICLE{chatterjee2019,
       author = {{Chatterjee}, K. and {Liska}, M. and {Tchekhovskoy}, A. and {Markoff}, S.~B.},
        title = "{Accelerating AGN jets to parsec scales using general relativistic MHD simulations}",
      journal = {\mnras},
         year = 2019,
        month = dec,
       volume = {490},
       number = {2},
        pages = {2200-2218},
          doi = {10.1093/mnras/stz2626},
archivePrefix = {arXiv},
       eprint = {1904.03243},
 primaryClass = {astro-ph.HE},
       adsurl = {https://ui.adsabs.harvard.edu/abs/2019MNRAS.490.2200C}
}

@ARTICLE{Jacquemin-Ide2021,
       author = {{Jacquemin-Ide}, J. and {Lesur}, G. and {Ferreira}, J.},
        title = "{Magnetic outflows from turbulent accretion disks. I. Vertical structure and secular evolution}",
      journal = {\aap},
         year = 2021,
        month = mar,
       volume = {647},
          eid = {A192},
        pages = {A192},
          doi = {10.1051/0004-6361/202039322},
archivePrefix = {arXiv},
       eprint = {2011.14782},
 primaryClass = {astro-ph.HE},
       adsurl = {https://ui.adsabs.harvard.edu/abs/2021A&A...647A.192J}
}

@ARTICLE{martel2022,
       author = {{Martel}, {\'E}tienne and {Lesur}, Geoffroy},
        title = "{Magnetised winds in transition discs. I. 2.5D global simulations}",
      journal = {\aap},
         year = 2022,
        month = nov,
       volume = {667},
          eid = {A17},
        pages = {A17},
          doi = {10.1051/0004-6361/202142946},
archivePrefix = {arXiv},
       eprint = {2205.02126},
 primaryClass = {astro-ph.EP},
       adsurl = {https://ui.adsabs.harvard.edu/abs/2022A&A...667A..17M}
}

@ARTICLE{RodriguezKamenetzky2017,
       author = {{Rodr{\'\i}guez-Kamenetzky}, Adriana and {Carrasco-Gonz{\'a}lez}, Carlos and {Araudo}, Anabella and {Romero}, Gustavo E. and {Torrelles}, Jos{\'e} M. and {Rodr{\'\i}guez}, Luis F. and {Anglada}, Guillem and {Mart{\'\i}}, Josep and {Perucho}, Manel and {Valotto}, Carlos},
        title = "{The Highly Collimated Radio Jet of HH 80-81: Structure and Nonthermal Emission}",
      journal = {\apj},
         year = 2017,
        month = dec,
       volume = {851},
       number = {1},
          eid = {16},
        pages = {16},
          doi = {10.3847/1538-4357/aa9895},
archivePrefix = {arXiv},
       eprint = {1711.02554},
 primaryClass = {astro-ph.HE},
       adsurl = {https://ui.adsabs.harvard.edu/abs/2017ApJ...851...16R}
}

@ARTICLE{moscadelli2021,
       author = {{Moscadelli}, L. and {Beuther}, H. and {Ahmadi}, A. and {Gieser}, C. and {Massi}, F. and {Cesaroni}, R. and {S{\'a}nchez-Monge}, {\'A}. and {Bacciotti}, F. and {Beltr{\'a}n}, M.~T. and {Csengeri}, T. and {Galv{\'a}n-Madrid}, R. and {Henning}, Th. and {Klaassen}, P.~D. and {Kuiper}, R. and {Leurini}, S. and {Longmore}, S.~N. and {Maud}, L.~T. and {M{\"o}ller}, T. and {Palau}, A. and {Peters}, T. and {Pudritz}, R.~E. and {Sanna}, A. and {Semenov}, D. and {Urquhart}, J.~S. and {Winters}, J.~M. and {Zinnecker}, H.},
        title = "{Multi-scale view of star formation in IRAS 21078+5211: from clump fragmentation to disk wind}",
      journal = {\aap},
         year = 2021,
        month = mar,
       volume = {647},
          eid = {A114},
        pages = {A114},
          doi = {10.1051/0004-6361/202039837},
archivePrefix = {arXiv},
       eprint = {2102.04872},
 primaryClass = {astro-ph.GA},
       adsurl = {https://ui.adsabs.harvard.edu/abs/2021A&A...647A.114M}
}

@ARTICLE{kovalev2020,
       author = {{Kovalev}, Y.~Y. and {Pushkarev}, A.~B. and {Nokhrina}, E.~E. and {Plavin}, A.~V. and {Beskin}, V.~S. and {Chernoglazov}, A.~V. and {Lister}, M.~L. and {Savolainen}, T.},
        title = "{A transition from parabolic to conical shape as a common effect in nearby AGN jets}",
      journal = {\mnras},
         year = 2020,
        month = jul,
       volume = {495},
       number = {4},
        pages = {3576-3591},
          doi = {10.1093/mnras/staa1121},
archivePrefix = {arXiv},
       eprint = {1907.01485},
 primaryClass = {astro-ph.GA},
       adsurl = {https://ui.adsabs.harvard.edu/abs/2020MNRAS.495.3576K}
}

@ARTICLE{Pushkarev2017,
       author = {{Pushkarev}, A.~B. and {Kovalev}, Y.~Y. and {Lister}, M.~L. and {Savolainen}, T.},
        title = "{MOJAVE - XIV. Shapes and opening angles of AGN jets}",
      journal = {\mnras},
         year = 2017,
        month = jul,
       volume = {468},
       number = {4},
        pages = {4992-5003},
          doi = {10.1093/mnras/stx854},
archivePrefix = {arXiv},
       eprint = {1705.02888},
 primaryClass = {astro-ph.HE},
       adsurl = {https://ui.adsabs.harvard.edu/abs/2017MNRAS.468.4992P}
}

@ARTICLE{Ferreira1993b,
       author = {{Ferreira}, J. and {Pelletier}, G.},
        title = "{Magnetized accretion-ejection structures. II. Magnetic channeling around compact objects}",
      journal = {\aap},
         year = 1993,
        month = sep,
       volume = {276},
        pages = {637},
       adsurl = {https://ui.adsabs.harvard.edu/abs/1993A&A...276..637F}
}

@ARTICLE{Ferreira1993a,
       author = {{Ferreira}, J. and {Pelletier}, G.},
        title = "{Magnetized accretion-ejection structures. 1. General statements}",
      journal = {\aap},
         year = 1993,
        month = sep,
       volume = {276},
        pages = {625},
       adsurl = {https://ui.adsabs.harvard.edu/abs/1993A&A...276..625F}
}

@ARTICLE{zapata2015,
       author = {{Zapata}, Luis A. and {Lizano}, Susana and {Rodr{\'\i}guez}, Luis F. and {Ho}, Paul T.~P. and {Loinard}, Laurent and {Fern{\'a}ndez-L{\'o}pez}, Manuel and {Tafoya}, Daniel},
        title = "{Kinematics of the Outflow from the Young Star DG Tau B: Rotation in the Vicinities of an Optical Jet}",
      journal = {\apj},
         year = 2015,
        month = jan,
       volume = {798},
       number = {2},
          eid = {131},
        pages = {131},
          doi = {10.1088/0004-637X/798/2/131},
archivePrefix = {arXiv},
       eprint = {1411.0173},
 primaryClass = {astro-ph.SR},
       adsurl = {https://ui.adsabs.harvard.edu/abs/2015ApJ...798..131Z}
}

@ARTICLE{laing2011,
       author = {{Laing}, R.~A. and {Guidetti}, D. and {Bridle}, A.~H. and {Parma}, P. and {Bondi}, M.},
        title = "{Deep imaging of Fanaroff-Riley Class I radio galaxies with lobes}",
      journal = {\mnras},
         year = 2011,
        month = nov,
       volume = {417},
       number = {4},
        pages = {2789-2808},
          doi = {10.1111/j.1365-2966.2011.19436.x},
archivePrefix = {arXiv},
       eprint = {1107.2511},
 primaryClass = {astro-ph.CO},
       adsurl = {https://ui.adsabs.harvard.edu/abs/2011MNRAS.417.2789L}
}

@ARTICLE{meskini2024,
       author = {{Meskini}, C. and {Sauty}, C. and {Marcowith}, A. and {Vlahakis}, N. and {Brunn}, V.},
        title = "{The role of heating in the formation and the dynamics of YSO jets. I. A parametric study}",
      journal = {\aap},
         year = 2024,
        month = jun,
       volume = {686},
          eid = {A287},
        pages = {A287},
          doi = {10.1051/0004-6361/202449219},
archivePrefix = {arXiv},
       eprint = {2403.10475},
 primaryClass = {astro-ph.HE},
       adsurl = {https://ui.adsabs.harvard.edu/abs/2024A&A...686A.287M}
}

@ARTICLE{burrows1996,
       author = {{Burrows}, Christopher J. and {Stapelfeldt}, Karl R. and {Watson}, Alan M. and {Krist}, John E. and {Ballester}, Gilda E. and {Clarke}, John T. and {Crisp}, David and {Gallagher}, III, John S. and {Griffiths}, Richard E. and {Hester}, J. Jeff and {Hoessel}, John G. and {Holtzman}, Jon A. and {Mould}, Jeremy R. and {Scowen}, Paul A. and {Trauger}, John T. and {Westphal}, James A.},
        title = "{Hubble Space Telescope Observations of the Disk and Jet of HH 30}",
      journal = {\apj},
         year = 1996,
        month = dec,
       volume = {473},
        pages = {437},
          doi = {10.1086/178156},
       adsurl = {https://ui.adsabs.harvard.edu/abs/1996ApJ...473..437B}
}

@ARTICLE{Algaba2017,
       author = {{Algaba}, J.~C. and {Nakamura}, M. and {Asada}, K. and {Lee}, S.~S.},
        title = "{Resolving the Geometry of the Innermost Relativistic Jets in Active Galactic Nuclei}",
      journal = {\apj},
         year = 2017,
        month = jan,
       volume = {834},
       number = {1},
          eid = {65},
        pages = {65},
          doi = {10.3847/1538-4357/834/1/65},
archivePrefix = {arXiv},
       eprint = {1611.04075},
 primaryClass = {astro-ph.HE},
       adsurl = {https://ui.adsabs.harvard.edu/abs/2017ApJ...834...65A}
}

@ARTICLE{giovannini2018,
       author = {{Giovannini}, G. and {Savolainen}, T. and {Orienti}, M. and {Nakamura}, M. and {Nagai}, H. and {Kino}, M. and {Giroletti}, M. and {Hada}, K. and {Bruni}, G. and {Kovalev}, Y.~Y. and {Anderson}, J.~M. and {D'Ammando}, F. and {Hodgson}, J. and {Honma}, M. and {Krichbaum}, T.~P. and {Lee}, S.-S. and {Lico}, R. and {Lisakov}, M.~M. and {Lobanov}, A.~P. and {Petrov}, L. and {Sohn}, B.~W. and {Sokolovsky}, K.~V. and {Voitsik}, P.~A. and {Zensus}, J.~A. and {Tingay}, S.},
        title = "{A wide and collimated radio jet in 3C84 on the scale of a few hundred gravitational radii}",
      journal = {Nature Astronomy},
         year = 2018,
        month = apr,
       volume = {2},
        pages = {472-477},
          doi = {10.1038/s41550-018-0431-2},
archivePrefix = {arXiv},
       eprint = {1804.02198},
 primaryClass = {astro-ph.GA},
       adsurl = {https://ui.adsabs.harvard.edu/abs/2018NatAs...2..472G}
}

@ARTICLE{okino2022,
       author = {{Okino}, Hiroki and {Akiyama}, Kazunori and {Asada}, Keiichi and {G{\'o}mez}, Jos{\'e} L. and {Hada}, Kazuhiro and {Honma}, Mareki and {Krichbaum}, Thomas P. and {Kino}, Motoki and {Nagai}, Hiroshi and {Bach}, Uwe and {Blackburn}, Lindy and {Bouman}, Katherine L. and {Chael}, Andrew and {Crew}, Geoffrey B. and {Doeleman}, Sheperd S. and {Fish}, Vincent L. and {Goddi}, Ciriaco and {Issaoun}, Sara and {Johnson}, Michael D. and {Jorstad}, Svetlana and {Koyama}, Shoko and {Lonsdale}, Colin J. and {Lu}, Ru-Sen and {Mart{\'\i}-Vidal}, Ivan and {Matthews}, Lynn D. and {Mizuno}, Yosuke and {Moriyama}, Kotaro and {Nakamura}, Masanori and {Pu}, Hung-Yi and {Ros}, Eduardo and {Savolainen}, Tuomas and {Tazaki}, Fumie and {Wagner}, Jan and {Wielgus}, Maciek and {Zensus}, Anton},
        title = "{Collimation of the Relativistic Jet in the Quasar 3C 273}",
      journal = {\apj},
         year = 2022,
        month = nov,
       volume = {940},
       number = {1},
          eid = {65},
        pages = {65},
          doi = {10.3847/1538-4357/ac97e5},
archivePrefix = {arXiv},
       eprint = {2112.12233},
 primaryClass = {astro-ph.HE},
       adsurl = {https://ui.adsabs.harvard.edu/abs/2022ApJ...940...65O}
}

@ARTICLE{boccardi2021,
       author = {{Boccardi}, B. and {Perucho}, M. and {Casadio}, C. and {Grandi}, P. and {Macconi}, D. and {Torresi}, E. and {Pellegrini}, S. and {Krichbaum}, T.~P. and {Kadler}, M. and {Giovannini}, G. and {Karamanavis}, V. and {Ricci}, L. and {Madika}, E. and {Bach}, U. and {Ros}, E. and {Giroletti}, M. and {Zensus}, J.~A.},
        title = "{Jet collimation in NGC 315 and other nearby AGN}",
      journal = {\aap},
         year = 2021,
        month = mar,
       volume = {647},
          eid = {A67},
        pages = {A67},
          doi = {10.1051/0004-6361/202039612},
archivePrefix = {arXiv},
       eprint = {2012.14831},
 primaryClass = {astro-ph.HE},
       adsurl = {https://ui.adsabs.harvard.edu/abs/2021A&A...647A..67B}
}

@ARTICLE{tseng2016,
       author = {{Tseng}, Chih-Yin and {Asada}, Keiichi and {Nakamura}, Masanori and {Pu}, Hung-Yi and {Algaba}, Juan-Carlos and {Lo}, Wen-Ping},
        title = "{Structural Transition in the NGC 6251 Jet: an Interplay with the Supermassive Black Hole and Its Host Galaxy}",
      journal = {\apj},
         year = 2016,
        month = dec,
       volume = {833},
       number = {2},
          eid = {288},
        pages = {288},
          doi = {10.3847/1538-4357/833/2/288},
archivePrefix = {arXiv},
       eprint = {1610.06351},
 primaryClass = {astro-ph.HE},
       adsurl = {https://ui.adsabs.harvard.edu/abs/2016ApJ...833..288T}
}

@ARTICLE{mizuno2014,
       author = {{Mizuno}, Yosuke and {Hardee}, Philip E. and {Nishikawa}, Ken-Ichi},
        title = "{Spatial Growth of the Current-driven Instability in Relativistic Jets}",
      journal = {\apj},
         year = 2014,
        month = apr,
       volume = {784},
       number = {2},
          eid = {167},
        pages = {167},
          doi = {10.1088/0004-637X/784/2/167},
archivePrefix = {arXiv},
       eprint = {1402.2370},
 primaryClass = {astro-ph.HE},
       adsurl = {https://ui.adsabs.harvard.edu/abs/2014ApJ...784..167M}
}

@ARTICLE{Baty2006,
       author = {{Baty}, H. and {Keppens}, R.},
        title = "{Kelvin-Helmholtz disruptions in extended magnetized jet flows}",
      journal = {\aap},
         year = 2006,
        month = feb,
       volume = {447},
       number = {1},
        pages = {9-22},
          doi = {10.1051/0004-6361:20053969},
       adsurl = {https://ui.adsabs.harvard.edu/abs/2006A&A...447....9B}
}

@ARTICLE{Baty2002,
       author = {{Baty}, H. and {Keppens}, R.},
        title = "{Interplay between Kelvin-Helmholtz and Current-driven Instabilities in Jets}",
      journal = {\apj},
         year = 2002,
        month = dec,
       volume = {580},
       number = {2},
        pages = {800-814},
          doi = {10.1086/343893},
       adsurl = {https://ui.adsabs.harvard.edu/abs/2002ApJ...580..800B}
}

@ARTICLE{Bodo1994,
       author = {{Bodo}, G. and {Massaglia}, S. and {Ferrari}, A. and {Trussoni}, E.},
        title = "{Kelvin-Helmholtz instability of hydrodynamic supersonic jets.}",
      journal = {\aap},
         year = 1994,
        month = mar,
       volume = {283},
        pages = {655-676},
       adsurl = {https://ui.adsabs.harvard.edu/abs/1994A&A...283..655B}
}

@ARTICLE{Freidberg1982,
       author = {{Freidberg}, J.~P.},
        title = "{Ideal magnetohydrodynamic theory of magnetic fusion systems}",
      journal = {Reviews of Modern Physics},
         year = 1982,
        month = jul,
       volume = {54},
       number = {3},
        pages = {801-902},
          doi = {10.1103/RevModPhys.54.801},
       adsurl = {https://ui.adsabs.harvard.edu/abs/1982RvMP...54..801F}
}

@ARTICLE{Agra-Amboage2014,
       author = {{Agra-Amboage}, V. and {Cabrit}, S. and {Dougados}, C. and {Kristensen}, L.~E. and {Ibgui}, L. and {Reunanen}, J.},
        title = "{Origin of the wide-angle hot H$_{2}$ in DG Tauri. New insight from SINFONI spectro-imaging}",
      journal = {\aap},
         year = 2014,
        month = apr,
       volume = {564},
          eid = {A11},
        pages = {A11},
          doi = {10.1051/0004-6361/201220488},
archivePrefix = {arXiv},
       eprint = {1402.1160},
 primaryClass = {astro-ph.SR},
       adsurl = {https://ui.adsabs.harvard.edu/abs/2014A&A...564A..11A}
}

@ARTICLE{Lee2017a,
       author = {{Lee}, Joyce W.~Y. and {Hull}, Charles L.~H. and {Offner}, Stella S.~R.},
        title = "{Synthetic Observations of Magnetic Fields in Protostellar Cores}",
      journal = {\apj},
         year = 2017,
        month = jan,
       volume = {834},
       number = {2},
          eid = {201},
        pages = {201},
          doi = {10.3847/1538-4357/834/2/201},
archivePrefix = {arXiv},
       eprint = {1611.08530},
 primaryClass = {astro-ph.GA},
       adsurl = {https://ui.adsabs.harvard.edu/abs/2017ApJ...834..201L}
}

@ARTICLE{Davis2011,
       author = {{Davis}, C.~J. and {Cervantes}, B. and {Nisini}, B. and {Giannini}, T. and {Takami}, M. and {Whelan}, E. and {Smith}, M.~D. and {Ray}, T.~P. and {Chrysostomou}, A. and {Pyo}, T.~S.},
        title = "{VLT integral field spectroscopy of embedded protostars: using near-infrared emission lines as tracers of accretion and outflow}",
      journal = {\aap},
         year = 2011,
        month = apr,
       volume = {528},
          eid = {A3},
        pages = {A3},
          doi = {10.1051/0004-6361/201015897},
       adsurl = {https://ui.adsabs.harvard.edu/abs/2011A&A...528A...3D}
}

@ARTICLE{Garofalo2019,
       author = {{Garofalo}, David and {Singh}, Chandra B.},
        title = "{FR0 Radio Galaxies and Their Place in the Radio Morphology Classification}",
      journal = {the Astrophysical Journal},
         year = 2019,
        month = feb,
       volume = {871},
       number = {2},
          eid = {259},
        pages = {259},
          doi = {10.3847/1538-4357/aaf056},
archivePrefix = {arXiv},
       eprint = {1811.05383},
 primaryClass = {astro-ph.HE},
       adsurl = {https://ui.adsabs.harvard.edu/abs/2019ApJ...871..259G}
}

@ARTICLE{Ghisellini2014,
       author = {{Ghisellini}, G. and {Tavecchio}, F. and {Maraschi}, L. and {Celotti}, A. and {Sbarrato}, T.},
        title = "{The power of relativistic jets is larger than the luminosity of their accretion disks}",
      journal = {\nat},
         year = 2014,
        month = nov,
       volume = {515},
       number = {7527},
        pages = {376-378},
          doi = {10.1038/nature13856},
archivePrefix = {arXiv},
       eprint = {1411.5368},
 primaryClass = {astro-ph.HE},
       adsurl = {https://ui.adsabs.harvard.edu/abs/2014Natur.515..376G}
}

@ARTICLE{Blandford2019,
       author = {{Blandford}, Roger and {Meier}, David and {Readhead}, Anthony},
        title = "{Relativistic Jets from Active Galactic Nuclei}",
      journal = {\araa},
         year = 2019,
        month = aug,
       volume = {57},
        pages = {467-509},
          doi = {10.1146/annurev-astro-081817-051948},
archivePrefix = {arXiv},
       eprint = {1812.06025},
 primaryClass = {astro-ph.HE},
       adsurl = {https://ui.adsabs.harvard.edu/abs/2019ARA&A..57..467B}
}

@ARTICLE{ellerbroeck2013,
       author = {{Ellerbroek}, L.~E. and {Podio}, L. and {Kaper}, L. and {Sana}, H. and {Huppenkothen}, D. and {de Koter}, A. and {Monaco}, L.},
        title = "{The outflow history of two Herbig-Haro jets in RCW 36: HH 1042 and HH 1043}",
      journal = {\aap},
         year = 2013,
        month = mar,
       volume = {551},
          eid = {A5},
        pages = {A5},
          doi = {10.1051/0004-6361/201220635},
archivePrefix = {arXiv},
       eprint = {1212.4144},
 primaryClass = {astro-ph.SR},
       adsurl = {https://ui.adsabs.harvard.edu/abs/2013A&A...551A...5E}
}

@article{suydam1958,
  title={Stability of a linear pinch},
  author={Suydam, Bergen R},
  journal={Journal of Nuclear Energy (1954)},
  volume={7},
  number={3-4},
  pages={275--276},
  year={1958},
  publisher={Citeseer}
}

@ARTICLE{marcel2022,
       author = {{Marcel}, G. and {Ferreira}, J. and {Petrucci}, P.-O. and {Barnier}, S. and {Malzac}, J. and {Marino}, A. and {Coriat}, M. and {Clavel}, M. and {Reynolds}, C. and {Neilsen}, J. and {Belmont}, R. and {Corbel}, S.},
        title = "{A unified accretion-ejection paradigm for black hole X-ray binaries. VI. Radiative efficiency and radio-X-ray correlation during four outbursts from GX 339-4}",
      journal = {\aap},
         year = 2022,
        month = mar,
       volume = {659},
          eid = {A194},
        pages = {A194},
          doi = {10.1051/0004-6361/202141375},
archivePrefix = {arXiv},
       eprint = {2109.13592},
 primaryClass = {astro-ph.HE},
       adsurl = {https://ui.adsabs.harvard.edu/abs/2022A&A...659A.194M}
}

@ARTICLE{marcel2019,
       author = {{Marcel}, G. and {Ferreira}, J. and {Clavel}, M. and {Petrucci}, P.-O. and {Malzac}, J. and {Corbel}, S. and {Rodriguez}, J. and {Belmont}, R. and {Coriat}, M. and {Henri}, G. and {Cangemi}, F.},
        title = "{A unified accretion-ejection paradigm for black hole X-ray binaries. IV. Replication of the 2010-2011 activity cycle of GX 339-4}",
      journal = {\aap},
         year = 2019,
        month = jun,
       volume = {626},
          eid = {A115},
        pages = {A115},
          doi = {10.1051/0004-6361/201935060},
archivePrefix = {arXiv},
       eprint = {1905.05057},
 primaryClass = {astro-ph.HE},
       adsurl = {https://ui.adsabs.harvard.edu/abs/2019A&A...626A.115M}
}

@ARTICLE{marcel2018a,
       author = {{Marcel}, G. and {Ferreira}, J. and {Petrucci}, P.-O. and {Belmont}, R. and {Malzac}, J. and {Clavel}, M. and {Henri}, G. and {Coriat}, M. and {Corbel}, S. and {Rodriguez}, J. and {Loh}, A. and {Chakravorty}, S.},
        title = "{A unified accretion-ejection paradigm for black hole X-ray binaries. III. Spectral signatures of hybrid disk configurations}",
      journal = {\aap},
         year = 2018,
        month = sep,
       volume = {617},
          eid = {A46},
        pages = {A46},
          doi = {10.1051/0004-6361/201833124},
archivePrefix = {arXiv},
       eprint = {1805.12407},
 primaryClass = {astro-ph.HE},
       adsurl = {https://ui.adsabs.harvard.edu/abs/2018A&A...617A..46M}
}

@ARTICLE{shakura1973,
       author = {{Shakura}, N.~I. and {Sunyaev}, R.~A.},
        title = "{Black holes in binary systems. Observational appearance.}",
      journal = {\aap},
         year = 1973,
        month = jan,
       volume = {24},
        pages = {337-355},
       adsurl = {https://ui.adsabs.harvard.edu/abs/1973A&A....24..337S}
}

@ARTICLE{Rezgui2025,
       author = {{Rezgui}, Ghassen and {Preiner}, Reinhold},
        title = "{The role of saturated thermal conduction in shaping electric current dynamics in jet-launching disks}",
      journal = {Journal of High Energy Astrophysics},
         year = 2025,
        month = jul,
       volume = {47},
          eid = {100394},
        pages = {100394},
          doi = {10.1016/j.jheap.2025.100394},
       adsurl = {https://ui.adsabs.harvard.edu/abs/2025JHEAp..4700394R}
}

@ARTICLE{Ferreira2006a,
       author = {{Ferreira}, J. and {Petrucci}, P.-O. and {Henri}, G. and {Saug{\'e}}, L. and {Pelletier}, G.},
        title = "{A unified accretion-ejection paradigm for black hole X-ray binaries. I. The dynamical constituents}",
      journal = {\aap},
         year = 2006,
        month = mar,
       volume = {447},
       number = {3},
        pages = {813-825},
          doi = {10.1051/0004-6361:20052689},
archivePrefix = {arXiv},
       eprint = {astro-ph/0511123},
 primaryClass = {astro-ph},
       adsurl = {https://ui.adsabs.harvard.edu/abs/2006A&A...447..813F}
}

@ARTICLE{Lee2020,
       author = {{Lee}, Chin-Fei},
        title = "{Molecular jets from low-mass young protostellar objects}",
      journal = {\aapr},
         year = 2020,
        month = mar,
       volume = {28},
       number = {1},
          eid = {1},
        pages = {1},
          doi = {10.1007/s00159-020-0123-7},
archivePrefix = {arXiv},
       eprint = {2002.05823},
 primaryClass = {astro-ph.GA},
       adsurl = {https://ui.adsabs.harvard.edu/abs/2020A&ARv..28....1L}
}

@ARTICLE{Roe1986,
       author = {{Roe}, P.~L.},
        title = "{Characteristic-based schemes for the euler equations}",
      journal = {Annual Review of Fluid Mechanics},
         year = 1986,
        month = jan,
       volume = {18},
        pages = {337-365},
          doi = {10.1146/annurev.fl.18.010186.002005},
       adsurl = {https://ui.adsabs.harvard.edu/abs/1986AnRFM..18..337R}
}

@ARTICLE{VanLeer1974,
       author = {{van Leer}, Bram},
        title = "{Towards the Ultimate Conservation Difference Scheme. II. Monotonicity and Conservation Combined in a Second-Order Scheme}",
      journal = {Journal of Computational Physics},
         year = 1974,
        month = mar,
       volume = {14},
       number = {4},
        pages = {361-370},
          doi = {10.1016/0021-9991(74)90019-9},
       adsurl = {https://ui.adsabs.harvard.edu/abs/1974JCoPh..14..361V}
}

@ARTICLE{VanLeer1977,
       author = {{van Leer}, Bram},
        title = "{Towards the Ultimate Conservative Difference Scheme. IV. A New Approach to Numerical Convection}",
      journal = {Journal of Computational Physics},
         year = 1977,
        month = mar,
       volume = {23},
        pages = {276},
          doi = {10.1016/0021-9991(77)90095-X},
       adsurl = {https://ui.adsabs.harvard.edu/abs/1977JCoPh..23..276V}
}

@ARTICLE{harten1983,
       author = {{Harten}, Ami},
        title = "{High Resolution Schemes for Hyperbolic Conservation Laws}",
      journal = {Journal of Computational Physics},
         year = 1983,
        month = mar,
       volume = {49},
       number = {3},
        pages = {357-393},
          doi = {10.1016/0021-9991(83)90136-5},
       adsurl = {https://ui.adsabs.harvard.edu/abs/1983JCoPh..49..357H}
}

@ARTICLE{jannaud2023,
       author = {{Jannaud}, T. and {Zanni}, C. and {Ferreira}, J.},
        title = "{Numerical simulations of MHD jets from Keplerian accretion disks. I. Recollimation shocks}",
      journal = {\aap},
         year = 2023,
        month = jan,
       volume = {669},
          eid = {A159},
        pages = {A159},
          doi = {10.1051/0004-6361/202244311},
archivePrefix = {arXiv},
       eprint = {2210.14809},
 primaryClass = {astro-ph.HE},
       adsurl = {https://ui.adsabs.harvard.edu/abs/2023A&A...669A.159J}
}

@ARTICLE{anderson2005,
       author = {{Anderson}, Jeffrey M. and {Li}, Zhi-Yun and {Krasnopolsky}, Ruben and {Blandford}, Roger D.},
        title = "{The Structure of Magnetocentrifugal Winds. I. Steady Mass Loading}",
      journal = {\apj},
         year = 2005,
        month = sep,
       volume = {630},
       number = {2},
        pages = {945-957},
          doi = {10.1086/432040},
archivePrefix = {arXiv},
       eprint = {astro-ph/0410704},
 primaryClass = {astro-ph},
       adsurl = {https://ui.adsabs.harvard.edu/abs/2005ApJ...630..945A}
}

@ARTICLE{Heyvaerts1989,
       author = {{Heyvaerts}, Jean and {Norman}, Colin},
        title = "{The Collimation of Magnetized Winds}",
      journal = {\apj},
         year = 1989,
        month = dec,
       volume = {347},
        pages = {1055},
          doi = {10.1086/168195},
       adsurl = {https://ui.adsabs.harvard.edu/abs/1989ApJ...347.1055H}
}

@ARTICLE{polko2010,
       author = {{Polko}, Peter and {Meier}, David L. and {Markoff}, Sera},
        title = "{Determining the Optimal Locations for Shock Acceleration in Magnetohydrodynamical Jets}",
      journal = {\apj},
         year = 2010,
        month = nov,
       volume = {723},
       number = {2},
        pages = {1343-1350},
          doi = {10.1088/0004-637X/723/2/1343},
archivePrefix = {arXiv},
       eprint = {1009.3031},
 primaryClass = {astro-ph.HE},
       adsurl = {https://ui.adsabs.harvard.edu/abs/2010ApJ...723.1343P}
}

@ARTICLE{blandford1977,
       author = {{Blandford}, R.~D. and {Znajek}, R.~L.},
        title = "{Electromagnetic extraction of energy from Kerr black holes.}",
      journal = {\mnras},
         year = 1977,
        month = may,
       volume = {179},
        pages = {433-456},
          doi = {10.1093/mnras/179.3.433},
       adsurl = {https://ui.adsabs.harvard.edu/abs/1977MNRAS.179..433B}
}

@ARTICLE{barniolduran2017,
       author = {{Barniol Duran}, Rodolfo and {Tchekhovskoy}, Alexander and {Giannios}, Dimitrios},
        title = "{Simulations of AGN jets: magnetic kink instability versus conical shocks}",
      journal = {\mnras},
         year = 2017,
        month = aug,
       volume = {469},
       number = {4},
        pages = {4957-4978},
          doi = {10.1093/mnras/stx1165},
archivePrefix = {arXiv},
       eprint = {1612.06929},
 primaryClass = {astro-ph.HE},
       adsurl = {https://ui.adsabs.harvard.edu/abs/2017MNRAS.469.4957B}
}

@ARTICLE{Fendt2006,
       author = {{Fendt}, Christian},
        title = "{Collimation of Astrophysical Jets: The Role of the Accretion Disk Magnetic Field Distribution}",
      journal = {\apj},
         year = 2006,
        month = nov,
       volume = {651},
       number = {1},
        pages = {272-287},
          doi = {10.1086/507976},
archivePrefix = {arXiv},
       eprint = {astro-ph/0511611},
 primaryClass = {astro-ph},
       adsurl = {https://ui.adsabs.harvard.edu/abs/2006ApJ...651..272F}
}

@ARTICLE{pudritz2006,
       author = {{Pudritz}, Ralph E. and {Rogers}, Conrad S. and {Ouyed}, Rachid},
        title = "{Controlling the collimation and rotation of hydromagnetic disc winds}",
      journal = {\mnras},
         year = 2006,
        month = feb,
       volume = {365},
       number = {4},
        pages = {1131-1148},
          doi = {10.1111/j.1365-2966.2005.09766.x},
archivePrefix = {arXiv},
       eprint = {astro-ph/0508295},
 primaryClass = {astro-ph},
       adsurl = {https://ui.adsabs.harvard.edu/abs/2006MNRAS.365.1131P}
}

@ARTICLE{marcel2018,
       author = {{Marcel}, G. and {Ferreira}, J. and {Petrucci}, P.-O. and {Henri}, G. and {Belmont}, R. and {Clavel}, M. and {Malzac}, J. and {Coriat}, M. and {Corbel}, S. and {Rodriguez}, J. and {Loh}, A. and {Chakravorty}, S. and {Drappeau}, S.},
        title = "{A unified accretion-ejection paradigm for black hole X-ray binaries. II. Observational signatures of jet-emitting disks}",
      journal = {\aap},
         year = 2018,
        month = jul,
       volume = {615},
          eid = {A57},
        pages = {A57},
          doi = {10.1051/0004-6361/201732069},
archivePrefix = {arXiv},
       eprint = {1803.04335},
 primaryClass = {astro-ph.HE},
       adsurl = {https://ui.adsabs.harvard.edu/abs/2018A&A...615A..57M}
}

@ARTICLE{tabone2020,
       author = {{Tabone}, B. and {Cabrit}, S. and {Pineau des For{\^e}ts}, G. and {Ferreira}, J. and {Gusdorf}, A. and {Podio}, L. and {Bianchi}, E. and {Chapillon}, E. and {Codella}, C. and {Gueth}, F.},
        title = "{Constraining MHD disk winds with ALMA. Apparent rotation signatures and application to HH212}",
      journal = {\aap},
         year = 2020,
        month = aug,
       volume = {640},
          eid = {A82},
        pages = {A82},
          doi = {10.1051/0004-6361/201834377},
archivePrefix = {arXiv},
       eprint = {2004.08804},
 primaryClass = {astro-ph.SR},
       adsurl = {https://ui.adsabs.harvard.edu/abs/2020A&A...640A..82T}
}

@ARTICLE{stute2008,
       author = {{Stute}, M. and {Tsinganos}, K. and {Vlahakis}, N. and {Matsakos}, T. and {Gracia}, J.},
        title = "{Stability and structure of analytical MHD jet formation models with a finite outer disk radius}",
      journal = {\aap},
         year = 2008,
        month = nov,
       volume = {491},
       number = {2},
        pages = {339-351},
          doi = {10.1051/0004-6361:200810499},
archivePrefix = {arXiv},
       eprint = {0809.1652},
 primaryClass = {astro-ph},
       adsurl = {https://ui.adsabs.harvard.edu/abs/2008A&A...491..339S}
}

@ARTICLE{walker2018,
       author = {{Walker}, R. Craig and {Hardee}, Philip E. and {Davies}, Frederick B. and {Ly}, Chun and {Junor}, William},
        title = "{The Structure and Dynamics of the Subparsec Jet in M87 Based on 50 VLBA Observations over 17 Years at 43 GHz}",
      journal = {\apj},
         year = 2018,
        month = mar,
       volume = {855},
       number = {2},
          eid = {128},
        pages = {128},
          doi = {10.3847/1538-4357/aaafcc},
archivePrefix = {arXiv},
       eprint = {1802.06166},
 primaryClass = {astro-ph.HE},
       adsurl = {https://ui.adsabs.harvard.edu/abs/2018ApJ...855..128W}
}

@ARTICLE{komissarov2009,
       author = {{Komissarov}, Serguei S. and {Vlahakis}, Nektarios and {K{\"o}nigl}, Arieh and {Barkov}, Maxim V.},
        title = "{Magnetic acceleration of ultrarelativistic jets in gamma-ray burst sources}",
      journal = {\mnras},
         year = 2009,
        month = apr,
       volume = {394},
       number = {3},
        pages = {1182-1212},
          doi = {10.1111/j.1365-2966.2009.14410.x},
archivePrefix = {arXiv},
       eprint = {0811.1467},
 primaryClass = {astro-ph},
       adsurl = {https://ui.adsabs.harvard.edu/abs/2009MNRAS.394.1182K}
}

@ARTICLE{komissarov2007,
       author = {{Komissarov}, Serguei S. and {Barkov}, Maxim V. and {Vlahakis}, Nektarios and {K{\"o}nigl}, Arieh},
        title = "{Magnetic acceleration of relativistic active galactic nucleus jets}",
      journal = {\mnras},
         year = 2007,
        month = sep,
       volume = {380},
       number = {1},
        pages = {51-70},
          doi = {10.1111/j.1365-2966.2007.12050.x},
archivePrefix = {arXiv},
       eprint = {astro-ph/0703146},
 primaryClass = {astro-ph},
       adsurl = {https://ui.adsabs.harvard.edu/abs/2007MNRAS.380...51K}
}

@ARTICLE{ferreira2004,
       author = {{Ferreira}, Jonathan and {Casse}, Fabien},
        title = "{Stationary Accretion Disks Launching Super-fast-magnetosonic Magnetohydrodynamic Jets}",
      journal = {\apjl},
         year = 2004,
        month = feb,
       volume = {601},
       number = {2},
        pages = {L139-L142},
          doi = {10.1086/381804},
archivePrefix = {arXiv},
       eprint = {astro-ph/0312157},
 primaryClass = {astro-ph},
       adsurl = {https://ui.adsabs.harvard.edu/abs/2004ApJ...601L.139F}
}

@ARTICLE{miyoshi2005,
       author = {{Miyoshi}, Takahiro and {Kusano}, Kanya},
        title = "{A multi-state HLL approximate Riemann solver for ideal magnetohydrodynamics}",
      journal = {Journal of Computational Physics},
         year = 2005,
        month = sep,
       volume = {208},
       number = {1},
        pages = {315-344},
          doi = {10.1016/j.jcp.2005.02.017},
       adsurl = {https://ui.adsabs.harvard.edu/abs/2005JCoPh.208..315M}
}

@INPROCEEDINGS{cabrit2007,
       author = {{Cabrit}, S.},
        title = "{The accretion-ejection connexion in T Tauri stars: jet models vs. observations}",
    booktitle = {Star-Disk Interaction in Young Stars},
         year = 2007,
       editor = {{Bouvier}, Jerome and {Appenzeller}, Immo},
       series = {IAU Symposium},
       volume = {243},
        month = may,
        pages = {203-214},
          doi = {10.1017/S1743921307009568},
       adsurl = {https://ui.adsabs.harvard.edu/abs/2007IAUS..243..203C}
}

@ARTICLE{corbel2003,
       author = {{Corbel}, S. and {Nowak}, M.~A. and {Fender}, R.~P. and {Tzioumis}, A.~K. and {Markoff}, S.},
        title = "{Radio/X-ray correlation in the low/hard state of GX 339-4}",
      journal = {\aap},
         year = 2003,
        month = mar,
       volume = {400},
        pages = {1007-1012},
          doi = {10.1051/0004-6361:20030090},
archivePrefix = {arXiv},
       eprint = {astro-ph/0301436},
 primaryClass = {astro-ph},
       adsurl = {https://ui.adsabs.harvard.edu/abs/2003A&A...400.1007C}
}

@ARTICLE{contopoulos1994,
       author = {{Contopoulos}, J. and {Lovelace}, R.~V.~E.},
        title = "{Magnetically Driven Jets and Winds: Exact Solutions}",
      journal = {\apj},
         year = 1994,
        month = jul,
       volume = {429},
        pages = {139},
          doi = {10.1086/174307},
       adsurl = {https://ui.adsabs.harvard.edu/abs/1994ApJ...429..139C}
}

@ARTICLE{matsakos2008,
       author = {{Matsakos}, T. and {Tsinganos}, K. and {Vlahakis}, N. and {Massaglia}, S. and {Mignone}, A. and {Trussoni}, E.},
        title = "{Two-component jet simulations. I. Topological stability of analytical MHD outflow solutions}",
      journal = {\aap},
         year = 2008,
        month = jan,
       volume = {477},
       number = {2},
        pages = {521-533},
          doi = {10.1051/0004-6361:20077907},
archivePrefix = {arXiv},
       eprint = {0710.3406},
 primaryClass = {astro-ph},
       adsurl = {https://ui.adsabs.harvard.edu/abs/2008A&A...477..521M}
}

@ARTICLE{Jacquemin-Ide2019,
       author = {{Jacquemin-Ide}, J. and {Ferreira}, J. and {Lesur}, G.},
        title = "{Magnetically driven jets and winds from weakly magnetized accretion discs}",
      journal = {\mnras},
         year = 2019,
        month = dec,
       volume = {490},
       number = {3},
        pages = {3112-3133},
          doi = {10.1093/mnras/stz2749},
archivePrefix = {arXiv},
       eprint = {1909.12258},
 primaryClass = {astro-ph.HE},
       adsurl = {https://ui.adsabs.harvard.edu/abs/2019MNRAS.490.3112J}
}

@ARTICLE{Boccardi2017,
       author = {{Boccardi}, B. and {Krichbaum}, T.~P. and {Ros}, E. and {Zensus}, J.~A.},
        title = "{Radio observations of active galactic nuclei with mm-VLBI}",
      journal = {\aapr},
         year = 2017,
        month = nov,
       volume = {25},
       number = {1},
          eid = {4},
        pages = {4},
          doi = {10.1007/s00159-017-0105-6},
archivePrefix = {arXiv},
       eprint = {1711.07548},
 primaryClass = {astro-ph.HE},
       adsurl = {https://ui.adsabs.harvard.edu/abs/2017A&ARv..25....4B}
}

@ARTICLE{Boccardi2016,
       author = {{Boccardi}, B. and {Krichbaum}, T.~P. and {Bach}, U. and {Mertens}, F. and {Ros}, E. and {Alef}, W. and {Zensus}, J.~A.},
        title = "{The stratified two-sided jet of <ASTROBJ>Cygnus A</ASTROBJ>. Acceleration and collimation}",
      journal = {\aap},
         year = 2016,
        month = jan,
       volume = {585},
          eid = {A33},
        pages = {A33},
          doi = {10.1051/0004-6361/201526985},
archivePrefix = {arXiv},
       eprint = {1509.06250},
 primaryClass = {astro-ph.HE},
       adsurl = {https://ui.adsabs.harvard.edu/abs/2016A&A...585A..33B}
}

@ARTICLE{Ferreira1997,
       author = {{Ferreira}, J.},
        title = "{Magnetically-driven jets from Keplerian accretion discs.}",
      journal = {\aap},
         year = 1997,
        month = mar,
       volume = {319},
        pages = {340-359},
          doi = {10.48550/arXiv.astro-ph/9607057},
archivePrefix = {arXiv},
       eprint = {astro-ph/9607057},
 primaryClass = {astro-ph},
       adsurl = {https://ui.adsabs.harvard.edu/abs/1997A&A...319..340F}
}

@INPROCEEDINGS{Frank2014,
       author = {{Frank}, A. and {Ray}, T.~P. and {Cabrit}, S. and {Hartigan}, P. and {Arce}, H.~G. and {Bacciotti}, F. and {Bally}, J. and {Benisty}, M. and {Eisl{\"o}ffel}, J. and {G{\"u}del}, M. and {Lebedev}, S. and {Nisini}, B. and {Raga}, A.},
        title = "{Jets and Outflows from Star to Cloud: Observations Confront Theory}",
    booktitle = {Protostars and Planets VI},
         year = 2014,
       editor = {{Beuther}, Henrik and {Klessen}, Ralf S. and {Dullemond}, Cornelis P. and {Henning}, Thomas},
        month = jan,
        pages = {451-474},
          doi = {10.2458/azu_uapress_9780816531240-ch020},
archivePrefix = {arXiv},
       eprint = {1402.3553},
 primaryClass = {astro-ph.SR},
       adsurl = {https://ui.adsabs.harvard.edu/abs/2014prpl.conf..451F}
}

@ARTICLE{Fichetdeclairfontaine2021,
       author = {{Fichet de Clairfontaine}, G. and {Meliani}, Z. and {Zech}, A. and {Hervet}, O.},
        title = "{Flux variability from ejecta in structured relativistic jets with large-scale magnetic fields}",
      journal = {\aap},
         year = 2021,
        month = mar,
       volume = {647},
          eid = {A77},
        pages = {A77},
          doi = {10.1051/0004-6361/202039654},
archivePrefix = {arXiv},
       eprint = {2101.06962},
 primaryClass = {astro-ph.HE},
       adsurl = {https://ui.adsabs.harvard.edu/abs/2021A&A...647A..77F}
}

@ARTICLE{casse2004,
       author = {{Casse}, Fabien and {Keppens}, Rony},
        title = "{Radiatively Inefficient Magnetohydrodynamic Accretion-Ejection Structures}",
      journal = {\apj},
         year = 2004,
        month = jan,
       volume = {601},
       number = {1},
        pages = {90-103},
          doi = {10.1086/380441},
archivePrefix = {arXiv},
       eprint = {astro-ph/0310322},
 primaryClass = {astro-ph},
       adsurl = {https://ui.adsabs.harvard.edu/abs/2004ApJ...601...90C}
}

@ARTICLE{casse2002,
       author = {{Casse}, Fabien and {Keppens}, Rony},
        title = "{Magnetized Accretion-Ejection Structures: 2.5-dimensional Magnetohydrodynamic Simulations of Continuous Ideal Jet Launching from Resistive Accretion Disks}",
      journal = {\apj},
         year = 2002,
        month = dec,
       volume = {581},
       number = {2},
        pages = {988-1001},
          doi = {10.1086/344340},
archivePrefix = {arXiv},
       eprint = {astro-ph/0208459},
 primaryClass = {astro-ph},
       adsurl = {https://ui.adsabs.harvard.edu/abs/2002ApJ...581..988C}
}

@ARTICLE{barnier2022,
       author = {{Barnier}, S. and {Petrucci}, P.-O. and {Ferreira}, J. and {Marcel}, G. and {Belmont}, R. and {Clavel}, M. and {Corbel}, S. and {Coriat}, M. and {Espinasse}, M. and {Henri}, G. and {Malzac}, J. and {Rodriguez}, J.},
        title = "{Clues on jet behavior from simultaneous radio-X-ray fits of GX 339-4}",
      journal = {\aap},
         year = 2022,
        month = jan,
       volume = {657},
          eid = {A11},
        pages = {A11},
          doi = {10.1051/0004-6361/202141182},
archivePrefix = {arXiv},
       eprint = {2109.02895},
 primaryClass = {astro-ph.HE},
       adsurl = {https://ui.adsabs.harvard.edu/abs/2022A&A...657A..11B}
}

@ARTICLE{gracia2006,
       author = {{Gracia}, J. and {Vlahakis}, N. and {Tsinganos}, K.},
        title = "{Jet simulations extending radially self-similar magnetohydrodynamics models}",
      journal = {\mnras},
         year = 2006,
        month = mar,
       volume = {367},
       number = {1},
        pages = {201-210},
          doi = {10.1111/j.1365-2966.2005.09945.x},
archivePrefix = {arXiv},
       eprint = {astro-ph/0507634},
 primaryClass = {astro-ph},
       adsurl = {https://ui.adsabs.harvard.edu/abs/2006MNRAS.367..201G}
}

@ARTICLE{murphy2010,
       author = {{Murphy}, G.~C. and {Ferreira}, J. and {Zanni}, C.},
        title = "{Large scale magnetic fields in viscous resistive accretion disks. I. Ejection from weakly magnetized disks}",
      journal = {\aap},
         year = 2010,
        month = mar,
       volume = {512},
          eid = {A82},
        pages = {A82},
          doi = {10.1051/0004-6361/200912633},
archivePrefix = {arXiv},
       eprint = {1003.4471},
 primaryClass = {astro-ph.SR},
       adsurl = {https://ui.adsabs.harvard.edu/abs/2010A&A...512A..82M}
}

@ARTICLE{stepanovs2016,
       author = {{Stepanovs}, Deniss and {Fendt}, Christian},
        title = "{An Extensive Numerical Survey of the Correlation Between Outflow Dynamics and Accretion Disk Magnetization}",
      journal = {\apj},
         year = 2016,
        month = jul,
       volume = {825},
       number = {1},
          eid = {14},
        pages = {14},
          doi = {10.3847/0004-637X/825/1/14},
archivePrefix = {arXiv},
       eprint = {1604.07313},
 primaryClass = {astro-ph.GA},
       adsurl = {https://ui.adsabs.harvard.edu/abs/2016ApJ...825...14S}
}

@ARTICLE{Sheikhnezami2012,
       author = {{Sheikhnezami}, Somayeh and {Fendt}, Christian and {Porth}, Oliver and {Vaidya}, Bhargav and {Ghanbari}, Jamshid},
        title = "{Bipolar Jets Launched from Magnetically Diffusive Accretion Disks. I. Ejection Efficiency versus Field Strength and Diffusivity}",
      journal = {\apj},
         year = 2012,
        month = sep,
       volume = {757},
       number = {1},
          eid = {65},
        pages = {65},
          doi = {10.1088/0004-637X/757/1/65},
archivePrefix = {arXiv},
       eprint = {1207.6086},
 primaryClass = {astro-ph.HE},
       adsurl = {https://ui.adsabs.harvard.edu/abs/2012ApJ...757...65S}
}

@ARTICLE{zanni2007,
       author = {{Zanni}, C. and {Ferrari}, A. and {Rosner}, R. and {Bodo}, G. and {Massaglia}, S.},
        title = "{MHD simulations of jet acceleration from Keplerian accretion disks. The effects of disk resistivity}",
      journal = {\aap},
         year = 2007,
        month = jul,
       volume = {469},
       number = {3},
        pages = {811-828},
          doi = {10.1051/0004-6361:20066400},
archivePrefix = {arXiv},
       eprint = {astro-ph/0703064},
 primaryClass = {astro-ph},
       adsurl = {https://ui.adsabs.harvard.edu/abs/2007A&A...469..811Z}
}

@ARTICLE{casse2000a,
       author = {{Casse}, Fabien and {Ferreira}, Jonathan},
        title = "{Magnetized accretion-ejection structures. IV. Magnetically-driven jets from resistive, viscous, Keplerian discs}",
      journal = {\aap},
         year = 2000,
        month = jan,
       volume = {353},
        pages = {1115-1128},
          doi = {10.48550/arXiv.astro-ph/9911471},
archivePrefix = {arXiv},
       eprint = {astro-ph/9911471},
 primaryClass = {astro-ph},
       adsurl = {https://ui.adsabs.harvard.edu/abs/2000A&A...353.1115C}
}

@ARTICLE{louvet2018,
       author = {{Louvet}, F. and {Dougados}, C. and {Cabrit}, S. and {Mardones}, D. and {M{\'e}nard}, F. and {Tabone}, B. and {Pinte}, C. and {Dent}, W.~R.~F.},
        title = "{The HH30 edge-on T Tauri star. A rotating and precessing monopolar outflow scrutinized by ALMA}",
      journal = {\aap},
         year = 2018,
        month = oct,
       volume = {618},
          eid = {A120},
        pages = {A120},
          doi = {10.1051/0004-6361/201731733},
archivePrefix = {arXiv},
       eprint = {1808.03285},
 primaryClass = {astro-ph.GA},
       adsurl = {https://ui.adsabs.harvard.edu/abs/2018A&A...618A.120L}
}

@ARTICLE{Ferreira1995,
       author = {{Ferreira}, J. and {Pelletier}, G.},
        title = "{Magnetized accretion-ejection structures. III. Stellar and extragalactic jets as weakly dissipative disk outflows.}",
      journal = {\aap},
         year = 1995,
        month = mar,
       volume = {295},
        pages = {807},
       adsurl = {https://ui.adsabs.harvard.edu/abs/1995A&A...295..807F}
}

@ARTICLE{Tchekhovskoy2016,
       author = {{Tchekhovskoy}, Alexander and {Bromberg}, Omer},
        title = "{Three-dimensional relativistic MHD simulations of active galactic nuclei jets: magnetic kink instability and Fanaroff-Riley dichotomy}",
      journal = {\mnras},
         year = 2016,
        month = sep,
       volume = {461},
       number = {1},
        pages = {L46-L50},
          doi = {10.1093/mnrasl/slw064},
archivePrefix = {arXiv},
       eprint = {1512.04526},
 primaryClass = {astro-ph.HE},
       adsurl = {https://ui.adsabs.harvard.edu/abs/2016MNRAS.461L..46T}
}

@ARTICLE{porth2010,
       author = {{Porth}, Oliver and {Fendt}, Christian},
        title = "{Acceleration and Collimation of Relativistic Magnetohydrodynamic Disk Winds}",
      journal = {\apj},
         year = 2010,
        month = feb,
       volume = {709},
       number = {2},
        pages = {1100-1118},
          doi = {10.1088/0004-637X/709/2/1100},
archivePrefix = {arXiv},
       eprint = {0911.3001},
 primaryClass = {astro-ph.HE},
       adsurl = {https://ui.adsabs.harvard.edu/abs/2010ApJ...709.1100P}
}

@ARTICLE{ustyugova1999,
       author = {{Ustyugova}, G.~V. and {Koldoba}, A.~V. and {Romanova}, M.~M. and {Chechetkin}, V.~M. and {Lovelace}, R.~V.~E.},
        title = "{Magnetocentrifugally Driven Winds: Comparison of MHD Simulations with Theory}",
      journal = {\apj},
         year = 1999,
        month = may,
       volume = {516},
       number = {1},
        pages = {221-235},
          doi = {10.1086/307093},
archivePrefix = {arXiv},
       eprint = {astro-ph/9812284},
 primaryClass = {astro-ph},
       adsurl = {https://ui.adsabs.harvard.edu/abs/1999ApJ...516..221U}
}

@ARTICLE{ustyugova1995,
       author = {{Ustyugova}, G.~V. and {Koldoba}, A.~V. and {Romanova}, M.~M. and {Chechetkin}, V.~M. and {Lovelace}, R.~V.~E.},
        title = "{Magnetohydrodynamic Simulations of Outflows from Accretion Disks}",
      journal = {\apjl},
         year = 1995,
        month = feb,
       volume = {439},
        pages = {L39},
          doi = {10.1086/187739},
       adsurl = {https://ui.adsabs.harvard.edu/abs/1995ApJ...439L..39U}
}

@ARTICLE{okamoto2003,
       author = {{Okamoto}, I.},
        title = "{Global Asymptotic Solutions for Magnetohydrodynamic Jets and Winds}",
      journal = {\apj},
         year = 2003,
        month = may,
       volume = {589},
       number = {1},
        pages = {671-676},
          doi = {10.1086/374356},
       adsurl = {https://ui.adsabs.harvard.edu/abs/2003ApJ...589..671O}
}

@ARTICLE{okamoto2001,
       author = {{Okamoto}, Isao},
        title = "{Magnetized centrifugal winds}",
      journal = {\mnras},
         year = 2001,
        month = oct,
       volume = {327},
       number = {1},
        pages = {55-68},
          doi = {10.1046/j.1365-8711.2001.04598.x},
       adsurl = {https://ui.adsabs.harvard.edu/abs/2001MNRAS.327...55O}
}

@ARTICLE{Ceccobello2018,
       author = {{Ceccobello}, C. and {Cavecchi}, Y. and {Heemskerk}, M.~H.~M. and {Markoff}, S. and {Polko}, P. and {Meier}, D.},
        title = "{A new method for extending solutions to the self-similar relativistic magnetohydrodynamic equations for black hole outflows}",
      journal = {\mnras},
         year = 2018,
        month = feb,
       volume = {473},
       number = {4},
        pages = {4417-4435},
          doi = {10.1093/mnras/stx2567},
archivePrefix = {arXiv},
       eprint = {1710.01070},
 primaryClass = {astro-ph.HE},
       adsurl = {https://ui.adsabs.harvard.edu/abs/2018MNRAS.473.4417C}
}

@ARTICLE{ostriker1997,
       author = {{Ostriker}, Eve C.},
        title = "{Self-similar Magnetocentrifugal Disk Winds with Cylindrical Asymptotics}",
      journal = {\apj},
         year = 1997,
        month = sep,
       volume = {486},
       number = {1},
        pages = {291-306},
          doi = {10.1086/304513},
archivePrefix = {arXiv},
       eprint = {astro-ph/9705226},
 primaryClass = {astro-ph},
       adsurl = {https://ui.adsabs.harvard.edu/abs/1997ApJ...486..291O}
}

@ARTICLE{mingo2019,
       author = {{Mingo}, B. and {Croston}, J.~H. and {Hardcastle}, M.~J. and {Best}, P.~N. and {Duncan}, K.~J. and {Morganti}, R. and {Rottgering}, H.~J.~A. and {Sabater}, J. and {Shimwell}, T.~W. and {Williams}, W.~L. and {Brienza}, M. and {Gurkan}, G. and {Mahatma}, V.~H. and {Morabito}, L.~K. and {Prandoni}, I. and {Bondi}, M. and {Ineson}, J. and {Mooney}, S.},
        title = "{Revisiting the Fanaroff-Riley dichotomy and radio-galaxy morphology with the LOFAR Two-Metre Sky Survey (LoTSS)}",
      journal = {\mnras},
         year = 2019,
        month = sep,
       volume = {488},
       number = {2},
        pages = {2701-2721},
          doi = {10.1093/mnras/stz1901},
archivePrefix = {arXiv},
       eprint = {1907.03726},
 primaryClass = {astro-ph.GA},
       adsurl = {https://ui.adsabs.harvard.edu/abs/2019MNRAS.488.2701M}
}

@ARTICLE{Park2019,
       author = {{Park}, Jongho and {Hada}, Kazuhiro and {Kino}, Motoki and {Nakamura}, Masanori and {Hodgson}, Jeffrey and {Ro}, Hyunwook and {Cui}, Yuzhu and {Asada}, Keiichi and {Algaba}, Juan-Carlos and {Sawada-Satoh}, Satoko and {Lee}, Sang-Sung and {Cho}, Ilje and {Shen}, Zhiqiang and {Jiang}, Wu and {Trippe}, Sascha and {Niinuma}, Kotaro and {Sohn}, Bong Won and {Jung}, Taehyun and {Zhao}, Guang-Yao and {Wajima}, Kiyoaki and {Tazaki}, Fumie and {Honma}, Mareki and {An}, Tao and {Akiyama}, Kazunori and {Byun}, Do-Young and {Kim}, Jongsoo and {Zhang}, Yingkang and {Cheng}, Xiaopeng and {Kobayashi}, Hideyuki and {Shibata}, Katsunori M. and {Lee}, Jee Won and {Roh}, Duk-Gyoo and {Oh}, Se-Jin and {Yeom}, Jae-Hwan and {Jung}, Dong-Kyu and {Oh}, Chungsik and {Kim}, Hyo-Ryoung and {Hwang}, Ju-Yeon and {Hagiwara}, Yoshiaki},
        title = "{Kinematics of the M87 Jet in the Collimation Zone: Gradual Acceleration and Velocity Stratification}",
      journal = {\apj},
         year = 2019,
        month = dec,
       volume = {887},
       number = {2},
          eid = {147},
        pages = {147},
          doi = {10.3847/1538-4357/ab5584},
archivePrefix = {arXiv},
       eprint = {1911.02279},
 primaryClass = {astro-ph.HE},
       adsurl = {https://ui.adsabs.harvard.edu/abs/2019ApJ...887..147P}
}

@ARTICLE{doi2018,
       author = {{Doi}, Akihiro and {Hada}, Kazuhiro and {Kino}, Motoki and {Wajima}, Kiyoaki and {Nakahara}, Satomi},
        title = "{A Recollimation Shock in a Stationary Jet Feature with Limb-brightening in the Gamma-Ray-emitting Narrow-line Seyfert 1 Galaxy 1H 0323+342}",
      journal = {\apjl},
         year = 2018,
        month = apr,
       volume = {857},
       number = {1},
          eid = {L6},
        pages = {L6},
          doi = {10.3847/2041-8213/aabae2},
archivePrefix = {arXiv},
       eprint = {1804.03776},
 primaryClass = {astro-ph.GA},
       adsurl = {https://ui.adsabs.harvard.edu/abs/2018ApJ...857L...6D}
}

@ARTICLE{lister2013,
       author = {{Lister}, M.~L. and {Aller}, M.~F. and {Aller}, H.~D. and {Homan}, D.~C. and {Kellermann}, K.~I. and {Kovalev}, Y.~Y. and {Pushkarev}, A.~B. and {Richards}, J.~L. and {Ros}, E. and {Savolainen}, T.},
        title = "{MOJAVE. X. Parsec-scale Jet Orientation Variations and Superluminal Motion in Active Galactic Nuclei}",
      journal = {\aj},
         year = 2013,
        month = nov,
       volume = {146},
       number = {5},
          eid = {120},
        pages = {120},
          doi = {10.1088/0004-6256/146/5/120},
archivePrefix = {arXiv},
       eprint = {1308.2713},
 primaryClass = {astro-ph.CO},
       adsurl = {https://ui.adsabs.harvard.edu/abs/2013AJ....146..120L}
}

@INPROCEEDINGS{laing1994,
       author = {{Laing}, R.~A. and {Jenkins}, C.~R. and {Wall}, J.~V. and {Unger}, S.~W.},
        title = "{Spectrophotometry of a Complete Sample of 3CR Radio Sources: Implications for Unified Models}",
    booktitle = {The Physics of Active Galaxies},
         year = 1994,
       editor = {{Bicknell}, Geoffrey V. and {Dopita}, Michael A. and {Quinn}, Peter J.},
       series = {Astronomical Society of the Pacific Conference Series},
       volume = {54},
        month = jan,
        pages = {201},
       adsurl = {https://ui.adsabs.harvard.edu/abs/1994ASPC...54..201L}
}

@ARTICLE{laing2014,
       author = {{Laing}, R.~A. and {Bridle}, A.~H.},
        title = "{Systematic properties of decelerating relativistic jets in low-luminosity radio galaxies}",
      journal = {\mnras},
         year = 2014,
        month = feb,
       volume = {437},
       number = {4},
        pages = {3405-3441},
          doi = {10.1093/mnras/stt2138},
archivePrefix = {arXiv},
       eprint = {1311.1015},
 primaryClass = {astro-ph.CO},
       adsurl = {https://ui.adsabs.harvard.edu/abs/2014MNRAS.437.3405L}
}

@ARTICLE{Fanaroff1974,
       author = {{Fanaroff}, B.~L. and {Riley}, J.~M.},
        title = "{The morphology of extragalactic radio sources of high and low luminosity}",
      journal = {\mnras},
         year = 1974,
        month = may,
       volume = {167},
        pages = {31P-36P},
          doi = {10.1093/mnras/167.1.31P},
       adsurl = {https://ui.adsabs.harvard.edu/abs/1974MNRAS.167P..31F}
}

@ARTICLE{Ferreira2006b,
       author = {{Ferreira}, J. and {Dougados}, C. and {Cabrit}, S.},
        title = "{Which jet launching mechanism(s) in T Tauri stars?}",
      journal = {\aap},
         year = 2006,
        month = jul,
       volume = {453},
       number = {3},
        pages = {785-796},
          doi = {10.1051/0004-6361:20054231},
archivePrefix = {arXiv},
       eprint = {astro-ph/0604053},
 primaryClass = {astro-ph},
       adsurl = {https://ui.adsabs.harvard.edu/abs/2006A&A...453..785F}
}

@ARTICLE{coriat2011,
       author = {{Coriat}, M. and {Corbel}, S. and {Prat}, L. and {Miller-Jones}, J.~C.~A. and {Cseh}, D. and {Tzioumis}, A.~K. and {Brocksopp}, C. and {Rodriguez}, J. and {Fender}, R.~P. and {Sivakoff}, G.~R.},
        title = "{Radiatively efficient accreting black holes in the hard state: the case study of H1743-322}",
      journal = {\mnras},
         year = 2011,
        month = jun,
       volume = {414},
       number = {1},
        pages = {677-690},
          doi = {10.1111/j.1365-2966.2011.18433.x},
archivePrefix = {arXiv},
       eprint = {1101.5159},
 primaryClass = {astro-ph.HE},
       adsurl = {https://ui.adsabs.harvard.edu/abs/2011MNRAS.414..677C}
}

@ARTICLE{fender2014,
       author = {{Fender}, Rob and {Gallo}, Elena},
        title = "{An Overview of Jets and Outflows in Stellar Mass Black Holes}",
      journal = {\ssr},
         year = 2014,
        month = sep,
       volume = {183},
       number = {1-4},
        pages = {323-337},
          doi = {10.1007/s11214-014-0069-z},
archivePrefix = {arXiv},
       eprint = {1407.3674},
 primaryClass = {astro-ph.HE},
       adsurl = {https://ui.adsabs.harvard.edu/abs/2014SSRv..183..323F}
}

@ARTICLE{gallo2004,
       author = {{Gallo}, E. and {Corbel}, S. and {Fender}, R.~P. and {Maccarone}, T.~J. and {Tzioumis}, A.~K.},
        title = "{A transient large-scale relativistic radio jet from GX 339-4}",
      journal = {\mnras},
         year = 2004,
        month = jan,
       volume = {347},
       number = {3},
        pages = {L52-L56},
          doi = {10.1111/j.1365-2966.2004.07435.x},
archivePrefix = {arXiv},
       eprint = {astro-ph/0311452},
 primaryClass = {astro-ph},
       adsurl = {https://ui.adsabs.harvard.edu/abs/2004MNRAS.347L..52G}
}

@ARTICLE{merloni2003,
       author = {{Merloni}, Andrea and {Heinz}, Sebastian and {di Matteo}, Tiziana},
        title = "{A Fundamental Plane of black hole activity}",
      journal = {\mnras},
         year = 2003,
        month = nov,
       volume = {345},
       number = {4},
        pages = {1057-1076},
          doi = {10.1046/j.1365-2966.2003.07017.x},
archivePrefix = {arXiv},
       eprint = {astro-ph/0305261},
 primaryClass = {astro-ph},
       adsurl = {https://ui.adsabs.harvard.edu/abs/2003MNRAS.345.1057M}
}

@ARTICLE{bollen2017,
       author = {{Bollen}, Dylan and {Van Winckel}, Hans and {Kamath}, Devika},
        title = "{Jet creation in post-AGB binaries: the circum-companion accretion disk around BD+46{\textdegree}442}",
      journal = {\aap},
         year = 2017,
        month = nov,
       volume = {607},
          eid = {A60},
        pages = {A60},
          doi = {10.1051/0004-6361/201731493},
archivePrefix = {arXiv},
       eprint = {1708.00202},
 primaryClass = {astro-ph.SR},
       adsurl = {https://ui.adsabs.harvard.edu/abs/2017A&A...607A..60B}
}

@ARTICLE{tudor2017,
       author = {{Tudor}, V. and {Miller-Jones}, J.~C.~A. and {Patruno}, A. and {D'Angelo}, C.~R. and {Jonker}, P.~G. and {Russell}, D.~M. and {Russell}, T.~D. and {Bernardini}, F. and {Lewis}, F. and {Deller}, A.~T. and {Hessels}, J.~W.~T. and {Migliari}, S. and {Plotkin}, R.~M. and {Soria}, R. and {Wijnands}, R.},
        title = "{Disc-jet coupling in low-luminosity accreting neutron stars}",
      journal = {\mnras},
         year = 2017,
        month = sep,
       volume = {470},
       number = {1},
        pages = {324-339},
          doi = {10.1093/mnras/stx1168},
archivePrefix = {arXiv},
       eprint = {1705.05071},
 primaryClass = {astro-ph.HE},
       adsurl = {https://ui.adsabs.harvard.edu/abs/2017MNRAS.470..324T}
}

@INPROCEEDINGS{bally2007,
       author = {{Bally}, J. and {Reipurth}, B. and {Davis}, C.~J.},
        title = "{Observations of Jets and Outflows from Young Stars}",
    booktitle = {Protostars and Planets V},
         year = 2007,
       editor = {{Reipurth}, Bo and {Jewitt}, David and {Keil}, Klaus},
        month = jan,
        pages = {215},
       adsurl = {htt@ARTICLE{2009MNRAS.400..820T,
       author = {{Tzeferacos}, P. and {Ferrari}, A. and {Mignone}, A. and {Zanni}, C. and {Bodo}, G. and {Massaglia}, S.},
        title = "{On the magnetization of jet-launching discs}",
      journal = {\mnras},
     keywords = {accretion, accretion discs, MHD, methods: numerical, ISM: jets and outflows},
         year = 2009,
        month = dec,
       volume = {400},
       number = {2},
        pages = {820-834},
          doi = {10.1111/j.1365-2966.2009.15502.x},
       adsurl = {https://ui.adsabs.harvard.edu/abs/2009MNRAS.400..820T},
      adsnote = {Provided by the SAO/NASA Astrophysics Data System}
}

ps://ui.adsabs.harvard.edu/abs/2007prpl.conf..215B}
}

@INPROCEEDINGS{ray2007,
       author = {{Ray}, T. and {Dougados}, C. and {Bacciotti}, F. and {Eisl{\"o}ffel}, J. and {Chrysostomou}, A.},
        title = "{Toward Resolving the Outflow Engine: An Observational Perspective}",
    booktitle = {Protostars and Planets V},
         year = 2007,
       editor = {{Reipurth}, Bo and {Jewitt}, David and {Keil}, Klaus},
        month = jan,
        pages = {231},
          doi = {10.48550/arXiv.astro-ph/0605597},
archivePrefix = {arXiv},
       eprint = {astro-ph/0605597},
 primaryClass = {astro-ph},
       adsurl = {https://ui.adsabs.harvard.edu/abs/2007prpl.conf..231R}
}

@ARTICLE{fichetdeclairfontaine2022,
       author = {{Fichet de Clairfontaine}, G. and {Meliani}, Z. and {Zech}, A.},
        title = "{Flare echoes from relaxation shocks in perturbed relativistic jets}",
      journal = {\aap},
         year = 2022,
        month = may,
       volume = {661},
          eid = {A54},
        pages = {A54},
          doi = {10.1051/0004-6361/202243119},
archivePrefix = {arXiv},
       eprint = {2203.02765},
 primaryClass = {astro-ph.HE},
       adsurl = {https://ui.adsabs.harvard.edu/abs/2022A&A...661A..54F}
}

@ARTICLE{matsakos2009,
       author = {{Matsakos}, T. and {Massaglia}, S. and {Trussoni}, E. and {Tsinganos}, K. and {Vlahakis}, N. and {Sauty}, C. and {Mignone}, A.},
        title = "{Two-component jet simulations. II. Combining analytical disk and stellar MHD outflow solutions}",
      journal = {\aap},
         year = 2009,
        month = jul,
       volume = {502},
       number = {1},
        pages = {217-229},
          doi = {10.1051/0004-6361/200811046},
archivePrefix = {arXiv},
       eprint = {0905.3519},
 primaryClass = {astro-ph.GA},
       adsurl = {https://ui.adsabs.harvard.edu/abs/2009A&A...502..217M}
}

@ARTICLE{vlahakis2000,
       author = {{Vlahakis}, N. and {Tsinganos}, K. and {Sauty}, C. and {Trussoni}, E.},
        title = "{A disc-wind model with correct crossing of all magnetohydrodynamic critical surfaces}",
      journal = {\mnras},
         year = 2000,
        month = oct,
       volume = {318},
       number = {2},
        pages = {417-428},
          doi = {10.1046/j.1365-8711.2000.03703.x},
archivePrefix = {arXiv},
       eprint = {astro-ph/0005582},
 primaryClass = {astro-ph},
       adsurl = {https://ui.adsabs.harvard.edu/abs/2000MNRAS.318..417V}
}

@ARTICLE{asada2012,
       author = {{Asada}, Keiichi and {Nakamura}, Masanori},
        title = "{The Structure of the M87 Jet: A Transition from Parabolic to Conical Streamlines}",
      journal = {\apjl},
         year = 2012,
        month = feb,
       volume = {745},
       number = {2},
          eid = {L28},
        pages = {L28},
          doi = {10.1088/2041-8205/745/2/L28},
archivePrefix = {arXiv},
       eprint = {1110.1793},
 primaryClass = {astro-ph.HE},
       adsurl = {https://ui.adsabs.harvard.edu/abs/2012ApJ...745L..28A}
}

@ARTICLE{cheung2007,
       author = {{Cheung}, C.~C. and {Harris}, D.~E. and {Stawarz}, {\L}.},
        title = "{Superluminal Radio Features in the M87 Jet and the Site of Flaring TeV Gamma-Ray Emission}",
      journal = {\apjl},
         year = 2007,
        month = jul,
       volume = {663},
       number = {2},
        pages = {L65-L68},
          doi = {10.1086/520510},
archivePrefix = {arXiv},
       eprint = {0705.2448},
 primaryClass = {astro-ph},
       adsurl = {https://ui.adsabs.harvard.edu/abs/2007ApJ...663L..65C}
}

@ARTICLE{ramsey2011,
       author = {{Ramsey}, Jon P. and {Clarke}, David A.},
        title = "{Simulating Protostellar Jets Simultaneously at Launching and Observational Scales}",
      journal = {\apjl},
         year = 2011,
        month = feb,
       volume = {728},
       number = {1},
          eid = {L11},
        pages = {L11},
          doi = {10.1088/2041-8205/728/1/L11},
archivePrefix = {arXiv},
       eprint = {1012.3723},
 primaryClass = {astro-ph.SR},
       adsurl = {https://ui.adsabs.harvard.edu/abs/2011ApJ...728L..11R}
}

@ARTICLE{Krasnopolsky2003,
       author = {{Krasnopolsky}, Ruben and {Li}, Zhi-Yun and {Blandford}, Roger D.},
        title = "{Magnetocentrifugal Launching of Jets from Accretion Disks. II. Inner Disk-driven Winds}",
      journal = {\apj},
         year = 2003,
        month = oct,
       volume = {595},
       number = {2},
        pages = {631-642},
          doi = {10.1086/377494},
archivePrefix = {arXiv},
       eprint = {astro-ph/0306519},
 primaryClass = {astro-ph},
       adsurl = {https://ui.adsabs.harvard.edu/abs/2003ApJ...595..631K}
}

@ARTICLE{ouyed1999,
       author = {{Ouyed}, Rachid and {Pudritz}, Ralph E.},
        title = "{Numerical simulations of astrophysical jets from Keplerian discs - III. The effects of mass loading}",
      journal = {\mnras},
         year = 1999,
        month = oct,
       volume = {309},
       number = {1},
        pages = {233-244},
          doi = {10.1046/j.1365-8711.1999.02828.x},
       adsurl = {https://ui.adsabs.harvard.edu/abs/1999MNRAS.309..233O}
}

@article{ouyed1997a,
	title = {Numerical {Simulations} of {Astrophysical} {Jets} from {Keplerian} {Disks}. {II}. {Episodic} {Outflows}},
	volume = {484},
	issn = {0004-637X},
	url = {http://adsabs.harvard.edu/abs/1997ApJ...484..794O},
	doi = {10.1086/304355},
	urldate = {2021-01-15},
	journal = {The Astrophysical Journal},
	author = {Ouyed, Rachid and Pudritz, Ralph E.},
	month = jul,
	year = {1997},
	pages = {794--809},
}

@ARTICLE{Blandford1982,
       author = {{Blandford}, R.~D. and {Payne}, D.~G.},
        title = "{Hydromagnetic flows from accretion disks and the production of radio jets.}",
      journal = {\mnras},
         year = 1982,
        month = jun,
       volume = {199},
        pages = {883-903},
          doi = {10.1093/mnras/199.4.883},
       adsurl = {https://ui.adsabs.harvard.edu/abs/1982MNRAS.199..883B}
}

@ARTICLE{porth2015,
       author = {{Porth}, Oliver and {Komissarov}, Serguei S.},
        title = "{Causality and stability of cosmic jets}",
      journal = {\mnras},
         year = 2015,
        month = sep,
       volume = {452},
       number = {2},
        pages = {1089-1104},
          doi = {10.1093/mnras/stv1295},
archivePrefix = {arXiv},
       eprint = {1408.3318},
 primaryClass = {astro-ph.HE},
       adsurl = {https://ui.adsabs.harvard.edu/abs/2015MNRAS.452.1089P}
}

@ARTICLE{ramsey2019,
       author = {{Ramsey}, Jon P. and {Clarke}, David A.},
        title = "{MHD simulations of the formation and propagation of protostellar jets to observational length-scales}",
      journal = {\mnras},
         year = 2019,
        month = apr,
       volume = {484},
       number = {2},
        pages = {2364-2387},
          doi = {10.1093/mnras/stz116},
archivePrefix = {arXiv},
       eprint = {1901.02845},
 primaryClass = {astro-ph.SR},
       adsurl = {https://ui.adsabs.harvard.edu/abs/2019MNRAS.484.2364R}
}

@ARTICLE{agra-amboage2011,
       author = {{Agra-Amboage}, V. and {Dougados}, C. and {Cabrit}, S. and {Reunanen}, J.},
        title = "{Sub-arcsecond [Fe ii] spectro-imaging of the DG Tauri jet. Periodic bubbles and a dusty disk wind?}",
      journal = {\aap},
         year = 2011,
        month = aug,
       volume = {532},
          eid = {A59},
        pages = {A59},
          doi = {10.1051/0004-6361/201015886},
archivePrefix = {arXiv},
       eprint = {1106.2690},
 primaryClass = {astro-ph.SR},
       adsurl = {https://ui.adsabs.harvard.edu/abs/2011A&A...532A..59A}
}

@ARTICLE{guedel2008,
       author = {{G{\"u}del}, M. and {Skinner}, S.~L. and {Audard}, M. and {Briggs}, K.~R. and {Cabrit}, S.},
        title = "{Discovery of a bipolar X-ray jet from the T Tauri star DG Tauri}",
      journal = {\aap},
         year = 2008,
        month = feb,
       volume = {478},
       number = {3},
        pages = {797-807},
          doi = {10.1051/0004-6361:20078141},
archivePrefix = {arXiv},
       eprint = {0712.1330},
 primaryClass = {astro-ph},
       adsurl = {https://ui.adsabs.harvard.edu/abs/2008A&A...478..797G}
}

@ARTICLE{lee2017,
       author = {{Lee}, Chin-Fei and {Ho}, Paul. T.~P. and {Li}, Zhi-Yun and {Hirano}, Naomi and {Zhang}, Qizhou and {Shang}, Hsien},
        title = "{A rotating protostellar jet launched from the innermost disk of HH 212}",
      journal = {Nature Astronomy},
         year = 2017,
        month = jul,
       volume = {1},
          eid = {0152},
        pages = {0152},
          doi = {10.1038/s41550-017-0152},
archivePrefix = {arXiv},
       eprint = {1706.06343},
 primaryClass = {astro-ph.GA},
       adsurl = {https://ui.adsabs.harvard.edu/abs/2017NatAs...1E.152L}
}

@ARTICLE{bonito2011,
       author = {{Bonito}, R. and {Orlando}, S. and {Miceli}, M. and {Peres}, G. and {Micela}, G. and {Favata}, F.},
        title = "{X-Ray Emission from Protostellar Jet HH 154: The First Evidence of a Diamond Shock?}",
      journal = {\apj},
         year = 2011,
        month = aug,
       volume = {737},
       number = {2},
          eid = {54},
        pages = {54},
          doi = {10.1088/0004-637X/737/2/54},
archivePrefix = {arXiv},
       eprint = {1105.4081},
 primaryClass = {astro-ph.SR},
       adsurl = {https://ui.adsabs.harvard.edu/abs/2011ApJ...737...54B}
}

@ARTICLE{lister2009,
       author = {{Lister}, M.~L. and {Cohen}, M.~H. and {Homan}, D.~C. and {Kadler}, M. and {Kellermann}, K.~I. and {Kovalev}, Y.~Y. and {Ros}, E. and {Savolainen}, T. and {Zensus}, J.~A.},
        title = "{MOJAVE: Monitoring of Jets in Active Galactic Nuclei with VLBA Experiments. VI. Kinematics Analysis of a Complete Sample of Blazar Jets}",
      journal = {\aj},
         year = 2009,
        month = dec,
       volume = {138},
       number = {6},
        pages = {1874-1892},
          doi = {10.1088/0004-6256/138/6/1874},
archivePrefix = {arXiv},
       eprint = {0909.5100},
 primaryClass = {astro-ph.CO},
       adsurl = {https://ui.adsabs.harvard.edu/abs/2009AJ....138.1874L}
}

@ARTICLE{tabone2018,
       author = {{Tabone}, B. and {Raga}, A. and {Cabrit}, S. and {Pineau des For{\^e}ts}, G.},
        title = "{Interaction between a pulsating jet and a surrounding disk wind. A hydrodynamical perspective}",
      journal = {\aap},
         year = 2018,
        month = jun,
       volume = {614},
          eid = {A119},
        pages = {A119},
          doi = {10.1051/0004-6361/201732031},
archivePrefix = {arXiv},
       eprint = {1801.03521},
 primaryClass = {astro-ph.SR},
       adsurl = {https://ui.adsabs.harvard.edu/abs/2018A&A...614A.119T}
}

@ARTICLE{ray2021,
       author = {{Ray}, T.~P. and {Ferreira}, J.},
        title = "{Jets from young stars}",
      journal = {\nar},
         year = 2021,
        month = dec,
       volume = {93},
          eid = {101615},
        pages = {101615},
          doi = {10.1016/j.newar.2021.101615},
archivePrefix = {arXiv},
       eprint = {2009.00547},
 primaryClass = {astro-ph.SR},
       adsurl = {https://ui.adsabs.harvard.edu/abs/2021NewAR..9301615R}
}

@ARTICLE{mignone2007,
       author = {{Mignone}, A. and {Bodo}, G. and {Massaglia}, S. and {Matsakos}, T. and {Tesileanu}, O. and {Zanni}, C. and {Ferrari}, A.},
        title = "{PLUTO: A Numerical Code for Computational Astrophysics}",
      journal = {\apjs},
         year = 2007,
        month = may,
       volume = {170},
       number = {1},
        pages = {228-242},
          doi = {10.1086/513316},
archivePrefix = {arXiv},
       eprint = {astro-ph/0701854},
 primaryClass = {astro-ph},
       adsurl = {https://ui.adsabs.harvard.edu/abs/2007ApJS..170..228M}
}

@ARTICLE{ferreira2022,
       author = {{Ferreira}, J. and {Marcel}, G. and {Petrucci}, P.-O. and {Rodriguez}, J. and {Malzac}, J. and {Belmont}, R. and {Clavel}, M. and {Henri}, G. and {Corbel}, S. and {Coriat}, M.},
        title = "{Are low-frequency quasi-periodic oscillations in accretion flows the disk response to jet instability?}",
      journal = {\aap},
         year = 2022,
        month = apr,
       volume = {660},
          eid = {A66},
        pages = {A66},
          doi = {10.1051/0004-6361/202040165},
archivePrefix = {arXiv},
       eprint = {2202.00438},
 primaryClass = {astro-ph.HE},
       adsurl = {https://ui.adsabs.harvard.edu/abs/2022A&A...660A..66F}
}

@ARTICLE{bromberg2016,
       author = {{Bromberg}, Omer and {Tchekhovskoy}, Alexander},
        title = "{Relativistic MHD simulations of core-collapse GRB jets: 3D instabilities and magnetic dissipation}",
      journal = {\mnras},
         year = 2016,
        month = feb,
       volume = {456},
       number = {2},
        pages = {1739-1760},
          doi = {10.1093/mnras/stv2591},
archivePrefix = {arXiv},
       eprint = {1508.02721},
 primaryClass = {astro-ph.HE},
       adsurl = {https://ui.adsabs.harvard.edu/abs/2016MNRAS.456.1739B}
}

\bsp	% typesetting comment
\label{lastpage}
\end{document}